\documentclass{article}

\usepackage[english]{babel}

\usepackage[letterpaper,top=2cm,bottom=2cm,left=3cm,right=3cm,marginparwidth=1.75cm]{geometry}

\usepackage{amsmath}
\usepackage{amssymb}
\usepackage{eurosym}
\usepackage{graphicx}
\usepackage[table]{xcolor}
\usepackage[colorlinks=true, allcolors=blue]{hyperref}
\usepackage{blindtext}
\usepackage[authoryear]{natbib}
\usepackage{float}
\usepackage{longtable}
\usepackage[table]{xcolor}
\usepackage{array}
\newcolumntype{L}[1]{>{\raggedright\arraybackslash}p{#1}}
\usepackage{enumitem}
\usepackage{booktabs}
\usepackage{xcolor}
\definecolor{DarkGreen}{RGB}{0,100,0} 
\usepackage{multirow}
\usepackage{array}
\usepackage{caption}
\usepackage{placeins}
\usepackage{geometry}
\usepackage{tabularx}
\usepackage{makecell}
\usepackage{float}
\usepackage{tabularx}
\usepackage{threeparttable}
\usepackage{array}
\usepackage{soul}
\usepackage{enumitem}
\title{The Impact of Renewable Energy Communities in the Italian Day-Ahead Electricity Market: A Scenario Analysis}
\author{
Maksym Koltunov\thanks{\hspace{0.2cm} Department of Economics, Business, Mathematics and Statistics "Bruno de Finetti" (DEAMS), University of Trieste, maksym.koltunov@phd.units.it}
\and Filippo Beltrami\thanks{\hspace{0.2cm} Corresponding author. Institute for Renewable Energy, Eurac Research,  filippo.beltrami@eurac.edu}
\and Luigi Grossi\thanks{\hspace{0.2cm} Department of Engineering for Industrial Systems and Technologies and Robust Statistics Academy, University of Parma, luigi.grossi@unipr.it}
\and Nicola Blasuttigh\thanks{\hspace{0.2cm} Department of Engineering and Architecture, and Center for
Energy, Environment and Transport Giacomo Ciamician, University of Trieste, nicola.blasuttigh@units.it}
}

\begin{document}
\maketitle
\begin{abstract}
This paper evaluates the economic impact of Renewable Energy Communities (RECs) on the Italian wholesale power market. Combining a bottom-up engineering approach with a short-run economic assessment, the study begins by mapping existing and emerging RECs in Italy. Key characteristics are identified, including installed capacity, institutional composition of members, types of renewable technologies, and geographical distribution across market zones. This mapping yields representative REC configurations, which are employed within a bottom-up engineering model to simulate energy injection and self-consumption profiles for different categories of members (residential, public, SMEs, non-profits, and standalone installations), accounting for variations in solar irradiance across Italian regions. These zonal results, aggregated hourly, feed into a synthetic counterfactual framework used to construct alternative deployment scenarios, including the 5 GW policy target for 2027, to assess the impact of RECs on the day-ahead power market. The results show that REC deployment tends to increase equilibrium quantities during daylight hours, while reducing them mostly in colder months. On average, market volumes vary within a range of -0.3\% to +1.16\%. The reduction effect is largely driven by the distinct load profiles of institutional members compared to residential users and becomes more pronounced as installed capacity increases. Overall, the coexistence of positive and negative effects suggests that REC deployment may contribute to lowering wholesale electricity prices.  Finally, results indicate that including standalone installations within heterogeneous REC configurations can enhance overall system performance.
\end{abstract}

\vskip 0.5 cm

\noindent \textit{Keywords: Energy Communities; Energy Policy; Energy Transition; Italian Electricity Market; Photovoltaic; Renewable Sources}
\vskip 0.5 cm
\noindent \textit{JEL Classification: Q42, Q47, Q55, Q56}
\vskip 0.5 cm
\noindent \textit{Word count: 13118}

\vskip 1cm

\normalsize
\pagebreak
\pagenumbering{arabic}
\section{Introduction}
\label{sec:intro} The current integration of Renewable Energy
Communities (RECs)\footnote{\nameref{sec: Appendix A} contains the
full list of acronyms used in this study.} in national electricity
systems has made it necessary to accurately describe its market
consequences. Depending on the historical path and national
legislation, energy communities could act on the market as utilities
do, thus having the opportunity to arbitrate between different
market prices. \textcolor{black}{Although regulatory developments
are progressively enabling  such direct participation
\textcolor{black}{(e.g., bidding)} of RECs and aggregators in
wholesale markets, these interactions are still evolving and not yet
fully established across all contexts.} Moreover, RECs typically
include a small number of users \textcolor{black} {who do not own
sufficient generating capacity to directly participate in wholesale
markets}. In such cases, members of RECs should be supplied by
external \textcolor{black}{providers} and considered as simple users
(or clients) of such utilities and/or external aggregators. These
users may exhibit various characteristics, acting as consumers,
prosumers, or prosumagers, thus \textcolor{black}{indirectly}
affecting the day-ahead market. \textcolor{black} {In the case of
Italy, energy injected into the grid by RECs is mostly reflected in
supply offers submitted by the GSE\footnote{GSE (in Italian,
\textit{Gestore dei Servizi Energetici}) is the energy system
manager, a public company that manages incentives in the energy
sector on the state’s behalf.} at a zero
price.}\footnote{\textcolor{black} {Supply offers from small power
plants below 10 MW (e.g., rooftop systems) are deemed non-relevant
by the market operator and are therefore included in the GSE
aggregated offer at a zero price. Nevertheless, the regulation
allows RECs and ESCOs/traders (acting on behalf of RECs) to submit
energy offers independently from GSE. However, at present, it is
practiced only in a very limited number of cases in Italy. The vast
majority of supply offers derived from REC injected energy is
submitted by the GSE (senior official from GME, personal
communication, March 2026).}}
\textcolor{black} {In turn, the reduction in loads due to self-consumption by RECs is reflected in demand bids of various load-serving entities that supply electricity to REC members.} Understanding the composition and behaviour of prosumers within RECs - including their seasonal and geographical heterogeneity - is essential for assessing how these configurations reshape wholesale \textcolor{black}{supply-demand} market signals and could reduce user dependence on the grid. Such insights are critical for designing policies that generate benefits not only for REC members but also for the broader energy system.\\

RECs are undoubtedly considered a unique instrument for a just energy transition. There are two core features that define a REC:
\begin{itemize}
    \item RECs are entities that should provide social, environmental and economic benefits to their members and respective local communities rather than solely for financial gain.
    \item RECs should be effectively controlled by shareholders or members who are located in close proximity to the renewable energy projects owned and developed by the REC’s legal entity.
\end{itemize}

Table \ref{tab:taxonomy_economic_impacts} depicts a taxonomy of economic impacts of RECs. These can be various, ranging from effects on members and investors, local and regional economies, and on the electricity market and its stakeholders.

The pre-Directives literature \citep{Brummer2017, Candelise2017,
heras-saizarbitoriaEmergenceRenewableEnergy2018,
holstenkampWhatAreCommunity2016, K.Huntala2016,
koltunovMultipleImpactsEnergy2021, Kooij2018, Magnani2016,
Magnusson2019, tricaricoCommunityEnergyEnterprises2018, Vernay2020,
Wirth2014, Walker2010, Berka2018, moroniLocalEnergyCommunities2019}
mainly explored micro-scale socio-economic impacts on REC members
with a specific focus on domains of management and sociology, while
rarely examining the meso\footnote{Pre-Directives, there are only
few economic studies with those exploring the impact of RECs on
regional economies using input-output models \citep{Lantz2011,
Okkonen2016, Torgerson200603, Phimister2012, Allan2011, Bere2015,
Entwistle2014}} and macro effects of RECs on the economy and
markets.

In contrast to the multiple benefits at the organizational and local levels, REC deployment could lead to adverse effects from an economic perspective at macro level. For example, the extra injection of intermittent renewable energy sources (RES) could cause higher ramp/start events of expensive peaking plants. An increase in distribution grid charges due to RECs in the presence of volumetric tariffs could adversely affect households non-participating in RECs. Post-Directives, several economic empirical studies have appeared on the effect of RECs on the electricity system \citep{backeImpactEnergyCommunities2022, sarfaraziAggregationHouseholdsCommunity2020, fuentesgonzalezCommunityEnergyProjects2022, disilvestreEnergySelfconsumersRenewable2021, boccardPowerTradingEnergy2025}. However, most post-Directive studies focus on the analysis of peer-to-peer trading, an innovative feature of REC that is yet rarely encountered in real-world applications\footnote{More details on RECs classifications can be found in Rossetto et al. \citeyearpar{rossetto1TaxonomyEnergy2022}, Kolesar \citeyearpar{kolesar7EnergyCommunities2022}, Koltunov et al. \citeyearpar{koltunovENERGYCOMMUNITIESINSTITUTIONS2025}} \citep{castelliniEnergyExchangeHeterogeneous2021, chenNovelModelRural2024, dongIncentivizingSustainablePractices2024, glachantNewTransactionsElectricity2021, hahnelBecomingProsumerRevealing2020, hahnelPricingDecisionsPeertopeer2022, nieto-martinCommunityEnergyRetail2019, sousaPeertopeerCommunitybasedMarkets2019}.\\

\begin{table}[h]
\caption{Taxonomy of economic impacts by RECs. Source: \citet{koltunovEnergyCommunitiesSocial2025}.}
\centering
\begin{tabular}{cccc}
    \hline
       Category & Impact on members  & Impact on local & Impact on market stakeholders
        \\
        of impact & and investors  & and regional economies &
         and the electricity system
        \\
        \hline
        \hline
         \textit{Scale} & Micro & Meso & Macro\\
        \hline
         \textit{Subject} & One EC  & Multiple ECs & All RECs deployed \\
          &   &  & at a country level\\
        \hline
         \textit{Objects} & Individual members, & Local and & Generators, retailers, \\
          & investors (if not a member) & regional economies  & DSOs, aggregators, \\
          & &  & non-member consumers, \\
          & &  & ESCOs, technology providers \\
    \hline
    \end{tabular}
    \label{tab:taxonomy_economic_impacts}
\end{table}

The macro impact of REC deployment on electricity systems remains an underexplored domain \citep{koltunovEnergyCommunitiesSocial2025}. Robinson and del Guayo \citeyearpar{robinson5AlignmentEnergy2022} propose two categories: systemic impact and stakeholder impact. The first category explores the impact of RECs on the electricity systems overall, which entails multiple spillovers on the operations of various actors. Examples of such systemic impacts include changes in distribution charges by Distribution System Operators (DSOs) \citep{BERG2024123060}, transmission charges by Transmission System Operators (TSOs), distribution and transmission system expansion, wholesale prices, changes in collected taxes and levies embedded in electricity tariffs, and market competitiveness. The second category examines the impact of RECs on specific stakeholder groups, such as non-members, retailers, generators, DSOs, aggregators, and technology providers. The present article belongs to the first category, as we focus on the merit-order effect and the impact of RECs on the Italian day-ahead (DA) power market equilibrium, which entails multiple spillovers on the operations of various stakeholders.\\

In terms of business model archetype, Italian RECs can be classified as a ``community collective generation''.\footnote{In this type of REC, generation facilities must be connected to the same voltage substation and all members should live in its proximity, while individual members (households/SMEs/public entities/non-profit organizations) maintain their own retailers that take care of their residual demand. This type of REC is the one that is mostly adopted in Italy.} This type of REC includes prosumers, consumers, and producers (i.e., the collective generation facility). During periods of excess of generation, RECs supply energy to the grid, thereby increasing renewable dispatch on the supply-side of the wholesale market. At the same time, these RECs aim for higher self-consumption, especially when the selling price is lower than the retail price, thereby decreasing the aggregated load on the demand-side of the market.

For the scope of this paper, we state our research question as follows:
\begin{itemize}[noitemsep]
\item{What is the impact of REC deployment on the Italian wholesale day-ahead market equilibrium?}
\end{itemize}

\begin{figure}[h]
\centering
\includegraphics[width=1\linewidth]{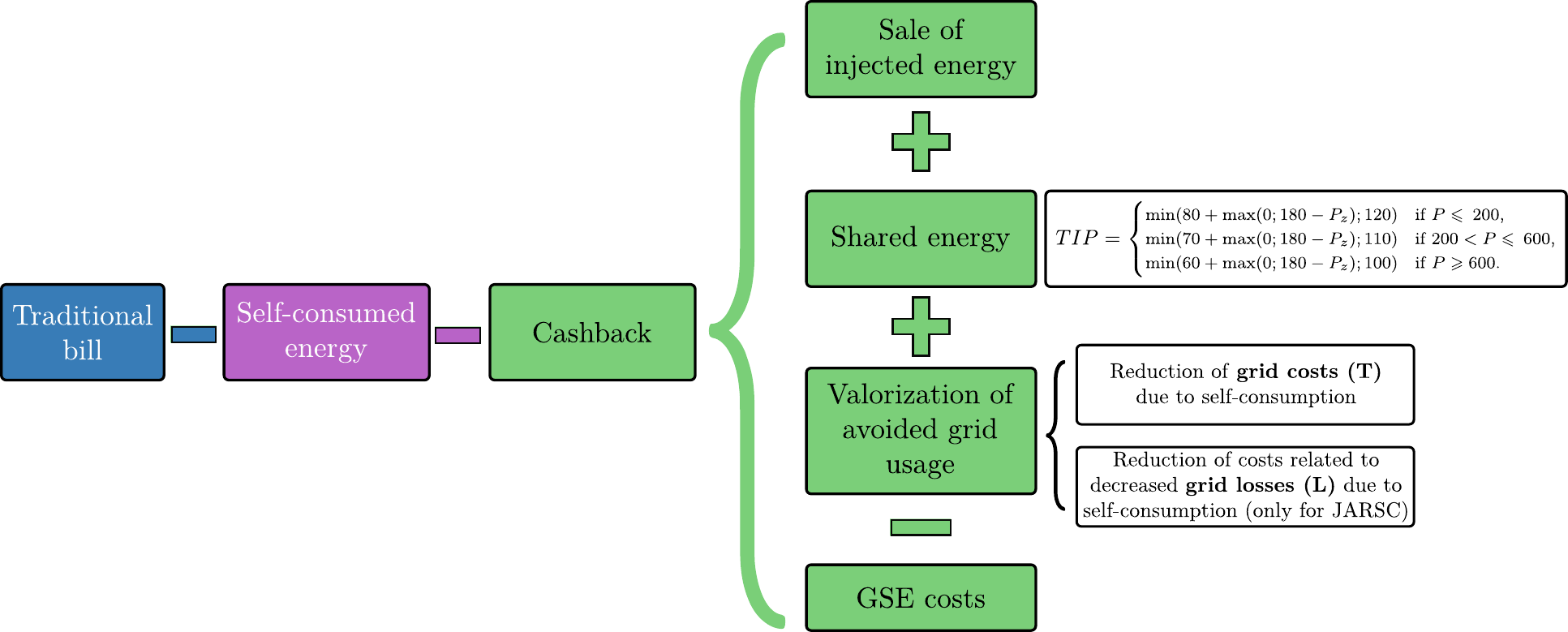}
\caption{A net cash flow for REC members in Italy (excl. explicit capital subsidy).}
\label{fig:scheme_benefit_CER}
\end{figure}

Europe has the largest number of RECs, with almost 4,000 initiatives and 900,000 members \citep{koltunov_mapping_2023}. European RECs deserve particular attention because of the scale of the phenomenon and their formal recognition at the EU level by the RED-II \citep{EU2018_REDII} and IEM Directives \citep{EU2019_IEM_Directive}, which triggered their further growth. 
Within Europe, we selected Italian RECs as a case study for several reasons. First, they have shown substantial growth during last few years.\footnote{In 2023, there were only 50 RECs aligned with new regulation in Italy \citep{koltunov_mapping_2023}.} Currently, there are around 362\footnote{Data includes RECs in both operational and design phases as of January 2025.} new RECs in Italy and the number continues to grow. The primary factors behind this notable deployment include the nuanced policy that allowed \textcolor{black}{new} institutional actors to participate in RECs\footnote{Historically, RECs were considered citizen-driven initiatives, which limited their deployment.} \textcolor{black}{and subsidies, specifically the CAPEX grant and}  the premium tariff. The latter is granted to REC plants of up to 1 MW capacity \textcolor{black}{if they are located within the boundaries of the same high/medium (primary) voltage substation} \citep{DecretoCACER2023}. 
Italy allocated €5.7 billion to support 5GW REC deployment: €3.5 billion for premium tariffs (financed via a levy on electricity consumption) and €2.2 billion from the National Recovery and Resilience Fund to cover up to 40\% of CAPEX in small \textcolor{black}{and medium} \footnote{Before July 2025, the capital contribution had been granted exclusively to small municipalities with up to 5,000 inhabitants. After July 2025, the threshold was increased to 50,000.} municipalities. These figures highlight the strategic importance of assessing the effectiveness and system-wide value of RECs. Second, the Italian wholesale day-ahead power market is well-suited for our research because of its transparency regarding hourly supply and demand bids placed by market operators. Third, Italian regulation is rather innovative and can be characterized as a “virtual scheme”, where REC members (prosumers and consumers) do not physically share energy in a microgrid.\footnote{In contrast, CEER \citeyearpar{ceerRegulatoryAspectsSelfConsumption2019} reports that “Some Member States, such as France and Austria, have developed a framework for collective self-consumption, where energy can be shared within a group of customers, without requiring direct involvement of a supplier." Finland adopted a concept of energy sharing that is typically limited to apartment blocks or single housing associations. Similar to Finland, German “collective self-consumers” concept relates largely to occupants of the same building or small groups rather than a rule allowing multi-building or multi-substation virtual sharing across a distribution grid.} Instead, they retain their existing retail contracts while sharing energy virtually, for which they are remunerated at the end of the year by the GSE  \citep{schiavo6VirtualModel2022}. This innovative regulation provides an interesting case for investigating its implications for the DA wholesale market. According to new rules \citep{TIAD2022}, REC members can benefit in four ways: energy self-consumption; a premium tariff for shared energy; valorization of avoided grid usage – calculated based on shared energy due to connection to the same primary substation; and energy sales \citep{blasuttigh_optimal_2025}. Figure \ref{fig:scheme_benefit_CER} illustrates the benefits while \nameref{sec: Appendix B} explains the formulas behind the cashback components. \\

Notably, only the sales of injected energy (by prosumers and producers) and the self-consumed energy (by prosumers) affect the wholesale market, with the former impacting the supply-side and the latter the demand-side. The incentivized ``shared energy'' is merely a virtual concept that does not directly trigger wholesale market changes. Therefore, we analyze only the operations of REC members that directly influence the market – namely, prosumers and producers – and disregard consumers.\footnote{When joining a REC, consumers do not necessarily change their consumption profile in the ``virtual scheme'' applied in Italy, therefore the wholesale market outcome does not amend because of them. Nonetheless, a REC manager/aggregator might advice consumers to change consumption in certain hours for the maximization of the common shared energy. Such advanced coordination requires smart infrastructure at the consumers' premises and could potentially affect the wholesale market situation. However, to the authors’ best knowledge, no RECs in Italy have yet employed such level of coordination. Therefore, this aspect lies beyond the scope of the current study and remains for further investigation.} From this perspective, our study pertains to the literature on the impact of prosumers on wholesale markets, although using Italian RECs as a specific case.\\

We apply an innovative mixed-model methodology. In the first stage of analysis, we design a bottom-up engineering model that simulates the hourly behavior of various typical prosumers and producers within a REC. Then, we project the model’s results onto several scenarios, which are based on the policy target for 2027 and on an actual REC deployment status in 2024.\footnote{\textcolor{black}{Our analysis relies on a quasi-experimental setting using 2024 as the reference year, as it represents the first full year following the implementation of the REC regulatory framework in Italy. The year 2027 is not used for forecasting purposes, but rather as a policy-based benchmark to calibrate alternative REC deployment "scenarios".}} The outcome of the first stage is then inserted into our synthetic model that simulates short-run schedule for the DA market on an hourly basis, from which we derive the effect of RECs on the merit order and aggregated demand.\\

\textcolor{black}{The Italian electricity market has recently (October 1st, 2025) transitioned to a 15-minute settlement frequency. This institutional evolution represents a relevant structural change that may affect the applicability of our findings in forward-looking contexts. The new temporal resolution could, in principle, lead to different market outcomes by more accurately capturing intra-hour variability in generation and demand, particularly in the presence of increasing shares of renewable energy and self-consumption. \textcolor{black}{For example, studies of other markets have shown that 15-minute bidding can increase the volume of intermittent renewable energy accepted in economic dispatch, especially during the ramp-up and ramp-down phases of solar power generation \citep{markle_huss2018contract} as well as smoothen the demand curve \citep{soonee2016subhourly}}. In addition, the shift to finer settlement intervals may induce changes in bidding strategies and, consequently, in the shape of the supply merit-order curve. Nevertheless, given the recent implementation of this reform in Italy, its empirical implications remain uncertain, as sufficiently detailed and consolidated data are not yet available, and market participants’ behavioral adjustments are still evolving.
The objective of this study is not to reproduce high-frequency dynamics or to model sub-hourly energy flows. Rather, the analysis focuses on the aggregate effects of RECs on wholesale market equilibria, with particular attention to seasonal and zonal heterogeneity. Accordingly, \textcolor{black}{hourly} results are presented at an aggregated level (seasonal and annual), which is consistent with the intended policy-oriented perspective.}\\

Hence, the novelty of our work is twofold. First, we provide a theoretical contribution to the understanding of the seasonal and hourly impact of prosumers with different load profiles on the wholesale power market. We take Italian RECs as a specific case study. The heterogeneous membership within RECs becomes typical in many EU countries, especially after the implementation of the RED-II and IEM Directives. However, the existing literature on modeling prosumers' behaviour mostly simulates the impacts of residential prosumers at a single geographical location (see review in \nameref{Appendix D}), whereas our study simulates RECs as a collection of prosumers with very different consumption profiles (public, residential, small and medium enterprises, non-profit organizations) and standalone producers at various climatic locations, which closely reflects the country-scale REC deployment situation. Our contribution is tested empirically for the entire year 2024 in the DA Italian pool using a synthetic approach to simulate real-world competition among market agents. By undertaking such a detailed modeling exercise, our theoretical contribution may achieve higher external validity, especially for jurisdictions and markets with similar technical definition of a REC (e.g., Spain, Greece, Czech Republic), climate conditions and generation profiles (e.g., Spain, Greece).\\

Second, we provide a novel mixed-model approach to assess the country-scale impact of prosumers and producers aggregated into a REC on wholesale market scheduling. The first stage of our methodology begins with detailed data collection, which incorporates information on the composition of prosumers within RECs, the capacities of generating plants, the load profiles of different categories of prosumers, and the geographical distribution of RECs. Using these real-world parameters, we design the engineering model to obtain the assumed prosumers’ generation and load profiles, which we then project to the policy-targeted deployment level and other predicted deployment levels. In the second stage, the methodology is expanded by incorporating the previous outputs into an economic synthetic model. In essence, our methodological approach can be used by policymakers and stakeholders to determine the actual and predicted effects of the growing presence of heterogeneous prosumers, including, but not limited to, those arranged within RECs, on wholesale markets. \textcolor{black}{Indeed, a key challenge in assessing the impact of RECs lies in disentangling their effects from concurrent macroeconomic and market-wide drivers. Electricity market outcomes are simultaneously influenced by multiple factors, including fuel price dynamics, demand fluctuations, weather conditions, and broader economic trends. Isolating the specific contribution of RECs is therefore complex.
In this paper, the analysis is conducted within a controlled microeconomic simulation framework that evaluates the effects of REC penetration under \textit{ceteris paribus} conditions. By construction, this approach does not aim to fully capture the joint influence of all relevant macroeconomic variables. Rather, it focuses on identifying the magnitude and direction of changes in wholesale market outcomes attributable to RECs, holding other factors constant.} \\

The remainder of the article is organized as follows. In Section \ref{sec:litrev}, we explore the existing literature on systemic impacts of RECs and studies that utilized the "synthetic approach” to modeling electricity markets. Section \ref{sec:methdata} explains the novel methodology employed in this paper, while Section \ref{sec:results} displays the major results. Section \ref{sec:disc} incorporates the results into the academic debate. Section \ref{sec:concl} concludes with final remarks and policy recommendations. \\

\section{Literature review}\label{sec:litrev}

This Section reviews the relevant literature on the topic, by exploring two main areas\footnote{We also discuss the studies that analyze the effect of prosumers on electricity systems in Section \ref{sec:disc}. However, since they do not constitute the central role to our paper, unlike studies that focus on the effect of REC deployment on the system, we omit reviewing them in the main text although including them in \nameref{Appendix D}.}:

\begin{enumerate}
    \item Studies that analyse the effect of RECs on electricity systems;
    \item Studies that employ a counterfactual approach for modeling the impact of specific power generation sources (e.g., RES) on the wholesale power market.
\end{enumerate}

\subsection{Impact of RECs on electricity systems}

Most studies use a narrative-based theoretical approach to discuss the systemic impacts of RECs \citep{biggar8EnergyCommunities2022, delpizzo18ItalianEnergy2022, paragElectricityMarketDesign2016a, robinson5AlignmentEnergy2022, disilvestreEnergySelfconsumersRenewable2021} and only few empirical economic studies exist \citep{backeImpactEnergyCommunities2022, sarfaraziAggregationHouseholdsCommunity2020, fuentesgonzalezCommunityEnergyProjects2022, disilvestreEnergySelfconsumersRenewable2021, boccardPowerTradingEnergy2025} due to the novelty and previously small scale of the phenomenon.  \textcolor{black}{Table \ref{tab:empirical-studies-recs} outlines key findings from both theoretical and available empirical literature. The theoretical studies suggest that RECs can improve system efficiency by stimulating embedded generation, reducing grid losses and transmission needs when sited in under-supplied areas, supporting ancillary services through aggregation and storage, enhancing energy efficiency efforts, and creating social and competitive benefits. However, the same literature warns that RECs can also increase system costs by weakening efficient tariff signals, contributing to network operators’ revenue decline, raising congestion and losses when located in over-supplied areas, and, in some cases, worsening privacy risks. As we see, the theoretical literature points to multiple potential benefits and detriments to the system and stakeholders\footnote{Since our study aims to analyze systemic impacts, we skip the literature review of impacts on the individual stakeholders in the main text. Instead, we include the review of these impacts in \nameref{sec:Appendix C}.} from the deployment of RECs. However, these potential benefits and detriments relate to a variety of REC types, and not all are necessarily applicable to the ‘community collective generation’ type deployed in Italy.\footnote{Koltunov \citeyearpar{Koltunov2025impactofREConElectricitySystem} can be consulted for an in-depth discussion on the impacts of individual REC business models on the electricity system and its stakeholders.} Nevertheless, the theoretical arguments have yet to be empirically validated. From Table \ref{tab:empirical-studies-recs}, we also observe that only a very scant number of empirical works perform this task. These studies found potentially ambiguous impacts of RECs on wholesale market quantities. Specifically, RECs may decrease or increase market quantities, depending on specific technology choices, the objectives of REC managers and retailers, and a business model chosen. In turn, the findings indicate that market prices decrease in systems with RECs.} \\

\textcolor{black}{Despite relevant findings, previous empirical studies have very limited external validity regarding impacts on wholesale markets due to objectives and scope that differ from ours. They either investigate a small number of nodes using hypothetical data, model a single REC rather than the deployed REC movement as a whole, or simulate only a very short time period. For example, quantities demanded by RECs in the static model of Fuentes Gonzalez et al. \citeyearpar{fuentesgonzalezCommunityEnergyProjects2022} are simulated for a single representative hour while prices are simply derived from linear demand functions. Consequently, this approach does not take into consideration a full interplay of technologies and variability over multiple time periods that are inherent in pay-as-clear zonal wholesale auctions of the real world. Sarfarazi et al. \citeyearpar{sarfaraziAggregationHouseholdsCommunity2020} focus primarily on a ``REC-retailer'' interactions on a micro scale and do not investigate REC deployment on a macro scale, merit-order effect (MOE) and complexities of a DA wholesale market. Boccard and Goetz \citeyearpar{boccardPowerTradingEnergy2025} investigate power exchanges of an individual REC with the grid and not impact on the wholesale market of many RECs. Zepter et al. \citeyearpar{zepter2019prosumer}  do not investigate the potential effect of RECs on market equilibrium as well as modeling only one week. Backe et al. \citeyearpar{backeImpactEnergyCommunities2022} do not quantify the MOE or REC interaction with the DA market, openly stating that this is beyond the scope of their paper. Hence, our study is the first to primarily investigate the impact of REC deployment on the equilibrium of the DA market. Moreover, we use real-world data on REC deployment and the bids of other market participants while simulating an entire year, allowing us to identify seasonal patterns. Unlike previous studies, we include standalone PV plants and institutional prosumers in our REC representation, thereby accurately reflecting the actual REC deployment situation. Taken together, these factors enable our study to achieve greater external validity than previous works.}\\

{
\color{black}
\arrayrulecolor{black}
\footnotesize
\setlength{\tabcolsep}{1.5pt}
\setlength{\LTpost}{0pt}
\renewcommand{\arraystretch}{1.02}
\begin{longtable}{@{}L{1.7cm}L{1.5cm}L{2cm}L{9.8cm}L{1.7cm}@{}}
\caption{\textcolor{black}{Theoretical and empirical studies on systemic impact of RECs.}}
\label{tab:empirical-studies-recs}\\
\toprule
\parbox[c]{1.7cm}{\textbf{Study}} &
\parbox[c]{1.5cm}{\textbf{Scope}} &
\parbox[c]{2.1cm}{\textbf{Methodology}} &
\parbox[c]{9.6cm}{\centering\textbf{Key findings}} &
\parbox[c]{1.6cm}{\textbf{Impact on\\markets studied?}} \\
\midrule
\endfirsthead
\multicolumn{5}{@{}l}{\textit{Table \thetable\ continued from previous page}}\\
\toprule
\textbf{Study} &
\textbf{Scope} &
\textbf{Methodology} &
\multicolumn{1}{c}{\textbf{Key findings}} &
\textbf{Impact on markets studied?} \\
\midrule
\endhead
\midrule
\multicolumn{5}{r@{}}{\textit{Continued on next page}}\\
\endfoot
\bottomrule
\endlastfoot
\multicolumn{5}{c}{\textit{Theoretical studies}} \\
\midrule
Biggar and Hesamzadeh \citeyearpar{biggar8EnergyCommunities2022} &
Not specified &
Theoretical (narrative) &
\textbf{Benefits:} Encouraging investment in embedded generation on a larger scale, including RECs, increases economic efficiency; Multiple social benefits derive from RECs.\newline
\textbf{Detriments:} A REC may arbitrate\textsuperscript{a} between retail and selling prices when acting as a unified entity purchasing part of its electricity from a single retailer, thus the effect of non-uniform tariffs (time-of-use and locational tariffs) is weakened, potentially increasing system operation costs and leading to further network expansion; When customers unite into RECs it may further amplify the impact of inefficient tariffs, wherein retail customers do not typically face the correct incentives to use energy according to market and grid situations; The reduction in network exploitation exacerbates network operator’s ‘death spiral’ effect\textsuperscript{b} &
Yes \\
\midrule
Robinson and Del Guayo \citeyearpar{robinson5AlignmentEnergy2022} &
Spain &
Theoretical (narrative) &
\textbf{Benefits:} Locating RECs in areas with under-supply of electricity reduces transmission grid costs and losses; RECs reduce the need for additional large-scale generation due to the merit-order effect, thereby lowering wholesale prices and transmission infrastructure needs; RECs can provide ancillary services when aggregated into virtual power plants, replacing more expensive conventional sources; multiple social benefits derive from RECs; RECs increase competitive pressure on other market agents.\newline
\textbf{Detriments:} Locating RECs in areas with over-supply of electricity increases transmission grid costs and losses due to greater electricity injections where congestion bottlenecks exist; Small-capacity renewable generators are costlier for the system because their LCOE is generally higher than that of large-capacity renewable generators. &
Yes \\
\midrule
Parag and Sovacool \citeyearpar{paragElectricityMarketDesign2016a} &
Not specified &
Theoretical (narrative) &
\textbf{Benefits:} New ESCOs and aggregators may emerge as facilitators between RECs and the grid, enhancing residential and commercial energy-efficiency efforts.\newline
\textbf{Detriments:} Locating RECs in areas with over-supply of electricity increases transmission grid costs and losses; RECs may also erode sensitive privacy protections. &
No \\
\midrule
\multicolumn{5}{c}{\textit{Empirical studies}} \\
\midrule
Fuentes Gonzalez et al. \citeyearpar{fuentesgonzalezCommunityEnergyProjects2022} &
Simplified Chilean market (only three nodes) &
Game theory, optimization modeling &
The nodal price can decrease when RECs are deployed. An increase in social welfare can also be observed when RECs are involved in energy production due to the avoidance of high generation cost at nodes without REC, as well as the less frequent need for transmission expansion. This leads to a reduction in the nodal price that triggers an increase in the quantity demanded in a node with a REC. &
Yes (indirectly) \\
\midrule
Sarfarazi et al. \citeyearpar{sarfaraziAggregationHouseholdsCommunity2020} &
``REC--retailer'' interactions on a micro scale in Germany &
Game theory, optimization modeling, dynamic programming &
The retailer's profit-maximization objective leads to welfare gains both for the system and a REC. Conversely, if a retailer pursues a REC self-sufficiency maximization objective, the REC reduces the energy injected at the supply side of the wholesale market while also consuming less electricity from the public grid and wholesale market. Consequently, less grid charges and levies are collected for the DSOs and general budget. &
Yes \\
\midrule
Boccard and Goetz \citeyearpar{boccardPowerTradingEnergy2025} &
Individual RECs in Northern Spain, Germany, Portugal &
Simulation &
In the scenario with heterogeneous public prosumers (schools and hospitals) that install PV systems covering 50\% and 100\% of the load, the power exchanges reduce by 8\% and 13\%, respectively, compared to a scenario without PV. Instead, 25\% PV coverage of public loads increases power exchanges with the grid by 10\%. Conversely, dynamics become reversed with residential prosumers, where lower PV coverage of aggregated REC loads reduces power exchanges while greater PV coverage increases it. &
Yes (indirectly) \\
\midrule
Zepter et al. \citeyearpar{zepter2019prosumer} &
Individual REC in London &
Optimization modeling, stochastic programming &
Peer-to-peer REC reduces the wholesale market exchange of prosumers. In contrast, exchange with the wholesale market is highest without P2P trading and BESS. In such cases, RECs can be detrimental for the system because they consume more energy during peak hours, thus increasing the overall system load, while injecting energy during off-peak hours, thus potentially causing local congestion. &
Yes \\
\midrule
Backe et al. \citeyearpar{backeImpactEnergyCommunities2022} &
Massive REC deployment in six EU countries &
Optimization modeling &
When RECs are aggregated and utilize BESS, they could reduce both electricity and heating costs and lessen the need for national capacity expansion by 50--60\,GW in six EU countries by 2060. &
No \\
\end{longtable}
{\footnotesize
\noindent\textsuperscript{a} Arbitrage effect emerges when collective generators and consumers/prosumers have distinct selling and buying prices. When they unite into an energy community, generators can supply energy directly to members of the REC while members remunerate the generators with a payment that is higher than the conventional selling price into the grid but lower than the conventional buying price for consumers. As a result, both parties benefit while the regulatory effect of the price vanishes. The buying price for consumers from local generators can be implicitly lowered compared to the buying price from a central market if the distribution/transmission charge is distance-based. \\
\noindent\textsuperscript{b} Death spiral effect describes a situation when distributed energy resources lead to prosumers paying less with
volumetric distribution charges. Consequently, system operators (DSOs/TSOs) become under-paid for its previous
infrastructure investment, which is usually higher than a decreased usage of a network due to self-consumption. System
operators shift these costs instead onto other consumers, which in turn stimulate them to transform into prosumers as
well, thereby shrinking the customer base iteratively.
\par}}

\subsection{Synthetic approach to simulating electricity markets}

\textcolor{black}{Simulation-based approaches have become a widely adopted tool for analysing electricity markets when the complexity of market interactions makes formal equilibrium modelling impractical. In contrast to equilibrium models which explicitly represent market-clearing conditions through systems of equations, simulation models provide greater flexibility in capturing heterogeneous market participants, technological changes and policy interventions \citep{ventosa2005}. A common framework within this literature is the synthetic supply approach, which evaluates the effects of a given market feature by comparing observed outcomes with a counterfactual scenario in which that feature is absent or modified.
The synthetic approach has been extensively employed to assess the market effects of renewable energy penetration. A consistent finding across countries is that renewable generation exerts a merit-order effect, displacing conventional generation and reducing wholesale electricity prices. Seminal work by \citet{sensfuss2008} showed that renewable electricity generation in Germany generated substantial consumer benefits through lower market prices. Similar conclusions emerge from studies on Italy and Spain, which find that renewable generation lowers wholesale prices and produces net welfare gains even after accounting for subsidy costs \citep{beltrami_value_2021, espinosa2018}. These studies demonstrate that counterfactual simulations are particularly well suited to quantifying the economic impact of renewable energy policies.
Beyond renewable deployment, the synthetic approach has also been used to investigate other market mechanisms. Several studies focus on market power, comparing observed prices with counterfactual outcomes under more competitive bidding behaviour. Evidence from Spain and Italy suggests that strategic conduct by dominant firms can significantly increase electricity prices and reduce consumer welfare \citep{ciarreta2010, rossetto2019}. Other applications examine technological innovations and alternative market designs. For example, \citet{beltrami2024hydro} evaluate the environmental effects of pumped-hydro storage, while \citet{ciarreta2024} analyse alternative market structures in Morocco and find substantial efficiency gains from market liberalisation and improved bidding arrangements.
The existing  literature demonstrates the versatility of synthetic simulation methods in evaluating how changes in technologies, market structures, and regulatory interventions affect electricity market outcomes. However, despite the growing policy relevance of RECs, little evidence exists on their implications for wholesale electricity markets. Current studies primarily focus on the benefits accruing to REC participants, such as self-consumption, investment incentives, and local welfare gains, while largely overlooking potential system-wide effects. This paper contributes to the literature by extending the synthetic simulation framework to the analysis of RECs. Combining a bottom-up engineering model of REC generation and consumption profiles with a counterfactual representation of the Italian day-ahead market, we quantify how different REC deployment scenarios affect equilibrium market volumes. Table \ref{tab:synthetic_literature} summarizes the main applications of synthetic simulation approaches in electricity market analysis.
}

\begin{table}[htbp]
\centering
{\color{black}
\caption{Synthetic simulation approaches in electricity market analysis.}
\label{tab:synthetic_literature}
\small
\renewcommand{\arraystretch}{1.25}
\begin{tabularx}{\textwidth}{L{2cm} L{1.1cm} L{2.5cm} X L{1.7cm}}
\toprule
\textbf{Study} &
\textbf{Scope} &
\textbf{Methodology} &
\textbf{Key findings} &
\textbf{Application} \\
\midrule

Sensfuß et al. \citeyearpar{sensfuss2008} &
Germany &
Synthetic supply counterfactual &
RES generation produces a substantial merit-order effect, lowering wholesale prices and generating net consumer benefits. &
RES deployment \\

Beltrami et al. \citeyearpar{beltrami_value_2021} &
Italy &
Synthetic supply counterfactual &
RES displace conventional generation and increase welfare even after accounting for subsidy costs. &
RES deployment \\

Espinosa and Pizarro-Irizar \citeyearpar{espinosa2018} &
Spain &
Synthetic supply counterfactual &
Subsidies initially generate net social benefits; reduced RES deployment weakens the merit-order effect. &
RES policy \\

Ciarreta and Espinosa \citeyearpar{ciarreta2010} &
Spain &
Counterfactual market simulation &
Strategic bidding by dominant firms significantly increases market prices. &
Market power \\

Rossetto et al. \citeyearpar{rossetto2019} &
Italy &
Counterfactual market simulation &
Market concentration leads to consumer welfare losses, especially during peak-demand periods. &
Market power \\

Beltrami \citeyearpar{beltrami2024hydro} &
Italy &
Synthetic market simulation &
Pumped-hydro storage generates net environmental benefits through reduced CO2 emissions. &
Energy storage \\

Ciarreta et al. \citeyearpar{ciarreta2024} &
Morocco &
Counterfactual market design simulation &
Alternative market designs improve dispatch efficiency and substantially reduce electricity prices. &
Market design \\

\midrule

\textbf{This paper} &
Italy &
Bottom-up engineering model and synthetic market simulation &
REC deployment alters wholesale market volumes and is expected to exert downward pressure on wholesale prices; effects depend on REC composition and installed capacity. &
RECs \\

\bottomrule
\end{tabularx}}
\end{table}

\section{Methods and data}\label{sec:methdata}

The first stage of our methodological approach begins with mapping all Italian RECs to derive input parameters for subsequent engineering modeling. The model’s output is used to project the REC deployment for a year 2027 according to three scenarios, which provides the input variables for the second stage. In the second stage, we first simulate a market equilibrium with RECs in 2027. Then, we construct a counterfactual scenario of market equilibrium without RECs. As the most recent annual data on day-ahead market bids is available only for 2024, and the policy target is set for 2027, we assume that the wholesale market conditions in 2027 are identical to those of 2024 for simplification purposes. This assumption does not compromise our modeling objective, which is to introduce a new methodology for estimating the impact of REC deployment on wholesale market equilibrium. Our aim is not to forecast the actual impact of REC deployment in the policy-target year. Nevertheless, policy-relevant insights can be drawn from the modeled scenarios. In the first stage of methodology, we used a combination of software applications: MS Excel (mapping), Matlab (engineering modeling), R (projection). In the second stage, R was utilized as the main modeling software.

\subsection{First stage: Mapping, Engineering modeling, Scenarios and Projections.}
\label{sec:mappingCER}
\subsubsection{Comprehensive Mapping of Italian RECs}
\label{sec:mappingCER_sub}
To accurately model the actual and potential impact of RECs, we began by collecting data on all operational and planned RECs in Italy.\footnote{The complete data on the mapping of Italian RECs are available in Supplementary Materials.} Due to the absence of a comprehensive dataset from a single source, we compiled our database from multiple sources. Table \ref{data sources} provides overview of these data sources.

{\begin{table}[htbp]
{\color{black}
\arrayrulecolor{black}
\caption{\textcolor{black}{Data sources.}}
\centering
\label{data sources}
\small
\renewcommand{\arraystretch}{1.55}
\setlength{\tabcolsep}{5pt}
\begin{tabular}{@{}L{0.28\textwidth}L{0.16\textwidth}L{0.13\textwidth}L{0.33\textwidth}@{}}
\toprule
\textbf{Data source} & \textbf{Time of the last update} & \textbf{Geographical scope} & \textbf{Dataset provider} \\
\midrule
\hline
Data portal of GSE
& October 2024
& Italy
& GSE -- state-owned renewable energy agency \\
Annual reports of Legambiente
& December 2024
& Italy
& Legambiente -- national environmental non-governmental organization (NGO) \\
Data portal NeXt ESG
& December 2024
& Italy
& NeXt -- the civil society network comprising the majority of Italian third-sector and public bodies working in the REC field \\
Data portal ``Sinergie Condivise''
& October 2024
& Piedmont region
& Fondazione Compagnia di San Paolo -- banking foundation, in collaboration with regional and municipal governments and universities in the Piedmont region \\
Business plans of individual RECs (publicly available)
& January 2025
& Italy
& Own (manual collection) \\
Academic publications
& January 2025
& Italy
& Own (manual collection) \\
Websites of REC developers
& January 2025
& Italy
& Own (manual collection) \\
Websites of news agencies, regional/municipal authorities, and individual initiatives
& January 2025
& Italy
& Own (manual collection) \\
\hline
\bottomrule
\end{tabular}}
\end{table}}

 We refer to Italian electricity market zones using acronyms: North -- NORD, Central North -- CNORD, Central South -- CSUD, Calabria -- CALA, South -- SUD, Sicily -- SICI, Sardinia -- SARD. The prosumer's categories (public, residential, SMEs, NPOs) were determined based on subjects that are entitled to become a member of REC by Italian law \citep{TIAD2022}, namely public entities, private citizens, small and medium enterprises, and non-profit organizations. Our final database contains 34 variables for 362 RECs in Italy, of which 184 are in the operational phase and 178 are in the design phase as of January 2025\footnote{\textcolor{black}{The full set of variables can be found in Supplementary Materials while their concise list is included in Table \ref{table: databases comparison}.}}. However, for the purposes of this study, we used only 20 variables\footnote{\textcolor{black}{Variables 4,5,6,8,9,11,13,15,16 from Table \ref{table: databases comparison}, including their sub-variables.}}. Zhu et al. \citeyearpar{zhuItalianRenewableEnergy2025} also built the database of new Italian RECs for the purpose of their study. In addition to the diverse scope of variables collected (due to different research objectives), another major difference lies in the development phase of RECs. Our database includes both operational RECs and those that were in the project design phase of development. Inclusion of RECs in the project design phase allowed us to have a more holistic perspective on future trends, which is discussed in further detail in Section \ref{sec:scen_projec_rec}. In turn, the database collected by GSE \citeyearpar{GSE_CER_Map_2025} is complete but contains a very small set of variables\footnote{The recent map published by RSE \citeyearpar{RSE_CER_Map_2025} is primarily based on the GSE data, although it contains more variables than the original database, but still does not include critical variables (i.e. related to categories of members) used in our research.}. However, neither of the other databases contain information about a category of members nor about the shares of installed PV capacity by prosumer categories. In contrast, our database contains these variables because it was collected specifically for the purpose of this study. A detailed comparison of three databases can be found in Table \ref{table: databases comparison}. \\

\begin{threeparttable}
\newcolumntype{L}{>{\hsize=1.0\hsize\raggedright\arraybackslash}X}
\newcolumntype{C}{>{\hsize=1.1\hsize\centering\arraybackslash}X}
\newcolumntype{R}{>{\hsize=0.8\hsize\centering\arraybackslash}X}
\centering
\caption{\textcolor{black}{Number of REC observations across databases by variable.}}
\small
\begin{tabularx}{\textwidth}{
  >{\raggedright\arraybackslash}X
  >{\centering\arraybackslash}m{3cm}
  >{\centering\arraybackslash}m{2.1cm}
  >{\centering\arraybackslash}m{2.1cm}
}
\hline
\textbf{\textcolor{black}{Name of a }variable\tnote{a}} & \textbf{\textcolor{black}{Number of available REC observations in }our database} & \textbf{\textcolor{black}{Number of available REC observations in }Zhu et al. \citeyearpar{zhuItalianRenewableEnergy2025}} & \textbf{\textcolor{black}{Number of available REC observations in }GSE \citeyearpar{GSE_CER_Map_2025}} \\
\hline
\hline
Total number of RECs, from which: & 362 & 212 & 344\\
\quad Operational RECs & 184 & 212 & 344\\
\quad Design-phase RECs & 178 & 0 & 0\\
\hline
\hline
1. Location \textcolor{black}{of a REC} & 362 & 212 & 344\\
\hline
2. Population of the municipality \textcolor{black}{where a REC is located} & -- & 85 & --\\
\hline
3. Climatic zone \textcolor{black}{where a REC is located} & -- & 85 & --\\
\hline
4. Market zone  \textcolor{black}{where a REC is located} & 362 & -- & --\\
\hline
\textcolor{black}{5. Data on }a generator capacity & 300 & 212 & 344 \\
\hline
\textcolor{black}{6. Data on a }type of technology installed/planned & 323 & 85 & --\\
\hline
\textcolor{black}{7. Data on a }number of members\tnote{b} & 283 & -- & 344 \\
\hline
\textcolor{black}{8. Data on a }category of members & 219 & -- & --\\
\hline
\textcolor{black}{9. Data on }shares of installed PV capacity of \textcolor{black}{different} categories of prosumers and producers \textcolor{black}{within an individual REC} & 271 & -- & -- \\
\hline
\textcolor{black}{10. Data on }technical indicators\tnote{c} & 147 & -- & -- \\
\hline
\textcolor{black}{11. Data on }BESS availability & 13 & 7 & --\\
\hline
\textcolor{black}{12. Data on }EVs and/or e-charging availability & 38 & 3 & --\\
\hline
\textcolor{black}{13. Data on }self-consumption level & 82 & -- & --\\
\hline
\textcolor{black}{14. Data on }economic indicators\tnote{d} & 105 & -- & --\\
\hline
15. Number of buildings/roofs by category of a prosumer \textcolor{black}{ and number of standalone facilities} & 158 & 55 & --\\
\hline
\textcolor{black}{16. Data on building type by use} & 132 & -- & --\\
\hline
17. Number of consumer buildings & not counted & 41 & --\\
\hline
\textcolor{black}{18. Data on }investment source & 252 & -- & --\\
\hline
\textcolor{black}{19. Data on }CAPEX costs & 106 & -- & --\\
\hline
\textcolor{black}{20. Data on }legal form & 180 & -- & 344\\
\hline
\textcolor{black}{21. Data on }REC builders and promoters & 204 & -- & --\\
\hline
\textcolor{black}{22. Data on a }DSO operating a primary cabin & -- & -- & 344 \\
\hline
\hline
Last update & January 2025 & February 2025 & October 2025\\
\hline
Data sources & Table \ref{data sources} & GSE, Legambiente, RSE, academic publications & own\\
\hline
\hline
\end{tabularx}
\begin{tablenotes}[flushleft]
\footnotesize
\item[a]\textcolor{black}{Several reported variables represent actually a set of variables in the actual database (i.e. variable 4 includes two sub-variables, variable 5 - three, variable 9 - five, variable 15 - five)}
\item[b] Members include technical users (prosumers, consumers, producers), and non-user members.
\item[c] Technical indicators collected from business plans include energy generated, injected, and shared as well as self-consumption level.
\item[d] Economic indicators collected from business plans include revenues from energy sales and energy sharing, savings from energy self-consumed.
\end{tablenotes}
\label{table: databases comparison}
\end{threeparttable} \\

As we see in Table~\ref{table: databases comparison}, the information on the market zone and the type of installed technology is the most complete (362 and 323 RECs respectively). Crucial data on the nominal capacity of power plants is available for 82.9\% (300) of RECs. The estimation of many parameters required data on the share of installed PV capacity by category of prosumers, which is available for 74.9\% (271) of RECs. The typology of prosumer buildings is available for 43.7\% (158) of RECs. In turn, the self-consumption level -- from which the range of 45--50--55 scenarios was assumed -- is available for 22.7\% (82) of RECs. Similarly, only 3.6\% (13) of RECs plan to install BESS.\\

\vspace{0.5cm}
From our database, we derived input parameters for subsequent methodological stages. \textcolor{black}{Specifically, Equation \ref{eq:cap1}, which gives the average PV capacity per prosumer/producer in each prosumer category, is the first parameter for constructing the realistic REC engineering model described in the following subsection. In turn, Equation \ref{eq:cap2} is used to derive Equation \ref{eq:cap1}. Finally, the following parameters - “average self-consumption level” and “most common building type” - are similarly required to simulate data-supported REC engineering model. Hence,} first input parameter can be calculated using the \textcolor{black}{Equation \ref{eq:cap1}}:
\begin{equation}
    P_{p,z}^{PV,avg,one} = \frac{P_{p,z}^{PV,avg,total}}{N_{p}^r} \label{eq:cap1}
\end{equation}
where
\begin{itemize}[noitemsep]
    \item $P_{p,z}^{PV,avg,one}$ is the average PV capacity per one prosumer/producer of category $p$ in electricity market zone $z$.
    \item $P_{p,z}^{PV,avg,total}$ is the total average PV capacity of all prosumers/producers of category $p$ within an individual REC in market zone $z$.
    \item $N_{p}^r$ is the average number of rooftop or standalone installations for prosumers/producers of category $p$ across the entire country\footnote{Due to the small sample sizes for the number of rooftop/standalone installations across individual zones in our database, we decided to estimate the average number of rooftop/standalone installations across the entire country, that is,  \( N_p^r \). However, if the sample size allows, the methodology should instead employ the averages across individual zones, that is,  \( N_{p,z}^r \) .}
    \item $p$ denotes a category of prosumer/producer, where $\mathcal{P} \in \{ \text{Public}, \text{Residential}, \text{SME}, \text{NPO},\\ \text{Standalone producing installation} \}$.
     \item $z$ denotes an electricity market zone, where $Z \in \{ \text{NORD}, \text{CNORD}, \text{CSUD}, \text{CALA},\\ \text{SUD}, \text{SICI}, \text{SARD} \}$.
    \item $one$ denotes one prosumer/producer.
\end{itemize}

In turn, the array of $P_{p,z}^{PV,avg,total}$ values was derived \textcolor{black}{using the Equation \ref{eq:cap2}}:
\begin{equation}
    P_{p,z}^{PV,avg,total} = \frac{1}{n_{p,z}} \sum_{n=1}^{n_{p,z}} P_{p,z}^{PV,ind} \left( \frac{Sh_{p,z}^{ind}}{100} \right) \label{eq:cap2}
\end{equation}
where
\begin{itemize}[noitemsep]
    \item $P_{p,z}^{PV,ind}$ is the PV capacity of all prosumers/producers of category $p$ within an individual REC located in market zone $z$.
    \item $ind$ refers to an individual REC.
    \item $Sh_{p,z}^{ind}$ is the capacity share of prosumers/producers of category $p$ within an individual REC in zone $z$.
    \item $n_{p,z}$ is the number of RECs that include at least one prosumer/producer of category $p$ in market zone $z$.
\end{itemize}

The self-consumption level of 49.1\% was identified as the average across all RECs. Consequently, our engineering model is based on a self-consumption range of 45\%--50\%--55\%, where the central value corresponds to the real-world situation, the lower bound reflects a slightly more pessimistic scenario, and the upper bound represents a more optimistic scenario that reflects the potential deployment of BESS. However, only 3.6\% of the RECs currently report having plans to install BESS. Although BESS would enable both a higher self-consumption rate and increased remuneration for shared energy, long payback periods inhibit their adoption.\\

Another input parameter is the ``most common building type'', which is essential to select realistic prosumer load profiles for the engineering model. For example, public prosumers may be represented by the load profile of a school or a sports facility, while SME prosumers may be modeled using the load profiles of commercial buildings or hotels, \textcolor{black}{etc.} This parameter was derived from our database by manually counting the types of buildings for different categories of prosumers. Clearly, different load profiles yield different results. Therefore, identifying the most common building type is critical to producing outputs that closely reflect reality without introducing excessive complexity by modeling all possible building types.

\subsubsection{Bottom-up Engineering modeling of RECs}
\label{bottom-up section}
\begin{figure}[t!]
\centering
\includegraphics[width=1\linewidth]{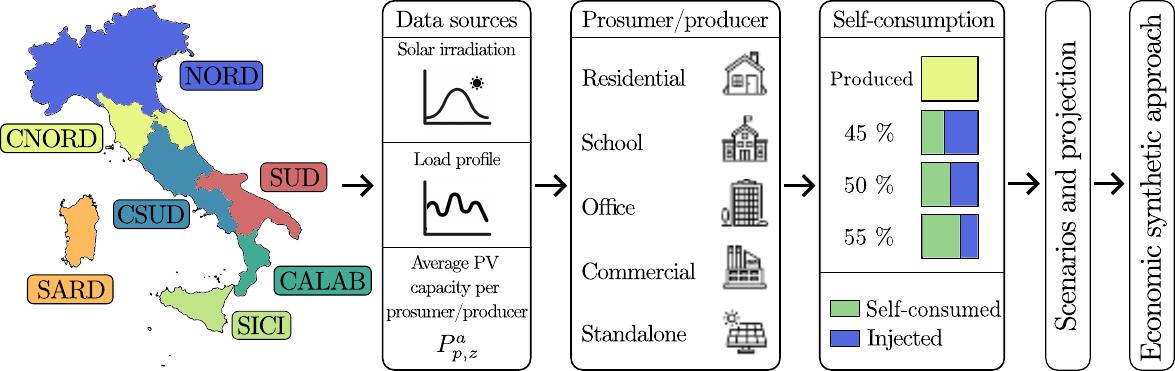}
\caption{\label{fig:bottom_up_eng}Graphical representation of the prosumer energy modeling framework used to simulate hourly energy flows for different prosumer/producer categories. The flowchart illustrates the input data (PV production and load profiles), the modeling process for each prosumer/producer category (residential, school, commercial, office, standalone), and the generation of annual hourly datasets for energy production, self-consumption, and grid injection across multiple market zones and self-consumption scenarios. These outputs are then used as inputs for the scenario-based projection and economic simulations in the day-ahead electricity market.}
\end{figure}

 Modeling RECs at a fine-grained level allows a detailed assessment of how different categories of users interact with their renewable energy systems over time. This insight is crucial for calculating quantities such as self-consumed energy and injected energy into the grid, which are useful for determining the change in energy volumes and thus for estimating economic market variables. An overview of the proposed bottom-up engineering methodology is shown in Fig. \ref{fig:bottom_up_eng}, which is explained in the following paragraphs.\\

To develop our behavioral energy model of prosumers, we designed a modeling scheme that represents five prosumer/producer categories, namely residential, schools, commercial, office, and standalone PV systems. Except for the standalone PV systems (which act purely as generators without any demand), all user categories are equipped with both PV production and electric load, connected to the national power grid.\\

From real-world data collected across operating RECs in Italy, as described in the previous Section \ref{sec:mappingCER_sub}, we estimated the average PV nominal capacity installed for each prosumer category within different RECs. These values were then used to scale hourly-based annual PV production profiles, obtained from location-specific PV yield profiles for seven representative cities covering the main Italian market zones: Milan for NORD, Florence for CNORD, Rome for CSUD, Brindisi for SUD, Catanzaro for CALA, Palermo for SICI, and Cagliari for SARD. \textcolor{black}{The exact installation coordinates were therefore not defined at the single-prosumer level, since all prosumers belonging to the same market zone were assigned the same representative irradiance profile.}\\

Consumption patterns for each prosumer were modeled using nPro \citep{wirtz_npro_2023}, a load profiling tool capable of generating hourly demand curves based on user-specific behavior. To explore different energy scenarios, the load profiles were scaled to achieve three predefined self-consumption levels: 45\%, 50\%, and 55\%. This approach resulted in the generation of annual hourly-based time series (8,760 values) for each prosumer/producer category and each scenario, capturing PV generation, electricity consumption, self-consumed energy and energy injected into the grid. With seven market zones and three levels of self-consumption considered, the final output of the engineering model consists of twenty-one structured datasets. Each dataset includes the energy behavior of the four prosumer categories and the standalone PV systems, whose entire production is fed into the grid. \textcolor{black}{Then, only self-consumed and injected energy profiles are used as input for the subsequent projection step, which is described in Section \ref{sec:scen_projec_rec}.}

\bigskip
\textit{\textbf{Photovoltaic Generation Profiles}}

To characterize the hourly production of PV systems across Italy, we made use of the PVGIS platform developed by the European Commission’s Joint Research Centre \citep{PVGIS}. For each of the seven reference city, we extracted location-specific PV yield data. Rather than relying on generic assumptions, we configured each PV system in PVGIS under optimal operating conditions. This meant that, for each site, the tilt angle of the PV modules was set to the value maximizing annual output, the azimuth orientation was chosen to face due south, which typically yields the best yearly performance in Italy and the PV technology considered was crystalline silicon, the most widespread module type. Also, system losses, such as reflecting factors, inverter efficiency, temperature effects, cable losses, dust, and shading, were included using an average value of 15\% \citep{ogliari_comparative_2023}.\\

For each location, the tool provides an hourly-based annual time series representing the energy yield per installed kilowatt-peak. These values were then scaled by the average installed PV capacity for each prosumer/producer category $P_{p,z}^{PV,avg,one}$, based on data from existing RECs, as in (\ref{eq:pv}):
\begin{equation}
    E^{PV,one}_{p,z}(t) = Y^{PV,one}_{p,z}(t) \cdot P_{p,z}^{PV,avg,one}
    \label{eq:pv}
\end{equation}

where
\begin{itemize}[noitemsep]
    \item $E^{PV,one}_{p,z}(t)$ [kWh] is the hourly-based annual profile of energy produced by the PV system of category \textit{p} in electricity market zone \textit{z} for one prosumer/producer.
    \item $Y^{PV,one}_{p,z}(t)$ [kWh/kWp] is the hourly-based annual yield profile of the reference city from PVGIS of category \textit{p} in electricity market zone \textit{z} for one prosumer/producer.
    \item \textit{t} is the time index $\in \{1,2,...,8760\}.$
\end{itemize}

By using PVGIS in this way, we were able to obtain consistent, reproducible photovoltaic production profiles that reflect regional climatic differences and common technical configurations, without introducing excessive complexity into the modeling process. It’s worth noting that, while real-world data from monitored PV systems might offer greater accuracy in principle, such datasets are often fragmented, inconsistent, or not openly available across all regions. Simulated data from PVGIS, on the other hand, ensures full spatial coverage and comparability, while still being grounded in satellite-based irradiance data and validated performance models.

\bigskip
\textit{\textbf{Load Consumption Profiles}}

To represent electricity consumption behavior for each prosumer category within the REC framework, we generated hourly demand profiles using nPro \citep{wirtz_npro_2023}, a profiling tool that provides synthetic yet behaviorally-informed load curves. The tool produces time series based on statistical models of daily and seasonal usage patterns for different consumer categories (residential, schools, commercial, office, and many others) under typical operational conditions. \footnote{\textcolor{black}{All institutional prosumers are assumed to own electrical loads related to heating, cooling and general electricity demand. Therefore, default nPro load profiles were retained without additional calibration. For residential prosumers, the default residential electricity demand profile was used and an additional cooling load was included to model summer air-conditioning, according to the actual status quo in Italy \citep{istat2022_consumi_energetici} and without explicitly distinguishing between single-family houses and apartments within multi-family buildings, consistently with the market-level scope of the study.}}However, while these base profiles are useful in capturing the temporal distribution of consumption during the year, they do not by default reflect a specific relationship with local PV generation. In particular, no predefined level of self-consumption (i.e., the portion of PV energy that is immediately consumed by the user) can be assumed unless demand and generation are explicitly aligned. \textcolor{black}{Since PV capacities were derived from the proposed database of real installed PV capacity values for REC members in Italy (see Table \ref{table: databases comparison}), the PV generation side was not treated as a free calibration parameter. Therefore, within the adopted framework, variability in the self-consumption level was introduced by acting on the local demand profiles, while keeping PV generation constrained by realistic installed-capacity data.}\\

To introduce variability in this key parameter and explore its influence on community-level energy flows, we implemented a simplified calibration procedure for the demand profiles. The idea was to scale the profiles vertically to increase their overall magnitude, without altering their temporal pattern. \textcolor{black}{This scaling should be interpreted as an engineering parametrization used to obtain alternative PV–demand matching conditions, rather than as an actual increase in the electricity demand submitted to the market model.} This approach changes the extent to which demand and generation tend to coincide and allows us to control the self-consumption ratio, a value that is typically chosen when designing PV systems, in a straightforward way.\\

Specifically, for each user category \textit{p} in electricity market zone \textit{z}, the original load profile generated by nPro was multiplied by a constant factor which was iteratively tuned until the ratio of self-consumed energy to total PV production matched a given target. \textcolor{black}{In practice, each standard hourly demand profile was paired with the corresponding hourly PV generation profile, and the scaling factor was adjusted until the intended annual self-consumption level was reached.} We considered three such targets (45\%, 50\% and 55\%) chosen to represent a plausible range of self-consumed energy. In this way, the original temporal shape of the standard load profile was preserved, while its overall magnitude was modified to generate alternative demand–PV matching conditions. \textcolor{black}{The resulting scaled profiles are used only within the bottom-up engineering model to determine, at hourly resolution, the split between self-consumed electricity and electricity injected into the grid. The scaled profiles are not transferred to the economic model as additional demand withdrawals, where demand bids are taken from the GME portal and kept unchanged across the self-consumption scenarios.}\\

However, to account for the likely evolution of REC configurations in the coming years, we also explored additional scenario featuring higher self-consumption levels (\textit{mixed scenario} in Table \ref{tab:scenarios}). This extended case is intended to represent more favourable future REC configurations characterized by improved demand–generation matching, potentially associated, at aggregate level, with greater flexibility and/or storage adoption. \textcolor{black}{In fact, BESS reduces the amount of surplus energy injected into the grid, thus increasing the local use of renewable electricity. Although storage systems are not explicitly represented through detailed mathematical models in the present framework, these higher self-consumption scenarios are intended as exploratory proxies for their aggregate effect on energy flows. In particular, the resulting behaviour is comparable to an optimization of self-consumption during central daylight hours through the use of storage systems.} Moreover, no external validation against metered customer-level data was performed, since the profiles were not designed to reproduce site-specific measured behaviour, but rather to provide synthetic and comparable inputs for aggregate scenario analysis.

\bigskip
\textit{\textbf{Calculation of Self-Consumed and Exported Energy}}

Once both the hourly PV generation and the electricity demand profiles were established for each prosumer/producer category and scenario, the next step involved calculating the two profiles that define the interaction between local generation and the grid: self-consumed energy and surplus energy injected into the main grid. These quantities were derived directly from the hourly time series of PV production $E^{PV,one}_{p,z}(t)$ and electrical load $E^{L,one}_{p,z}(t)$ of category \textit{p} in electricity market zone \textit{z} computed as explained in the previous sub-section. In particular, the self-consumed energy $E^{self,one}_{p,z}(t)$ of category \textit{p} in electricity market zone \textit{z} for one prosumer/producer at each hour is computed as in Eq.(\ref{eq:self-cons}):
\begin{equation}
    E^{self,one}_{p,z}(t) = min(E^{PV,one}_{p,z}(t),E^{L,one}_{p,z}(t))
    \label{eq:self-cons}
\end{equation}
This corresponds to the portion of the PV production that is immediately used to meet on-site demand. When the load exceeds the available PV power, all PV generation is consumed locally. Conversely, if generation exceeds demand, only part of it is self-consumed. In this case, the surplus energy exported to the grid $E^{exp,one}_{p,z}(t)$ of category \textit{p} in electricity market zone \textit{z} is calculated as in (\ref{eq:export}):
\begin{equation}
    E^{exp,one}_{p,z}(t) = max(E^{PV,one}_{p,z}(t)-E^{L,one}_{p,z}(t),0)
    \label{eq:export}
\end{equation}

In other words, the exported energy is the excess generation that is not used locally and is therefore injected into the public distribution network. These expressions were applied element-wise over the full annual time series, resulting in two additional vectors of hourly values for each prosumer/producer and scenario.\\

In the end, for each prosumer category, the combined simulation produces four time series per user: hourly PV production, electricity consumption, self-consumed energy and energy fed into the grid. Instead, for standalone PV producers, the energy injected into the grid is equal to the PV energy produced, hour-by-hour.

\subsubsection{Scenarios and Projection of REC Deployment} \label{sec:scen_projec_rec}

The core objective of the simulation framework is to evaluate the market impact of RECs within a range of realistic deployment pathways. The calibration of the scenarios is mainly based on the fulfillment - or its lack thereof - of Italy's policy target of 5 GW of installed REC-linked RES capacity by 2027, as well as the current levels of REC deployment (29.646 MW in operation phase; 367.514 MW of PV capacity in operation and design phase). \textcolor{black}{We assume uniform proliferation speed from 2024 to 2027\footnote{\textcolor{black}{A uniform proliferation means that each year an equal amount of new REC-driven capacity is deployed. This assumption is chosen for simplification purpose because the objective of this paper is to assess the possibility of the effect of RECs on the wholesale market equilibrium and the effect's direction across the seasons rather than to perform a specific forecasting exercise}.}} Three main narrative trajectories are hereby defined:

\begin{itemize}
     \item \textbf{Business-as-usual (BU) Scenario:} this scenario assumes no significant acceleration in REC deployment beyond the current trend, with 29.646 MW in operational phase in 2024. \textcolor{black}{The authorization procedure for commissioning new REC-pertaining plants by the GSE extends well beyond one year in many cases.}\footnote{\textcolor{black}{Based on personal communication with REC representatives across multiple occasions, including industrial events.}} \textcolor{black}{Hence, this scenario could be considered pessimistic, though realistic given the current status quo.} For 2027, we again assume a \textcolor{black}{uniform} proliferation of the 2024 operational capacity, reaching 0.119 GW.
     \item \textbf{Half-way (HW) Scenario:} we assume a \textcolor{black}{uniform} proliferation of the 2024 deployed capacity, with 0.368 GW in operational and design phase, reaching 1.47 GW by 2027. \textcolor{black}{This scenario is slightly more optimistic, while still realistic and supported by the collected data. It reflects an assumption on improved speed of commissioning procedures.}
    \item \textbf{Policy Scenario:} this scenario assumes the full achievement of the 5 GW target by 2027, reflecting an optimistic roll-out of RECs in terms of both regulatory support and investment mobilization.

\end{itemize}

\begin{table}[h!]
\caption{List of simulated scenarios.}
\centering
\small
\begin{tabular}{lllllll}
\hline
 \textbf{No.} & \textbf{Scenario name} & \textbf{Scenario code} & \textbf{Year} & \textbf{\% of sc assumed} & \textbf{\makecell[l]{GW of REC \\installed \\capacity}} & \textbf{\makecell[l]{Policy \\target \\achieving}} \\
\hline
1 & \makecell[l]{Policy \\} & \textit{sc45.2027} & 2027 & 45\% & 5 & Yes \\
2 & \makecell[l]{Policy \\} & \textit{sc50.2027} & 2027 & 50\% & 5 & Yes \\
3 & \makecell[l]{Policy \\} & \textit{sc55.2027} & 2027 & 55\% & 5 & Yes \\
4 & \makecell[l]{Half-way \\} & \textit{sc45.HW.2027} & 2027 & 45\% & 1.47 & No \\
5  & \makecell[l]{Half-way \\} & \textit{sc50.HW.2027} & 2027 & 50\% & 1.47 & No \\
6  & \makecell[l]{Half-way \\} & \textit{sc55.HW.2027} & 2027 & 55\% & 1.47 & No \\
7  & \makecell[l]{Business-as-usual} & \textit{sc45.BU.2027} & 2027 & 45\% & 0.119 & No \\
8  & \makecell[l]{Business-as-usual} & \textit{sc50.BU.2027} & 2027 & 50\% & 0.119 & No \\
9  & \makecell[l]{Business-as-usual} & \textit{sc55.BU.2027} & 2027 & 55\% & 0.119 & No \\
\hline
10 & \makecell[l]{Mixed scenario} & \textit{sc\_mix1.2027} & 2027 & \makecell[l]{Public: 50\% \\ Residential: 45\% \\ SME: 55\% \\NPO: 50\% \\ Standalone: no sc} & & \makecell[l]{No} \\
\hline \\
\end{tabular}
\label{tab:scenarios}
\end{table}

Within each of these trajectories, scenarios are further differentiated based on assumed self-consumption rates, defined as the share of generated renewable energy consumed by prosumers within a REC rather than injected into the grid. This parameter is a critical driver in the modeling framework, as it reflects the efficiency and “virtuosity” of RECs in optimizing on-site energy use - thus affecting both demand-side and supply-side market dynamics.
Additionally, the mixed scenario applies differentiated self-consumption rates by prosumer category, thereby enhancing the realism and policy relevance of the simulated outcomes. A summary of all scenarios is provided in Table~\ref{tab:scenarios}. \\

\textcolor{black}{Even though we assume a uniform proliferation speed of the total REC-driven capacity from 2024 to 2027 we differentiate on the categories of prosumers and on the market zones. This differentiation is supported by the collected data for the sector in 2024. Therefore,} the scenario-based projection inputs are drawn from (i) distribution parameters derived from our database (ii) the engineering model, which provided two key variables for one typical prosumer/producer within a REC: annual energy self-consumed and energy injected into the grid. The first \textcolor{black}{projection} parameter, ``Zonal Shares of RECs,'' represents the percentage distribution of RECs across electricity market zones:

\begin{align}
    Sh_z^{total} &= \frac{100}{n} \sum_{n_z=1}^{n_z} n_{ind} \label{eq:zonal-share}
\end{align}
where
\begin{itemize}[noitemsep]
    \item \( Sh_z^{total} \) denotes the zonal share of RECs,
    \item \( n_{ind} \) is an individual renewable energy community,
    \item \( n_z \) is the total number of RECs in each zone \( z \), where \( z \in Z \),
    \item \( n \) is the total number of RECs in Italy, \( n = 362 \).
\end{itemize}

The ``average capacity share of all REC prosumers/producers'' is the second projection parameter. It is calculated using the following equation:

\begin{align}
    Sh_{p,z}^{avg} &= \frac{1}{n_{p,z}} \sum_{n_{p,z}=1}^{n_{p,z}} Sh_{p,z}^{ind} \label{eq:avg-capacity-share}
\end{align}
where
\begin{itemize}[noitemsep]
    \item \( Sh_{p,z}^{avg} \) is the average capacity share of all REC prosumers/producers of category \( p \) in market zone \( z \).
    \item $Sh_{p,z}^{ind}$ is the capacity share of prosumers/producers of category $p$ within an individual REC in zone $z$.
    \item \( n_{p,z} \) is the number of RECs where a prosumer category \( p \) is present in zone \( z \)
\end{itemize}

The third \textcolor{black}{projection} parameter, ``total REC deployed PV capacity per prosumer category \( p \) and zone \( z \)'' was calculated for each scenario using the equation:

\begin{align}
    P_{p,z}^{PV,total} &= \frac{P_{scen}^{PV,total} \times Sh_z^{total} \times Sh_{p,z}^{avg}}{100} \label{eq:total-capacity}
\end{align}
where
\begin{itemize}[noitemsep]
    \item \( P_{p,z}^{PV,total} \) is the total REC deployed PV capacity per prosumer category \( p \) in market zone \( z \),
    \item \( P_{scen}^{PV,total} \) is the scenario-based total REC deployed PV capacity.
\end{itemize}

The fourth \textcolor{black}{projection} parameter, the ``number of PV plants per prosumer category \( p \) in market zone \( z \)'' can be calculated using:

\begin{align}
    Plant_{p,z} &= \frac{P_{p,z}^{PV,total}}{P_{p,z}^{PV,avg,one}} \label{eq:number-of-plants}
\end{align}
where
\begin{itemize}[noitemsep]
    \item \( {Plant_{p,z}} \) is the number of PV plants per prosumer category \( p \) in market zone \( z \).
    \item \( {P_{p,z}^{PV,avg,one}} \) is the average PV capacity installed by one REC per prosumer category \( p \) in market zone \( z \).
\end{itemize}

By taking the outputs of the engineering model - \( E_{p,z}^{\mathrm{exp,one}}(t) \) and \( E_{p,z}^{\mathrm{self,one}}(t) \) - and knowing \( Plant_{p,z} \), we can derive the projected energy injected into the grid and the projected energy self-consumed by different categories of prosumers/producers:

\begin{align}
    E_{p,z}^{exp}(t)  &= E_{p,z}^{\mathrm{exp,one}}(t) \times Plant_{p,z} \label{eq:energy-export} \\
    E_{p,z}^{self}(t) &= E_{p,z}^{\mathrm{self,one}}(t) \times Plant_{p,z} \label{eq:energy-self}
\end{align}
where
\begin{itemize}[noitemsep]
    \item \( E_{p,z}^{exp}(t) \) is the projected energy injected into the grid during hour \( t \) by all photovoltaic plants of prosumer category \( p \) in market zone \( z \),
    \item \( E_{p,z}^{\mathrm{exp,one}}(t) \) is the energy injected into the grid during hour \( t \) by one photovoltaic plant of prosumer category \( p \) in market zone \( z \),
    \item \( E_{p,z}^{self}(t) \) is the projected energy self-consumed during hour \( t \) by prosumer category \( p \) in market zone \( z \),
    \item \( E_{p,z}^{\mathrm{self,one}}(t) \) is the energy self-consumed during hour \( t \) by one prosumer of category \( p \) in market zone \( z \),
    \item \( t \) is a specific hour in a year, \( t \in T \) and \( T = \{1, 2, \ldots, 8760\} \).
\end{itemize}

Finally, we derive two arrays of projected variables: (i) energy injected into the grid by RECs, and (ii) energy self-consumed by RECs, using the following equations:

\begin{align}
    E_{z}^{exp}(t) &= \sum_{p \in \mathcal{P}} E_{p,z}^{exp}(t) \label{eq:total-export} \\
    E_{z}^{self}(t) &= \sum_{p \in \mathcal{P}} E_{p,z}^{self}(t) \label{eq:total-self}
\end{align}
where
\begin{itemize}[noitemsep]
    \item \( E_{z}^{exp}(t) \) is the projected energy injected into the grid by all RECs during hour \( t \) in market zone \( z \),
    \item \( E_{z}^{self}(t) \) is the projected energy self-consumed by all RECs during hour \( t \) in market zone \( z \).
\end{itemize}

\subsection{Second stage: Economic modeling}

To assess the short-run economic impact of RECs on the Italian wholesale power market, we apply the empirical hourly counterfactual simulation proposed by \citet{beltrami_value_2021}, further extended by \citet{beltrami2024hydro}. This methodology builds on publicly available data from Gestore del Mercato Elettrico (GME) – namely, the \textit{``Offerte Pubbliche''} – which report all price–quantity bids submitted to the Italian day-ahead auction. These data allow us to reconstruct the hourly merit-order demand and supply curves for each electricity market zone.\\

For a given hour and zone, each supply bid is characterized by an offered quantity $q_i^s$ and an associated bid price $p_i^s$, while demand bids are defined by requested quantities $q_j^d$ at bid prices $p_j^d$. Following the merit-order principle, supply bids are ranked in ascending price order and aggregated into a stepwise supply curve:
\begin{align}
S_{actual}(Pr) = \sum_{i : p_i^s \leq Pr} q_i^s
\end{align}

Analogously, demand bids are ordered in descending price order and
aggregated into a stepwise demand curve:
\begin{align}
D_{actual}(Pr) = \sum_{j : p_j^d \geq Pr} q_j^d
\end{align}
By construction, both curves are piecewise-constant functions, reflecting
the discrete nature of bidding in day-ahead auctions. The market-clearing price $Pr_{actual}$ is
defined as the minimum price such that:
\begin{align}
S_{actual}(Pr_{actual}) \geq D_{actual}(Pr_{actual}),
\end{align}

with the corresponding cleared quantity given by
$Q_{actual} = D_{actual}(Pr_{actual})$.
\\

Crucially, the counterfactual simulation aims to reproduce the hypothetical configuration of the day-ahead market in the absence of REC-driven generation and self-consumption. In this framework, the
observed demand and supply curves, $D_{actual}(Pr)$ and
$S_{actual}(Pr)$, constitute the empirical baseline from which synthetic
curves are constructed to evaluate the equilibrium effects induced by
RECs. As stated above, RECs typically self-consume a share of their
electricity production. This self-consumed electricity is not
visible in the observed market demand. \textcolor{black}{In a counterfactual scenario where REC-supported distributed generation and associated self-consumption are absent, the corresponding net demand reduction would need to be satisfied by the wholesale market.} \\

From now on, to maintain the focus on the synthetic counterfactual approach, we employ a simplified notation consistent with Subsection  \ref{sec:mappingCER} In particular, $\Delta Q_{\text{REC,d}}$ corresponds to $E_{z}^{self}(t)$, while $\Delta Q_{\text{REC,s}}$ corresponds to $E_{z}^{exp}(t)$ in Eq. (\ref{eq:total-export}) and (\ref{eq:total-self}).\footnote{Time and zonal indices are omitted for expositional clarity, while the strategy applies to all zones and settlement periods.} Let then \( \Delta Q_{\text{REC,d}} \) denote the total quantity of REC self-consumption. To account for this, the demand curve is adjusted by shifting it horizontally to the right:

\begin{align}
D_{synt}(Pr) = D(Pr) + \Delta Q_{\text{REC,d}}
\end{align}

On the supply side, RECs inject electricity into the market through surplus generation.\footnote{The GSE is responsible for informing the market operator, GME, about distributed generation volumes. The REC-driven quantities are then placed on the supply curve at zero price, along with the other distributed renewable generation.} Removing RECs implies that this
contribution is also removed from the market supply. Let \( \Delta
Q_{\text{REC,s}} \) represent the total quantity of REC-generated
electricity that would have been offered to the market. The adjusted
supply curve is thus defined as:

\begin{align}
S_{synt}(Pr) = S(Pr) - \Delta Q_{\text{REC,s}}
\end{align}

This corresponds to a horizontal shift of the supply curve to the
left.

The counterfactual equilibrium price and quantity in the absence of
RECs are denoted by \( Pr_{synt} \) and \( Q_{synt} \), respectively.
These are determined by the intersection of the adjusted demand and
supply curves:

\begin{align}
D_{synt}(Pr_{synt}) = S_{synt}(Pr_{synt})
\end{align}

Substituting the shifted functions, the equilibrium condition
becomes:

\begin{align}
D(Pr_{synt}) + \Delta Q_{\text{REC,d}} = S(Pr_{synt}) - \Delta
Q_{\text{REC,s}}
\end{align}

This can be rearranged as:

\begin{align}
D(Pr_{synt}) + \Delta Q_{\text{REC,d}} + \Delta Q_{\text{REC,s}} =
S(Pr_{synt})
\end{align}

\begin{figure}[h!]
\centering
\includegraphics[width=0.8\linewidth]{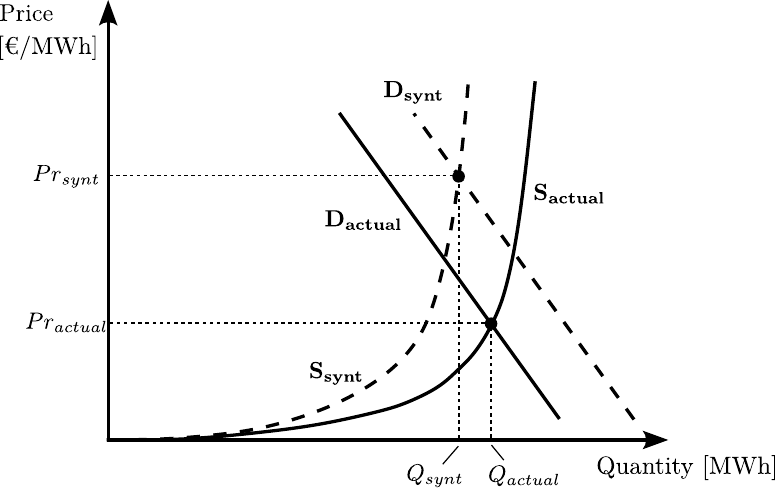}
\caption{\label{fig:cap_curves} Graphical representation of the market equilibrium taking into account the effect of RECs. The curves $D_{\text{actual}}$ and $S_{\text{actual}}$ represent the actual demand and supply with RECs, leading to a market equilibrium at price $Pr_{\text{actual}}$ and quantity $Q_{\text{actual}}$. The dashed curves $D_{\text{synt}}$ and $S_{\text{synt}}$ refer to synthetic demand and supply, resulting in an alternative equilibrium at price $Pr_{\text{synt}}$ and quantity $Q_{\text{synt}}$.}
\end{figure}

This framework captures the dual effect of RECs on market
equilibrium (see Figure \ref{fig:cap_curves}). The removal of self-consumption by RECs increases observed demand, while the removal of injection by RECs reduces available supply, thus producing a rightward shift in the demand curve and a leftward shift in the supply curve. As a result, the counterfactual equilibrium price \( Pr_{synt} \) is expected to be higher than the baseline price \( Pr_{actual} \). The change in equilibrium quantity,
\( Q_{synt} - Q_{actual} \), depends on the relative elasticities of
supply and demand. This formalization enables a quantitative
evaluation of the role of RECs \textcolor{black}{in the change in equilibrium volumes and prices} in wholesale power auctions.\\

Specifically, we modify the observed merit-order curves as follows:
\begin{itemize}
    \item Supply curve shift. In hours where RECs inject renewable electricity into the grid, we assume that this volume would not be available under the counterfactual scenario. Accordingly, the supply curve is shifted leftward to reflect the reduction in total market supply, primarily from RES. This leads to a counterfactual configuration in which electricity prices would be higher, ceteris paribus.
    \item Demand curve shift. In hours where RECs self-consume a portion of their generation, the equivalent electricity demand is effectively removed from the market. In the counterfactual scenario, where self-consumption does not occur, we expand the demand curve rightward (upward shift in the merit-order framework), capturing the higher residual market demand that would otherwise materialize.
\end{itemize}

This dual (contemporaneous) adjustment is performed hourly and for each market zone, generating synthetic demand and supply curves representing a no-REC baseline condition in the short-term. \textcolor{black}{In doing so, we abstract from potential feedback effects arising from active market participation of RECs and aggregators with the market, consistently with the short-term nature of our counterfactual framework.} The resulting counterfactual quantity outcomes are then compared with the observed market outcomes to estimate the impact of REC operations. To quantify the magnitude of REC effects on market outcomes, we define the
\textit{hourly percentage impact on equilibrium quantities} as the relative
difference between the observed equilibrium quantity in the presence of RECs
and the counterfactual equilibrium quantity in their absence.
Formally, for each hour and market zone, the indicator is computed as:
\begin{align}
\text{Impact}_{Q,h} = \frac{Q_{actual,h} - Q_{synt,h}}{Q_{synt,h}} \times 100,
\label{eq:impact}
\end{align}

\noindent where $Q_{actual,h}$ denotes the cleared quantity observed in the day-ahead
market with RECs, while $Q_{synt,h}$ represents the synthetic equilibrium
quantity obtained under the no-REC counterfactual scenario.\\

A positive value of the computed impact in Eq. (\ref{eq:impact}) indicates that REC deployment is
associated with higher market-cleared volumes relative to the counterfactual
baseline, whereas a negative value signals a reduction in traded quantities,
typically driven by higher levels of REC self-consumption. This metric is
computed on an hourly basis and then analyzed across months, seasons, and
policy scenarios, as reported in Section \ref{sec:results}.\\

We acknowledge that this simulation is based on several assumptions. First, the counterfactual is computed under a \textit{ceteris paribus} condition, assuming all other market dynamics unchanged.\footnote{This assumption is particularly relevant when comparing the effects of RECs on demand and supply across the different scenarios, and becomes evident mostly in interpreting the results from the Policy Scenario compared to BU and HW. In our setting, we implicitly assume that the installed capacity of other RES technologies (beyond RECs) remains fixed. This does not reflect the actual dynamics of the Italian power system, where RES capacity additions are currently progressing at sustained rates as well as does not reflect the demand growth. Nevertheless, this choice is consistent with our primary goal, i.e. to isolate the specific marginal contribution of REC deployment, without counfouding their individual effect with broader renewable expansion trends.} Second, each market zone is treated as a “closed system”, abstracting from inter-zonal electricity flows. While this assumption may limit economic and policy implications, it remains defensible for two main reasons: (1) RES-generated power is typically prioritized in dispatch due to the merit-order principle, making its theoretical removal analytically legitimate; (2) the current scale of REC operations in Italy is still limited, rendering its impact negligible in terms of strategic bidding by large market players deeply affecting the market clearing dynamics.
Overall, this synthetic control approach allows us to isolate and quantify the localized, short-term effects of REC deployment on market outcomes, such as zonal traded quantities.


\section{Results}\label{sec:results}

\subsection{Descriptive \textcolor{black}{results} on REC \textcolor{black}{deployment}}

\begin{figure}[h!]
    \centering
    \includegraphics[width=0.5\textwidth]{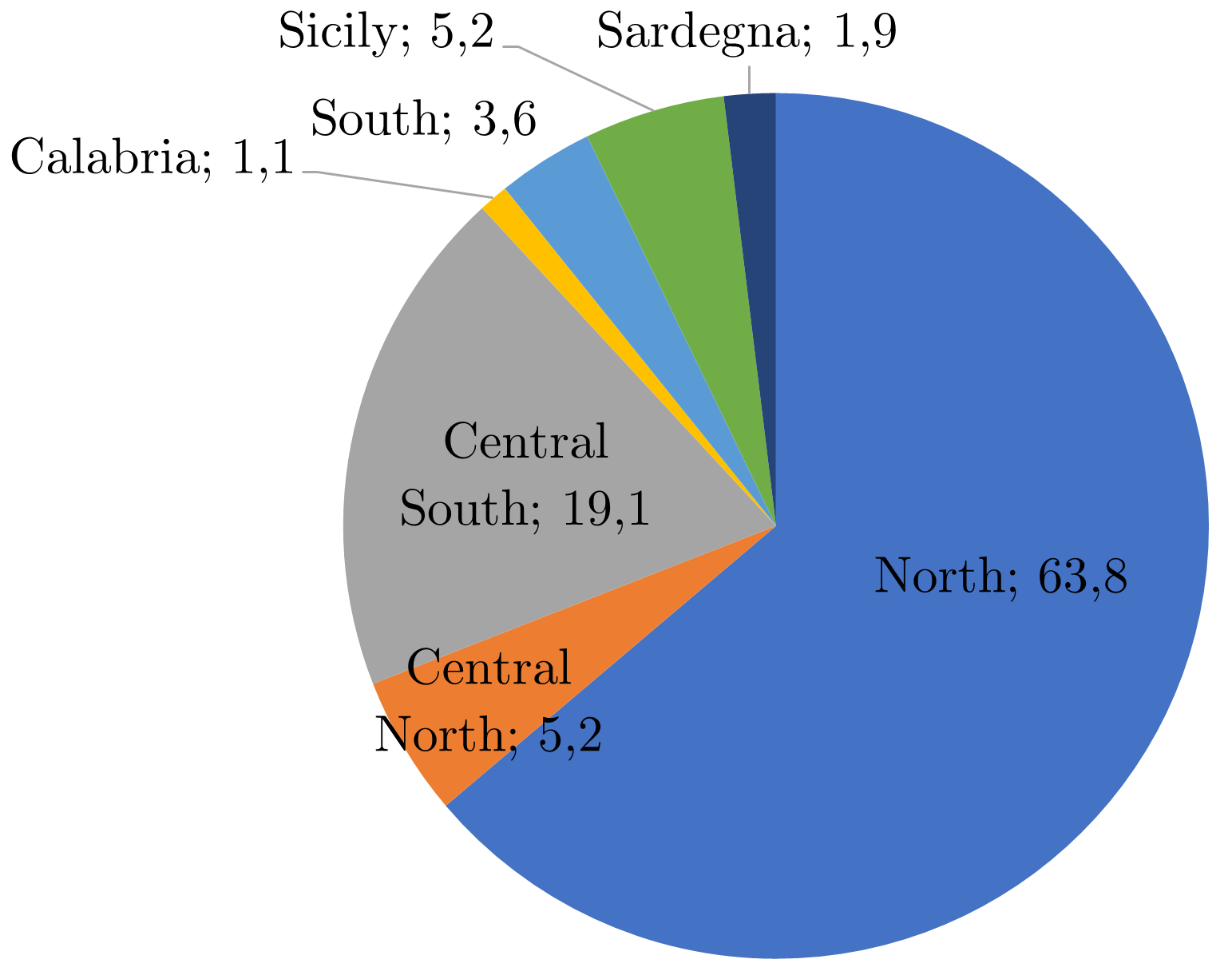}
    \caption{Zonal shares of RECs, $Sh_z^{\text{total}}$.}
    \label{fig:zonal_shares}
\end{figure}

Figure~\ref{fig:zonal_shares} shows that most RECs are situated in the North market zone (63.8\%) and the Central-South zone (19.1\%). In contrast, RECs are scarcely present in the rest of Italy. This deployment pattern may be associated with the general distribution of economic activity across the country. Since the second half of the 20\textsuperscript{th} century, Northern Italy and the regions surrounding Rome have exhibited high levels of industrial and entrepreneurial activity. Consequently, the greater availability of expertise and financial resources for REC establishment has supported their rapid proliferation in these two market zones.\\

\begin{figure}[h!]
    \centering
    \includegraphics[width=\textwidth]{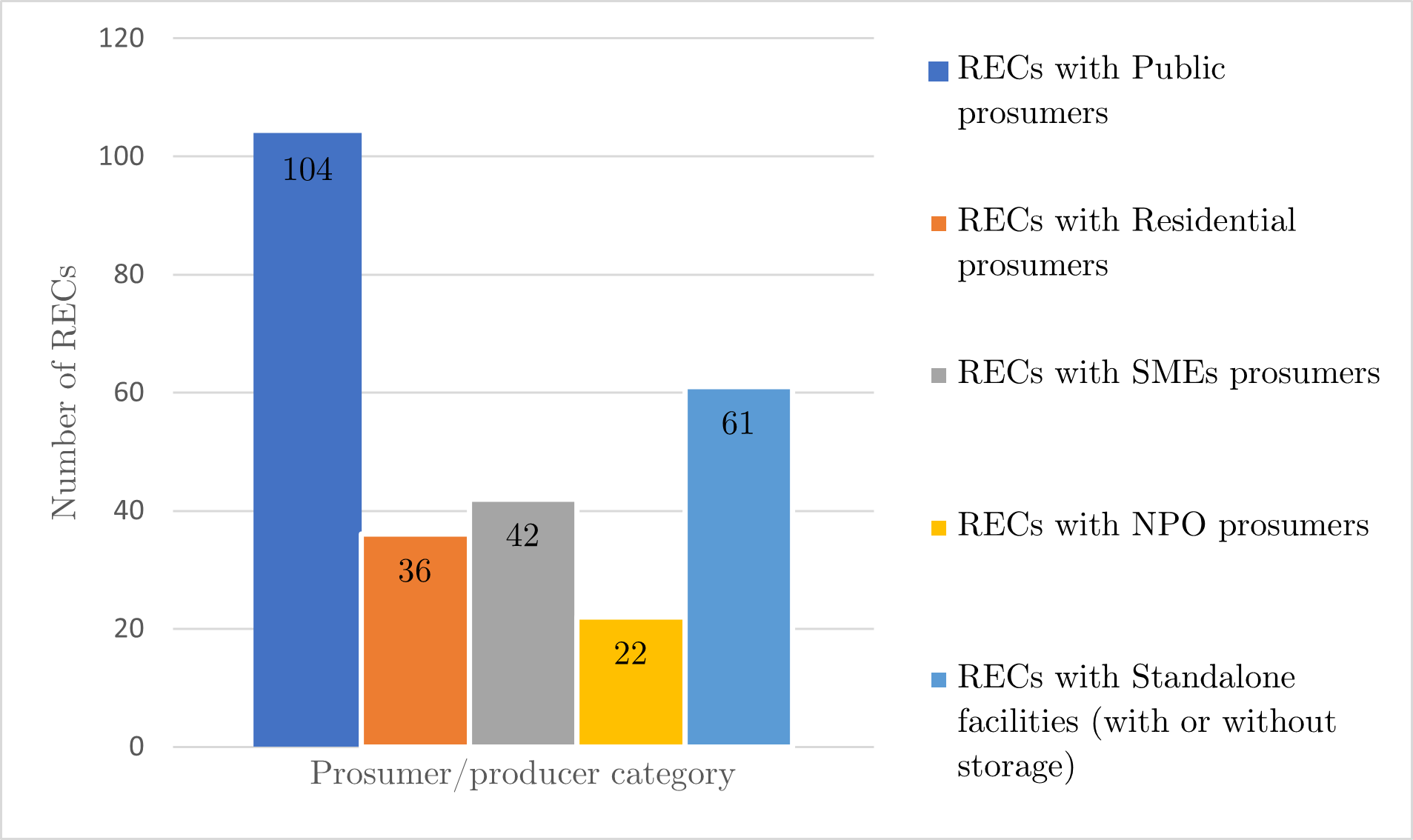}
    \caption{Distribution of RECs across prosumer/producer categories}
    \label{fig:prosumer_distribution}
\end{figure}

\noindent In Figure~\ref{fig:prosumer_distribution},\footnote{Figure 6 contains information not for all 362 RECs from our database but for RECs with the available information.} we observe that the most common prosumer/producer category is public prosumers, typically represented by municipal buildings. Standalone systems are reported for 61 RECs, while rooftop installations owned by SMEs and residential prosumers are reported for 42 and 36 RECs, respectively. In contrast, only 22 RECs report involving NPO prosumers. The higher number of RECs with public prosumers may be explained by the fact that municipalities are, by far, the most common promoters of RECs in Italy, for whom targeted \textcolor{black}{national} and regional grants are also more accessible. Moreover, public buildings consume most of their load during daylight and off-peak hours (see Fig. \ref{fig:avg_daily_energy_jan}, \ref{fig:avg_daily_energy_apr}), making them ideal candidates for maximizing the state incentive on shared energy (see \nameref{sec: Appendix B} and Koltunov et al. \citeyearpar{koltunov2025financing}). Besides, standalone plants also allow for maximization of profits due to their greater average installed capacity, especially when balanced with the daytime loads of public, SME, and NPO prosumers and consumers. Although residential prosumers have not yet actively participated in RECs with rooftop systems, many residential consumers\footnote{REC consuming members are not the focus of this study. Therefore, even though data on REC consumers is available, we do not report it here.} are members of RECs in Italy\footnote{The elaborate discussion on equity concerns and private citizen participation in Italian RECs can be found in Koltunov \citeyearpar{koltunovENERGYCOMMUNITIESINSTITUTIONS2025}.}.\\

\begin{figure}[h!]
    \centering
    \includegraphics[width=\textwidth]{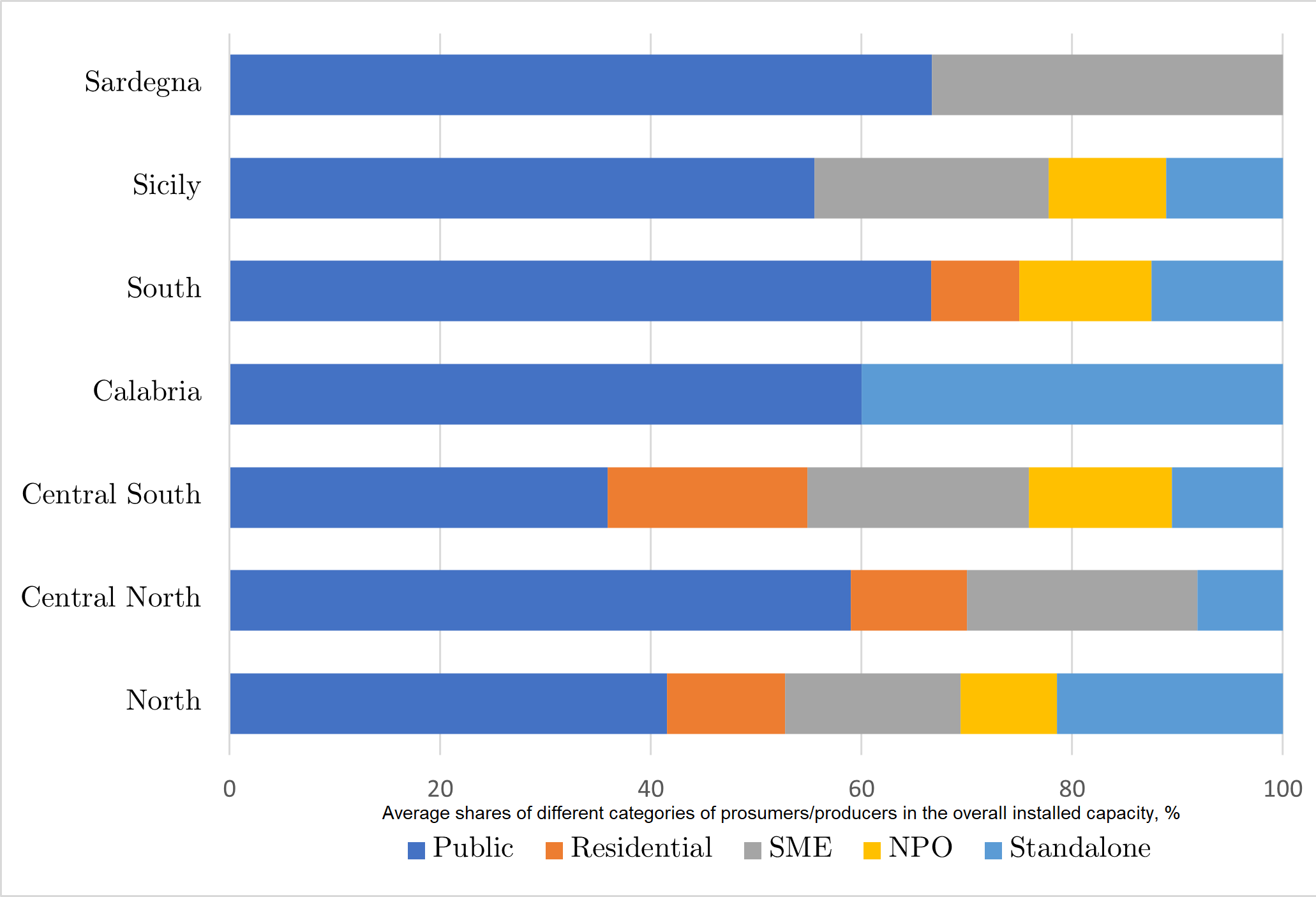}
    \caption{\textcolor{black}{Average shares of different categories of prosumers/producers in the overall installed capacity} across market zones, \( Sh^{a}_{p,z} \)}
    \label{fig:avg_capacity_shares}
\end{figure}

\begin{figure}[h!]
    \centering
    \includegraphics[width=\textwidth]{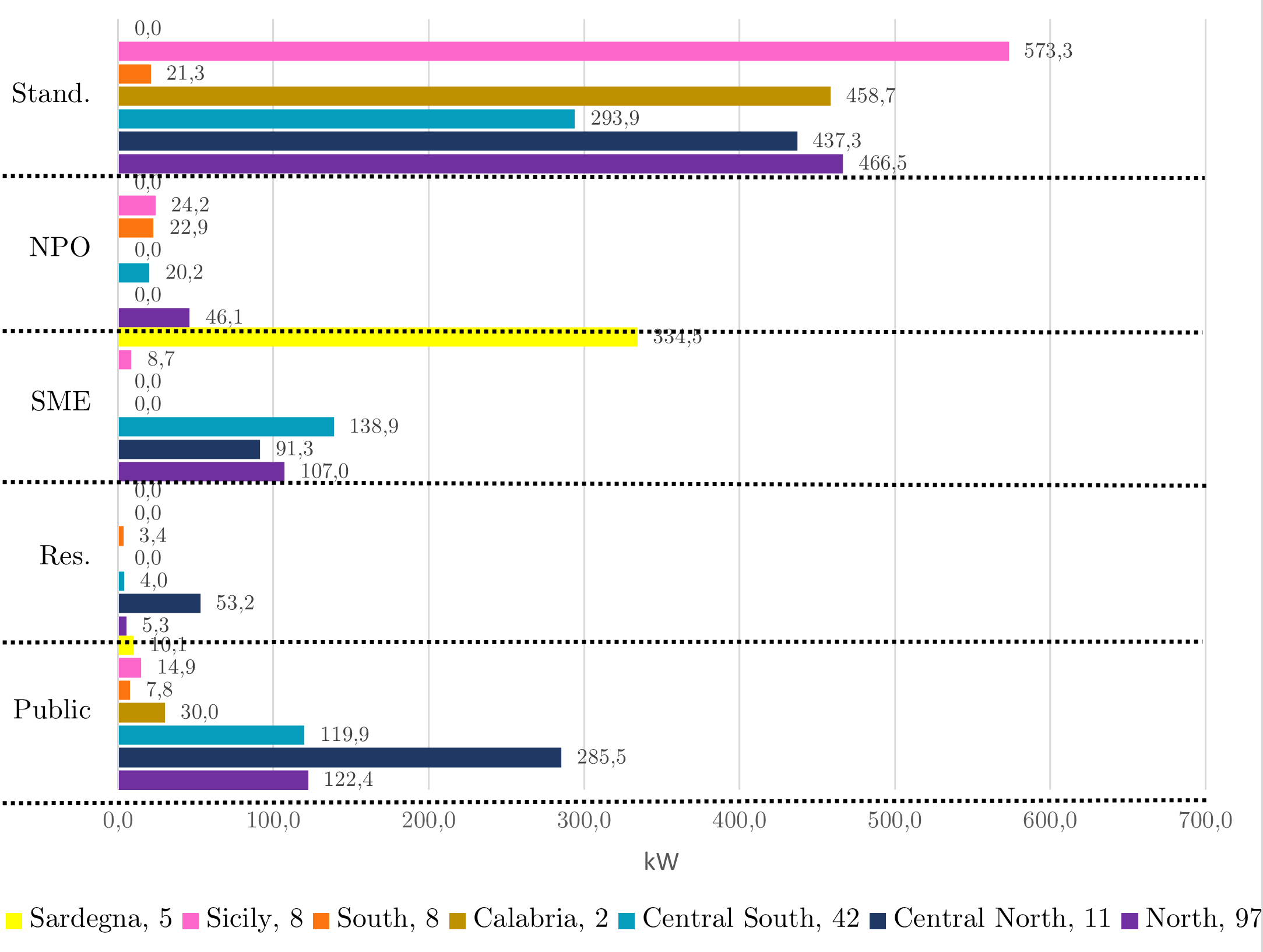}
    \caption{Average PV capacity per one Prosumer/Producer, \( P^{a.one}_{p,z} \).\\
    {Note}: The number after the zone name indicates the sample size (REC observations).}
    \label{fig:avg_capacity_one}
\end{figure}

Figure~\ref{fig:avg_capacity_shares} shows the average \textcolor{black}{shares of different categories of prosumers/producers in the overall installed capacity} across seven market zones. Photovoltaic installations of all categories are present only in three zones: North, Central South and South. The Central South and North market zones have the most proportionate prosumer/producer distributions of installed capacity. This situation is possibly associated with the high number of RECs located in these zones (see Fig.~\ref{fig:avg_capacity_one}). Moreover, the strong economic activity in these regions may contribute to a greater diversity of stakeholders who possess the financial, administrative, technical, legal, and social engagement resources and skills necessary to participate in RECs \citep{koltunov2025financing, musolinoThreeCaseStudies2023}. \textcolor{black}{In addition, Castellini et al. \citeyearpar{castellini2025paving} found that the Central Southern and Northern regions distribute the larger regional funding to RECs. Another reason may be due to greater number of awareness-raising campaigns in these regions.} In contrast, a more disproportionate distribution of capacity across prosumer/producer categories---where public bodies dominate---can be observed in the market zones of Southern Italy (South, Sardegna, Sicily, Calabria). Importantly, the absence of rooftop plants for certain member categories does not automatically exclude them from REC membership. All categories of members may also participate in RECs as consumers without owning rooftop systems.\\

\begin{figure}[h!]
    \centering
    \includegraphics[width=\textwidth]{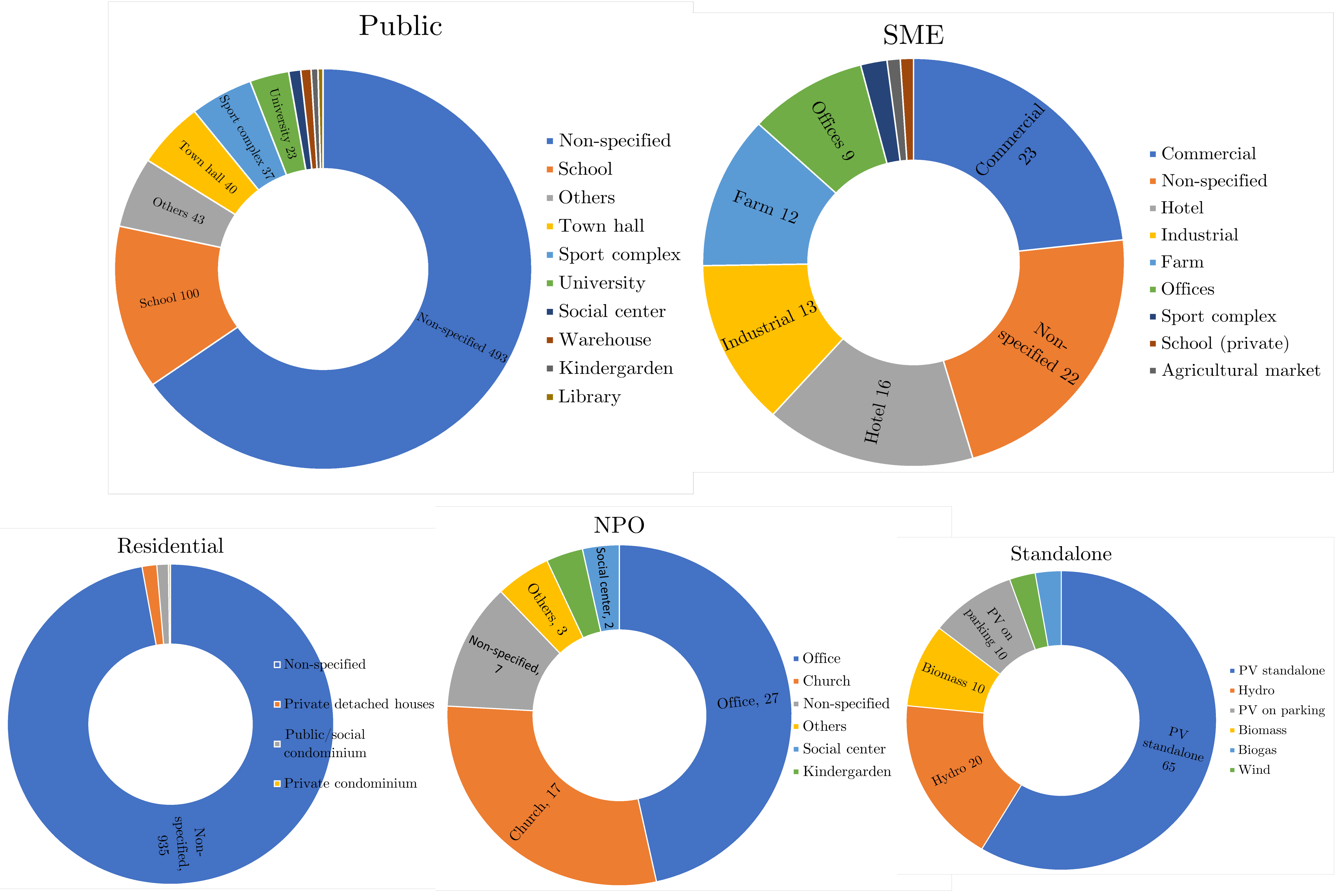}
    \caption{Building/installation types by prosumer/producer categories. \\
    {Note 1}: The numbers after building types indicate the number of individual buildings. \\
    {Note 2}: ``Others'' in the Public category include police stations, autodromes, waste management companies, cemeteries, public utility facilities, etc. ``Others'' in the NPO category include social canteens, sports and educational centers, and social farms.}
    \label{fig:building_types}
\end{figure}

Figure~\ref{fig:avg_capacity_one} demonstrates the ``Average PV capacity per one Prosumer/Producer``, \( P^{a.one}_{p,z} \), which is one of the two real-world input parameters for the engineering model. The average capacity of a single photovoltaic plant varies across market zones. For public and SME prosumers, the average capacity in the North and Central zones (Central North, Central South) is significantly higher than in the Southern zones (Calabria, South, Sicily).\footnote{Only 5 observations were used to estimate the average SME capacity in Sardegna. Therefore, 334.5 kW of the average capacity installed by SMEs in Sardegna is based on a limited sample and should be interpreted with caution. Similarly, the very large capacity of public prosumers in the Central North zone, 285.5 kw, may be related to the small observational sample.} Another notable observation is that NPOs have significantly smaller installed capacities on average compared to SMEs and public prosumers. Typically, non-profit organizations have more limited access to private financing than SMEs, while state capital subsidies are reserved exclusively for small municipalities \citep{koltunov2025financing}. Finally, zero values are mostly observed in the Southern zones, possibly due to the small number of REC observations.\\

In Figure~\ref{fig:building_types}, we observe the types of buildings---where rooftop installations have been constructed---estimated for prosumer categories. This information was used to identify the ``most common building type.'' For public prosumers, data on specific building types is unavailable for most RECs; consequently, the most common identified building type is a school. For SMEs, it is a commercial building, typically a supermarket or shopping mall. Hotels and industrial buildings are also relatively common in the SME category. The type of almost all residential buildings is unspecified (935 buildings), followed by private detached houses (14 buildings). Offices and churches are the most common building types for the NPO category. Finally, 65 standalone photovoltaic plants have been reported, while non-photovoltaic technologies have been used in a much smaller number of facilities.\\

We also analyzed the average number of buildings participating as prosumers in a single REC, \( N^{r}_{p} \). On average, 4.4 public buildings with rooftop plants participate in a REC. Only 2.3 SME buildings and the same number of NPO buildings participate as prosumers in a REC. In contrast, approximately 17.7 residential houses participate with rooftop systems in a REC. However, as shown in Figure~\ref{fig:avg_capacity_one}, residential prosumers have, on average, much smaller generating capacities. The average number of standalone plants per REC is 1.7. \\

\begin{figure}[b!]
\centering
\includegraphics[width=1\linewidth]{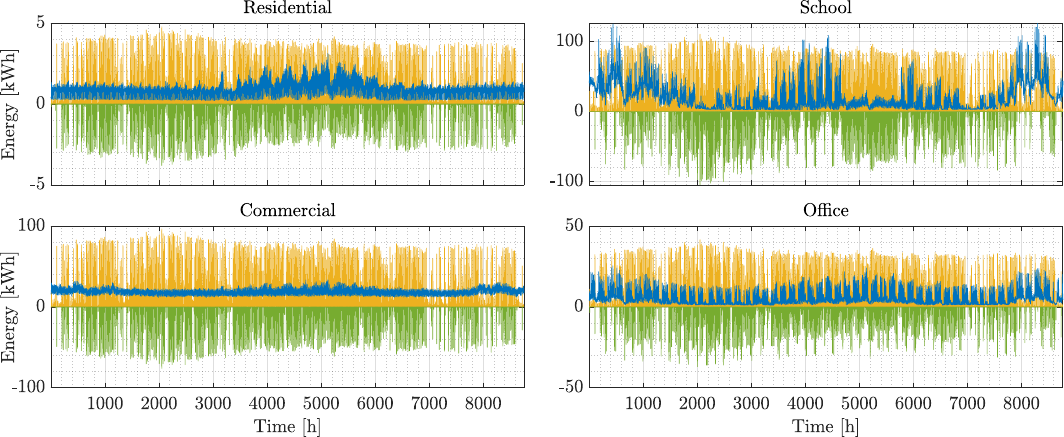}
\caption{\label{fig:time_series_example} Hourly energy profiles for Milan (NORD) over a full year for four prosumer user categories. Each subplot shows PV generation (yellow line), electricity consumption (blue line), and energy injected into the grid (green line).}
\end{figure}

\begin{figure}[b!]
\centering
\includegraphics[width=1\linewidth]{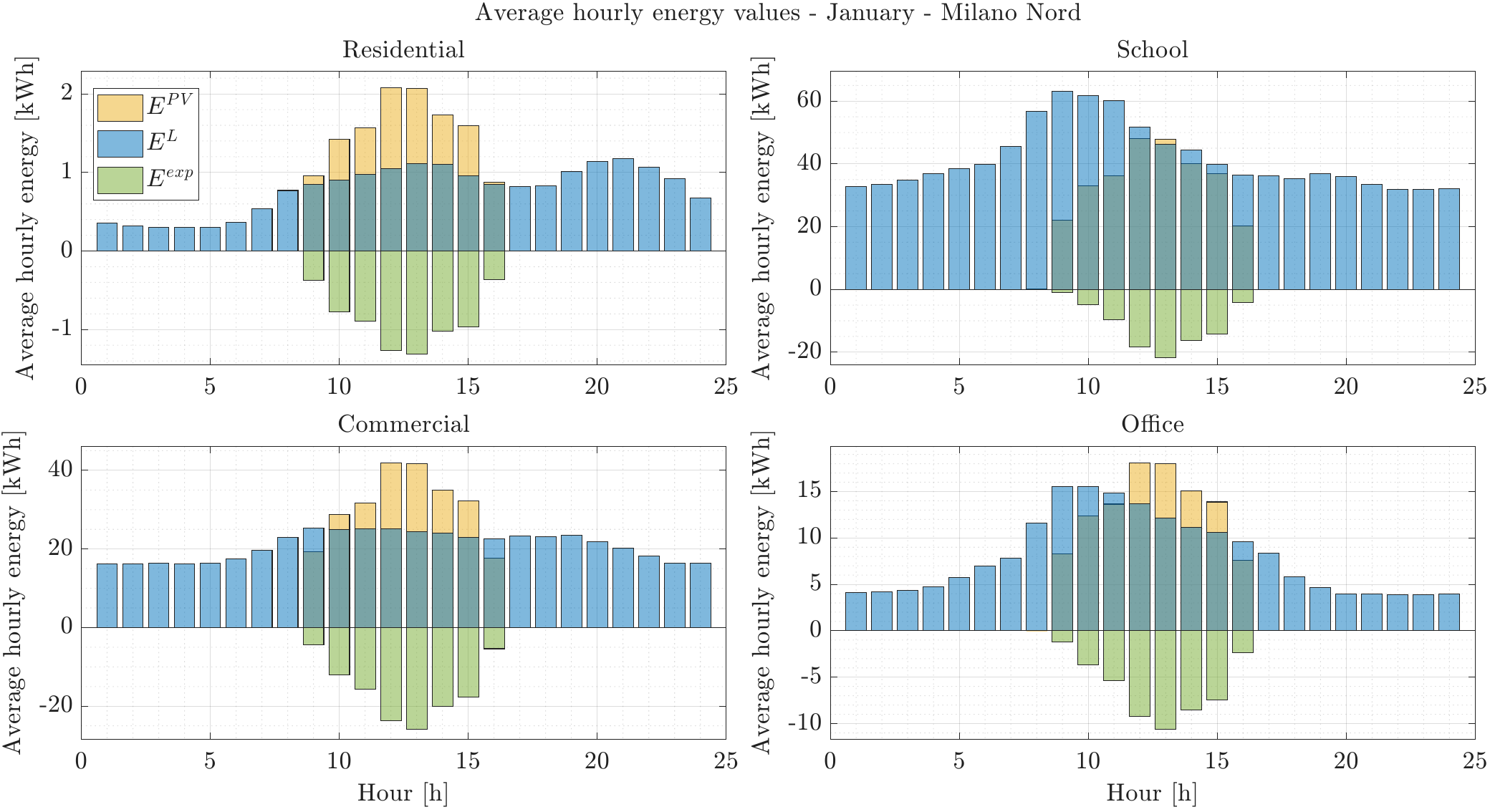}
\caption{\label{fig:avg_daily_energy_jan} Hourly energy profiles for Milan (NORD) over an average day in January for four prosumer user categories. Each subplot shows PV generation (yellow bars), electricity consumption (blue bars), and energy injected into the grid (green bars).}
\end{figure}

\begin{figure}[b!]
\centering
\includegraphics[width=1\linewidth]{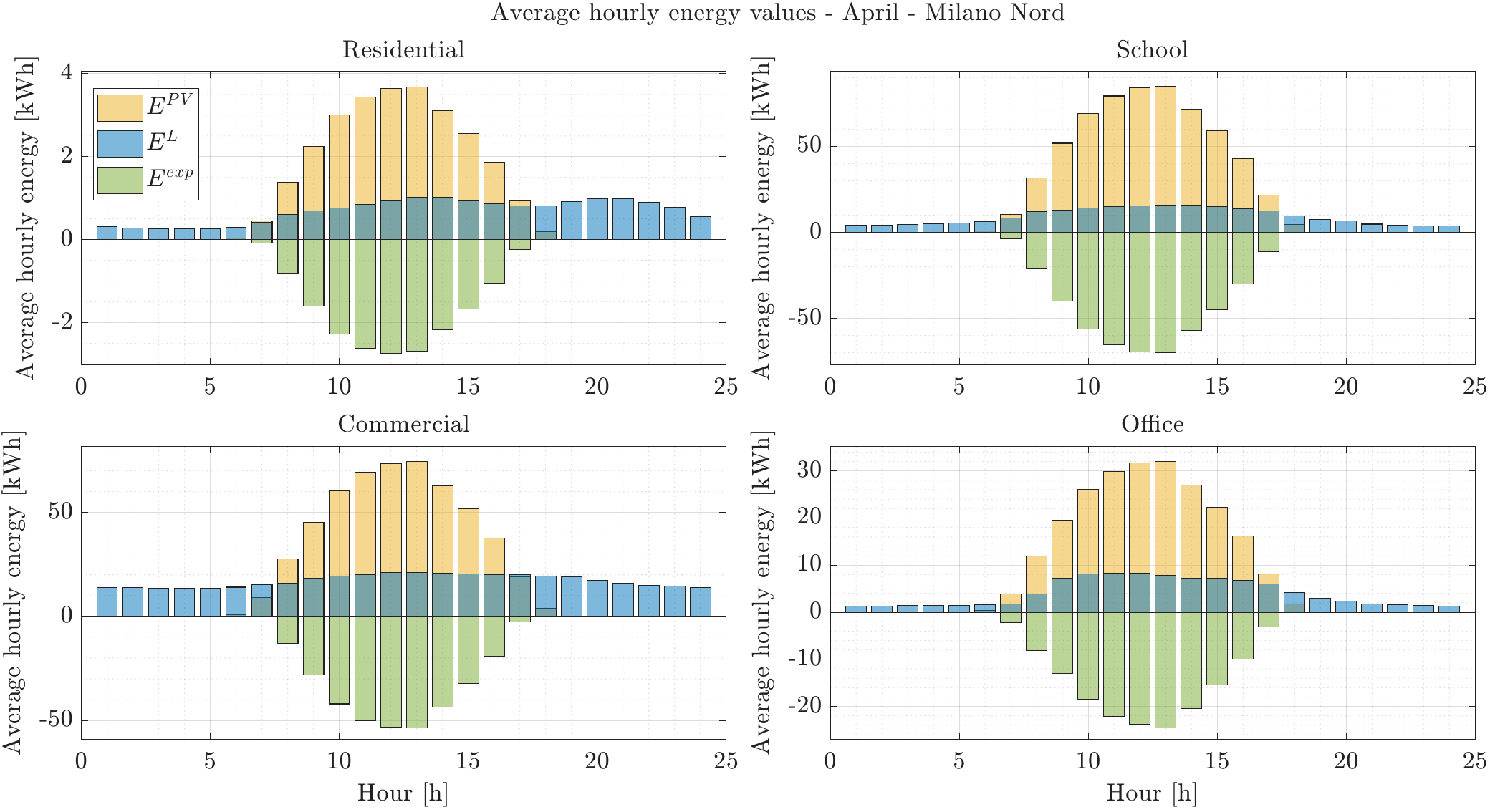}
\caption{\label{fig:avg_daily_energy_apr} Hourly energy profiles for Milan (NORD) over an average day in April for four prosumer user categories. Each subplot shows PV generation (yellow bars), electricity consumption (blue bars), and energy injected into the grid (green bars).}
\end{figure}

\textcolor{black}{Based on the methodology of the bottom-up engineering model (see Section \ref{bottom-up section}), descriptive statistics from our database, along with data from \textit{nPro}, were simulated typical energy profiles of individual RECs.} Figure \ref{fig:time_series_example} shows such energy profiles \textcolor{black}{for four most common buildings representing different} REC prosumer categories in Milan (NORD). 
In each subplot, three key variables are shown over the full time horizon of one year. The yellow curve represents the electricity generated by the PV system, while the blue line indicates the hourly electricity consumption associated with the user. The green line, plotted below the horizontal axis, corresponds to the surplus energy that is not self-consumed and is instead injected into the grid. The hourly profile of self-consumed energy is intentionally not shown to make the graphs easier to read. It can be observed that energy behavior varies significantly across prosumer categories. Residential prosumers exhibit relatively low and stable generation throughout the year, with a load profile that allows only limited self-consumption, especially during daytime hours. School buildings show a highly intermittent load, with pronounced reductions in summer months and strong peaks during winter, reflecting the academic calendar. Commercial prosumers show a more consistent and uniform demand profile, enabling better alignment between PV production and consumption. In addition, office buildings exhibit similar patterns \textcolor{black}{to the commercial building}, although with lower overall demand and more visible weekday-weekend variability. In all cases, the amount of \textcolor{black}{injected energy} is visibly related to the time mismatch between photovoltaic generation and load demand.\\

Figures \ref{fig:avg_daily_energy_jan} and \ref{fig:avg_daily_energy_apr} display the same hourly profiles albeit during an average day in January and April. We see that in January (Figure \ref{fig:avg_daily_energy_jan}) a highest self-consumption (intersection between yellow and blue bars) is exhibited by a school. Moreover, its consumption (blue bars) surpasses its production during midday off-peak hours (10 a.m.--15 p.m.), thereby potentially offsetting an extra generation from REC's residential and commercial prosumers. The load profile of an office is very similar to the school's profile, though on a smaller scale (vertical axes). In April (Figure \ref{fig:avg_daily_energy_apr}), we see that the absence of heating and cooling needs entails a more moderate demand for energy. Consequently, a lot of excessive energy goes into the grid. In both graphs, we observe that residential  prosumers exhibit much smaller energy flows \textcolor{black}{(scale of the vertical axis)} and a different load profile than other prosumers. In addition to evening peak difference, a peak midday consumption of an office, school and commercial buildings occurs at 9 a.m.--12 a.m., whereas of a residential building at 13 p.m.--14 p.m., although more pronounced in January than in April. \\

\bigskip
\subsection{\textcolor{black}{Results from the counterfactual analysis}}

This subsection displays the core results of the paper, presenting the main outputs in weekly, monthly, and quarterly resolutions. The results are reported exclusively for the NORD and CSUD zones.\footnote{\textcolor{black}{We focused the analysis on the NORD and CSUD market zones because these are the Italian bidding zones with the highest current and projected deployment of RECs, making them the most relevant cases for assessing equilibrium impacts on the day-ahead market. From a methodological perspective, the simulation framework requires sufficiently continuous and reliable zonal data series to ensure convergence of the equilibrium algorithm. For the remaining five Italian market zones, the combination of limited RECs penetration and discontinuities in the available zonal data prevented the model from producing stable and economically meaningful equilibrium outcomes. Including such zones would therefore have introduced results with low statistical and interpretative reliability.}}
The analysis focuses on the comparative dynamics between observed market outcomes and their synthetic counterfactuals, highlighting the extent to which REC-driven injections and self-consumption might influence zonal market equilibria in terms of traded volumes. 
To improve the readability of the plots that represent the percentage
impact of RECs on equilibrium quantities in the DA market, we applied a smoothing procedure to all Figures from \ref{fig:NORD_ADP_rel_diff} to \ref{fig:NORD_POLICY_rel_diff_by_month_weekend}
using a centered simple moving average. Specifically, for each hour
of the day, we replaced the original values with the average
computed over a 7-hour symmetric window, using the \texttt{rollmean}
function from the \texttt{zoo} package in R. The smoothing was
applied selectively, only to positive values, while non-positive
entries were left unchanged. This method reduces short-term
fluctuations and highlights general trends across the different
scenarios and seasons.

\subsubsection{Results for main scenarios}

\begin{figure}[htbp]
\centering
\includegraphics[width=1\linewidth]{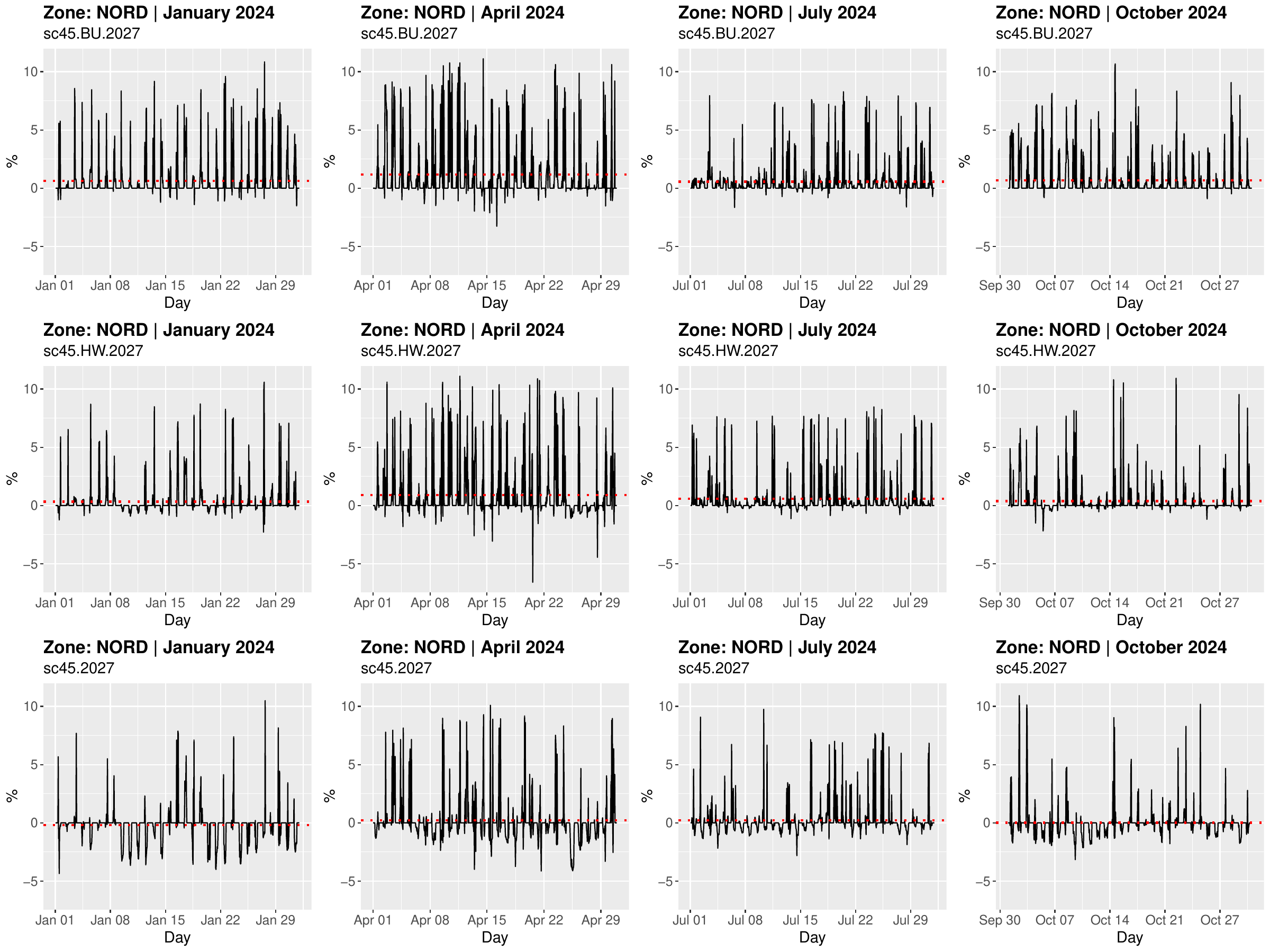}
\caption{\label{fig:NORD_12_windows} Hourly percentage impact on equilibrium quantities from deployment of RECs in the North market zone.}
\end{figure}

Figure \ref{fig:NORD_12_windows} reports the hourly percentage impact on market equilibrium quantities\footnote{For reference through the Section, the hourly percentage impact on equilibrium quantities is defined in Eq. (\ref{eq:impact}).} for the NORD zone, taking four representative months (January, April, July and October), and showcasing the results by assumed scenarios (see Table \ref{tab:scenarios}). In terms of magnitude, as expected, all boxes display a limited effect of RECs self-consumption and injected supply. The range of the impact lies, on average, between -0.19$\%$ (January, Policy Scenario) and 1.16$\%$ (April, BU Scenario). As concerns the BU scenario, the results show a consistent - despite limited - net positive effect on market volumes, which oscillates between 0.56$\%$ (July) and, again, 1.16$\%$ (April). Similarly, as concerns the HW scenario, the results show a net positive effect, which ranges between 0.33$\%$ (January) and 0.89$\%$ (April). When looking at the Policy scenario, the results show a slightly different effect, being closely aligned to the zero-line. The monthly average impact for October stands at 0.01$\%$, while the one for April and July would result into 0.021$\%$ and 0.02$\%$, respectively. However, in January, the effect is negative, pointing to the hypothesis that, given the current market structure, the amount of self-consumption by RECs would outweigh the amount of energy injected by RECs into the grid in the policy scenario with a higher number of prosumers in the system\footnote{In Policy scenario in NORD: 89583 total prosumers from four categories. In HW scenario: 26338 total prosumers from four categories. In BU scenario: 2125 total prosumers from four categories.} (demand side) and greater capacity deployed \footnote{5 GW compared to 1.47 GW and 0.119 GW} (supply side), thus indicating that RECs would eventually slightly reduce the amount of equilibrium volumes traded on the wholesale power market. In particular, this might be explained by the larger frequency of relevant downward spikes from self-consumed energy that dominate when observing the impacts for the policy scenario case. \\

\begin{figure}[htbp]
\centering
\includegraphics[width=1\linewidth]{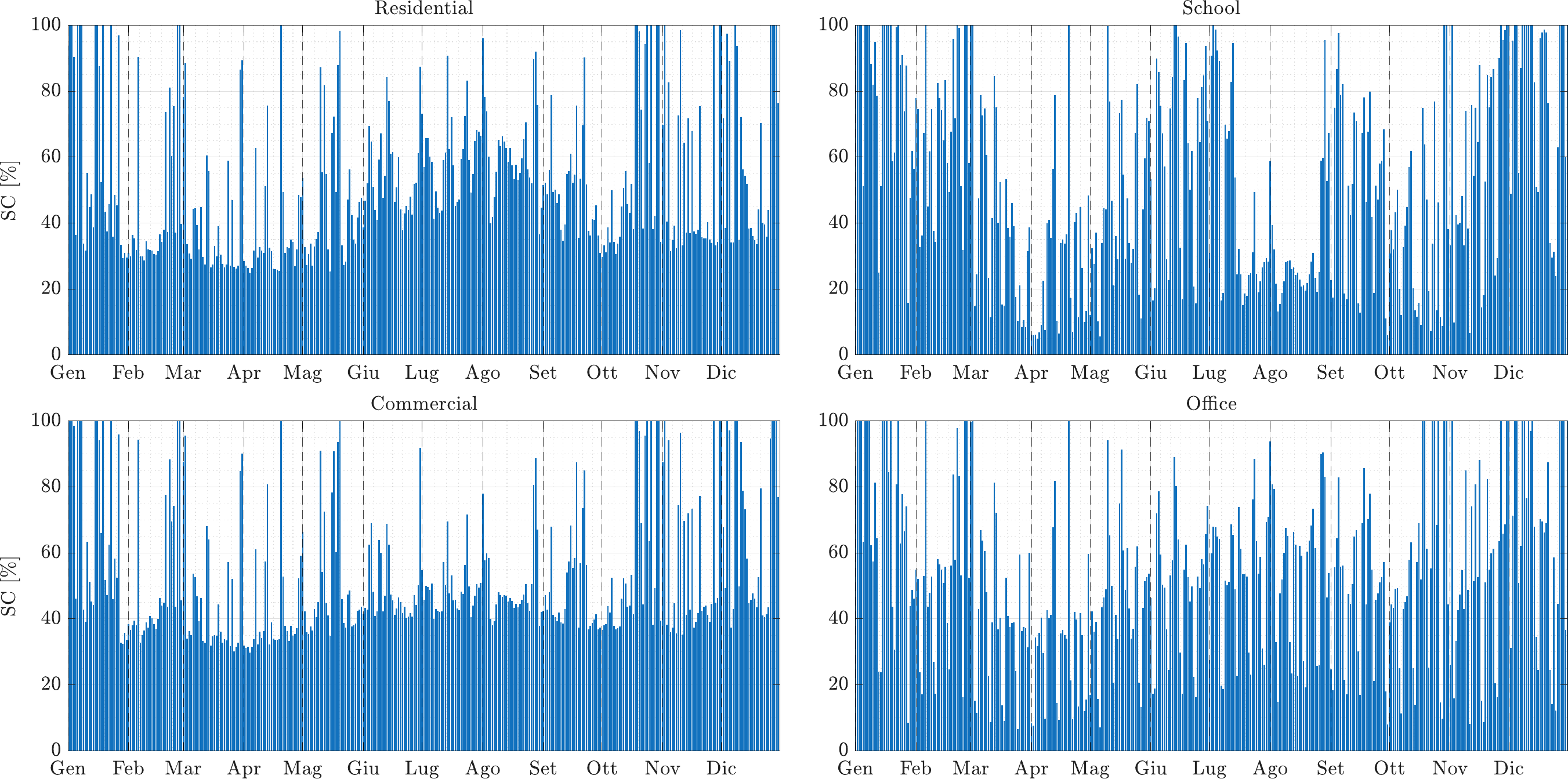}
\caption{\label{fig:SC-of-indiv-prosumer-categories} Daily self-consumption rates for all prosumer categories in 12 months of the year.}
\end{figure}

In Figure \ref{fig:NORD_12_windows}, the notable difference between the magnitude of the equilibrium effects in winter (smaller spike density) and other seasons (greater spike density) can be observed. A combination of the low solar irradiance while high-self consumption in winter induces this trend. First, lower solar irradiation in winter lead to fewer quantities of energy to be injected into the grid by RECs. Second, self-consumption rate by RECs is higher during winter. Figure \ref{fig:SC-of-indiv-prosumer-categories} illustrates daily self-consumption levels for all prosumer categories in different months. All categories self-consume on average more energy in January than in April, July or October. As a result, the RECs' self-consumption effect on the demand side of the market is relatively bigger than the RECs' injection effect on the supply side of the market in winter than in other seasons. In our engineering model, we added electrified heating and cooling to the load profiles of public (schools), SME (commercial), and NPO (offices) categories\footnote{However, residential prosumers own just electrified cooling systems in our model and not heating systems, similar to the actual status quo in Italian households.}. Therefore, in Figure \ref{fig:SC-of-indiv-prosumer-categories}, we observe that heating needs increase self-consumption during winter, mostly, for only three, albeit REC dominating (Figure \ref{fig:avg_capacity_shares}), categories of prosumers. \\

\begin{figure}[htbp]
\centering
\includegraphics[width=1\linewidth]{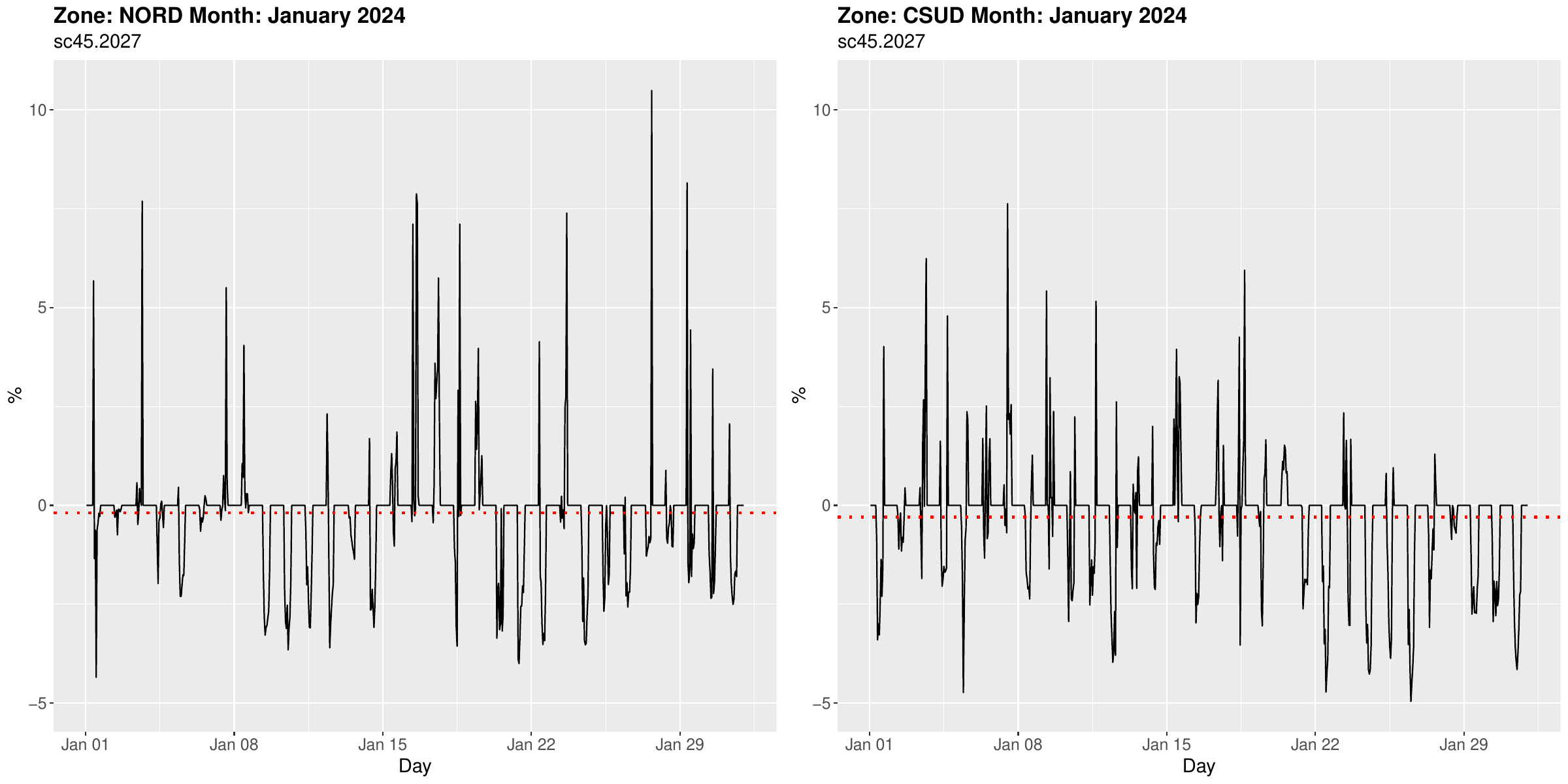}
\caption{\label{fig:impact_POLICY_scenario_NORD_vs_CSUD}
Comparison between NORD and CSUD: hourly percentage impact of RECs on equilibrium quantities, by assuming a homogeneous 45$\%$ self-consumption rate for all categories of prosumers. Month: January 2024. Scenario: \textit{sc45.2027}.}
\end{figure}

This observed "REC winter effect" derived for the Policy scenario (\textit{sc45.2027}) - is further explored in Figure \ref{fig:impact_POLICY_scenario_NORD_vs_CSUD}, which compares NORD and CSUD market zones for the same time interval (January 2024). Despite diverse impacts due to the configuration of each market zone and specific generation mix, the empirical finding of the negative average effect for NORD is confirmed for CSUD, too, under the assumption of homogeneous 45$\%$ self-consumption rate applied to all categories of prosumers. \\

\begin{figure}[htbp]
\centering
\includegraphics[width=1\linewidth]{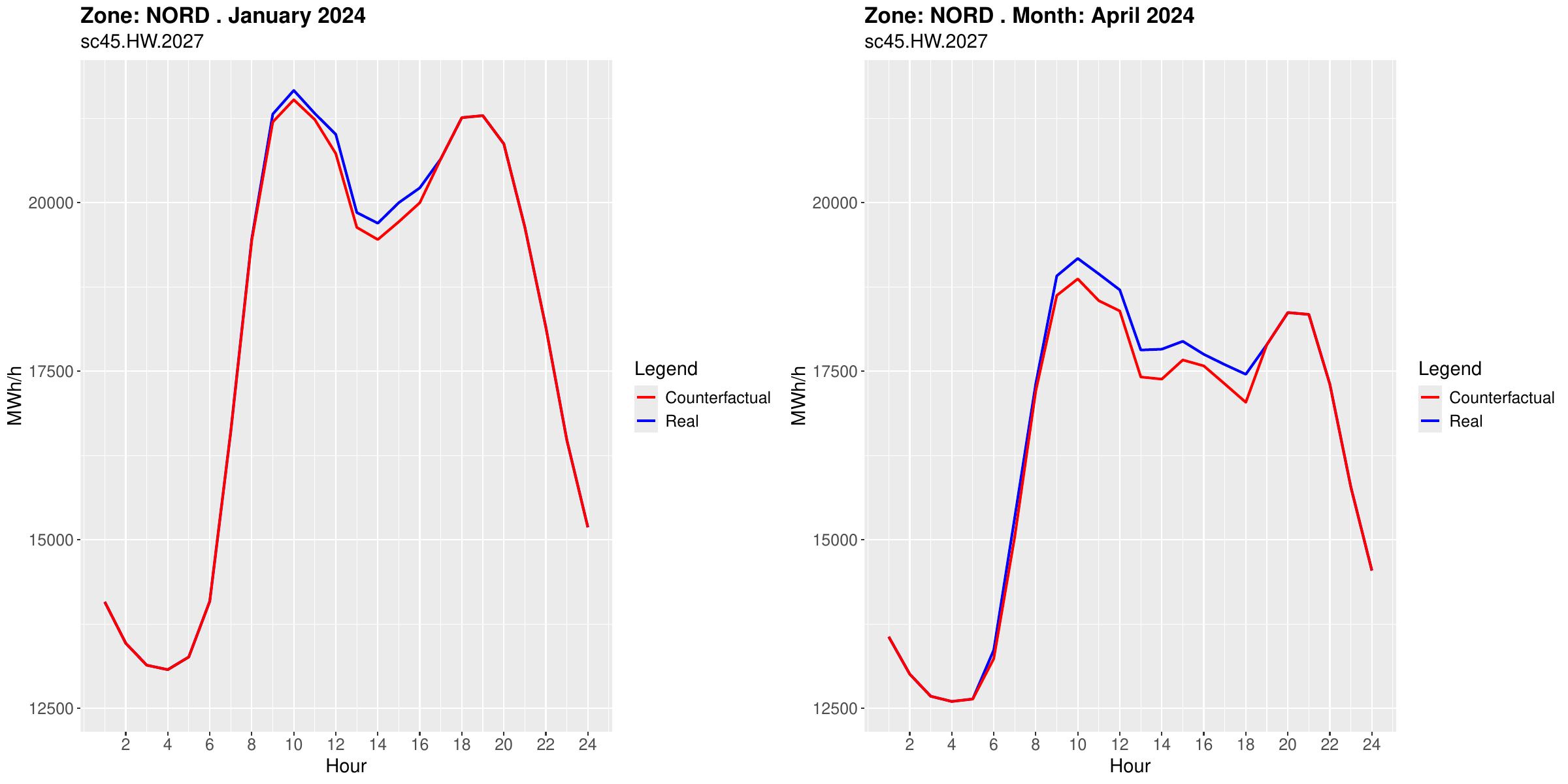}
\caption{\label{fig:impact_HALFWAY_NORD_Jan24_Apr24_hourly_profile}Profile of average hourly impact on quantities by RECs in NORD for both actual and counterfactual scenarios, assuming a homogeneous 45$\%$ self-consumption rate for all categories of prosumers. Periods: January and April 2024. Scenario: \textit{sc45.HW.2027}.}
\end{figure}

By focusing on the NORD zone and delving into seasonal variations, Figure \ref{fig:impact_HALFWAY_NORD_Jan24_Apr24_hourly_profile} displays the average profile of actual and counterfactual market quantities by hourly settlement period for January and April 2024. Under the HW scenario assumption, RECs deliver a net positive effect on actual market quantities only for central hours of the day in January (left panel). In contrast, during a typical spring month (April), the positive effect on market quantities is already visible since 6am, lasting until 7.30pm, in line with the fading sunlight (right panel). Thus, the scale of the impact of RECs is, as expected, stronger in April, because of the larger penetration of RES from RECs and their more effective displacement of expensive thermal generation due to the merit-order effect. \\

We further investigate the latter finding by analyzing the percentage
\textit{relative difference} between \emph{average} real and counterfactual
hourly quantity profiles, as stemming from the outcomes displayed in
Figure~\ref{fig:impact_HALFWAY_NORD_Jan24_Apr24_hourly_profile}, and by applying the smoothing procedure described at the beginning of this Section.
Similarly to the hourly percentage impact on equilibrium quantities defined
in Eq.~(\ref{eq:impact}), this indicator is computed as the relative deviation
between average observed and synthetic quantities over the selected time
window. \\

Formally, for each settlement hour $h$, let $\bar{Q}_{actual,h}$ and
$\bar{Q}_{synt,h}$ denote the average cleared quantities in the actual and
counterfactual markets, respectively, computed over all days in the
reference period. The relative difference is then defined as:
\begin{align}
\text{RelDiff}_{h} =
\frac{\bar{Q}_{actual,h} - \bar{Q}_{synt,h}}
{\bar{Q}_{synt,h}} \times 100.
\end{align}

By construction, a positive relative difference indicates that, on average,
market equilibrium volumes are higher in the presence of RECs compared to
the no-REC baseline, while a negative value signals a contraction in traded
quantities, typically associated with a prevailing effect of REC
self-consumption. This metric enables a transparent assessment of both the
magnitude and the temporal (intra-day and seasonal) heterogeneity of
REC-driven deviations from counterfactual market equilibria.\\

\begin{figure}[htbp]
\centering
\includegraphics[width=1\linewidth]{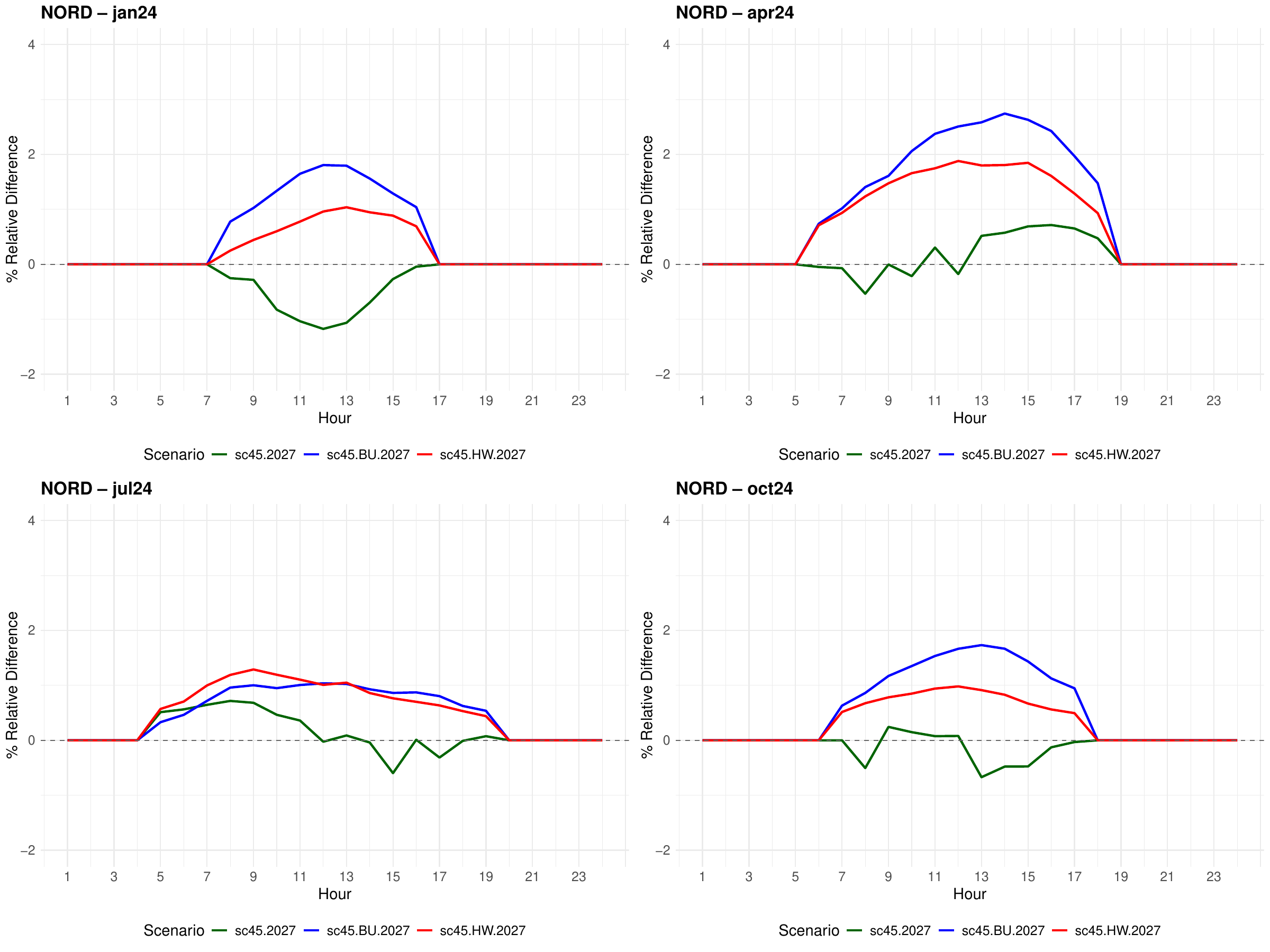}
\caption{\label{fig:NORD_ADP_rel_diff} Percentage relative difference of average hourly impact on quantities from RECs in the NORD. The outcomes are disentangled by settlement period, and displayed by month for each designed scenario.}
\end{figure}

Yet, Figure \ref{fig:NORD_ADP_rel_diff} shows that both the HW (\textit{sc45.HW.2027}) and the BU (\textit{sc45.BU.2027}) scenarios display a positive relative percentage effect of RECs on actual market quantities. Conversely, the Policy scenario (\textit{sc45.2027}) falls under the zero line for several settlement periods, thus suggesting that RECs have the potential to reduce traded market volumes in the Italian DA power market under large levels of deployment. Nevertheless, the magnitude of such effect has a strong seasonal dependency. Indeed, the largest magnitude of relative reduction in DA equilibrium quantities occurs in January during off-peak hours (10 a.m.--14 p.m.), as a result of the aforementioned "REC winter effect". These reductions during off-peak hours are related to loads from institutional REC members (public, SME, NPO). \\

\begin{figure}[htbp]
\centering
\includegraphics[width=0.7\linewidth]{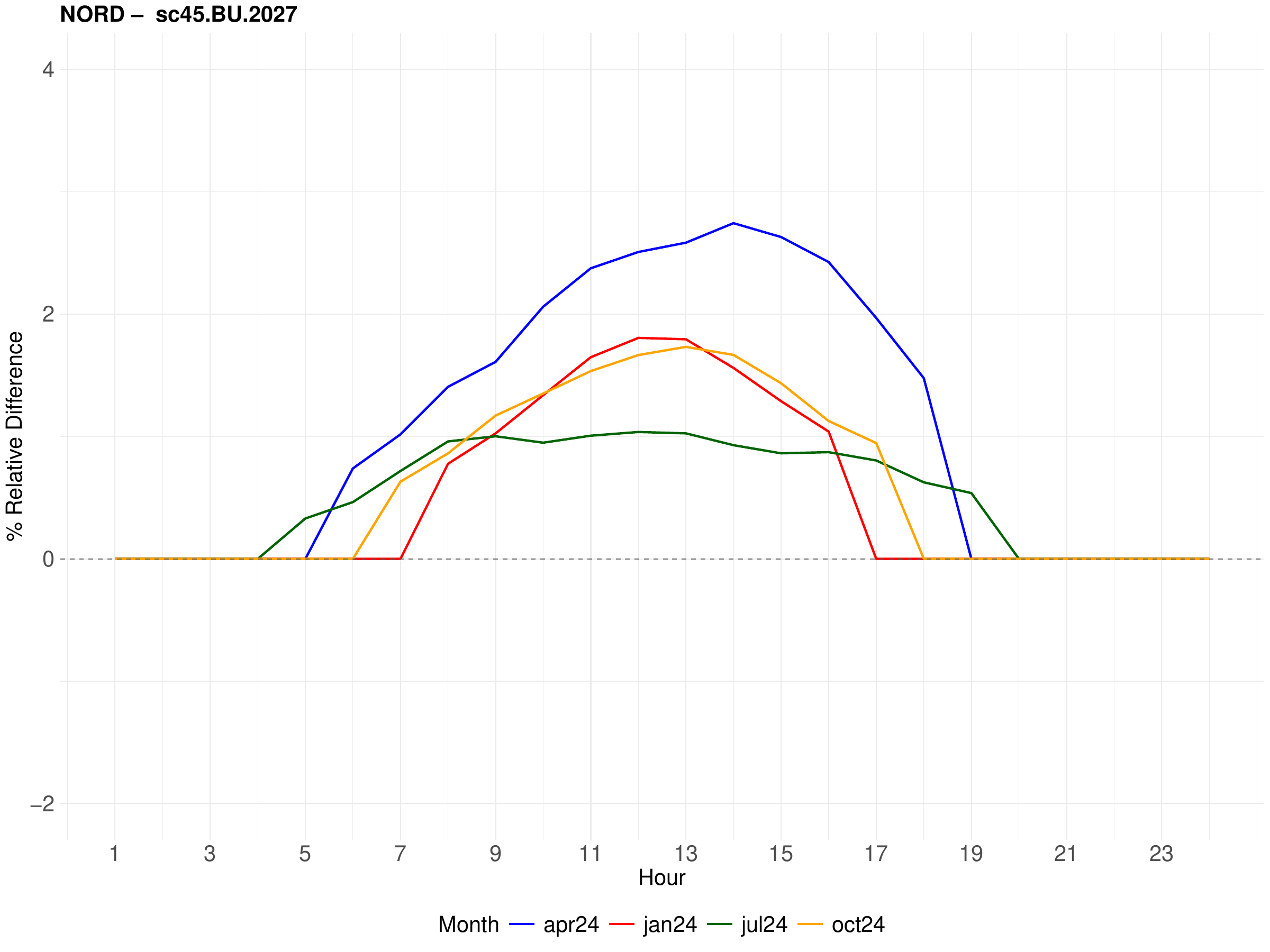}
\caption{\label{fig:NORD_BAU_rel_diff_by_month} Percentage relative difference between actual and counterfactual average hourly quantities in the NORD for each month, by hour. Scenario: \textit{sc45.BU.2027}.}
\end{figure}

\begin{figure}[htbp]
\centering
\includegraphics[width=0.7\linewidth]{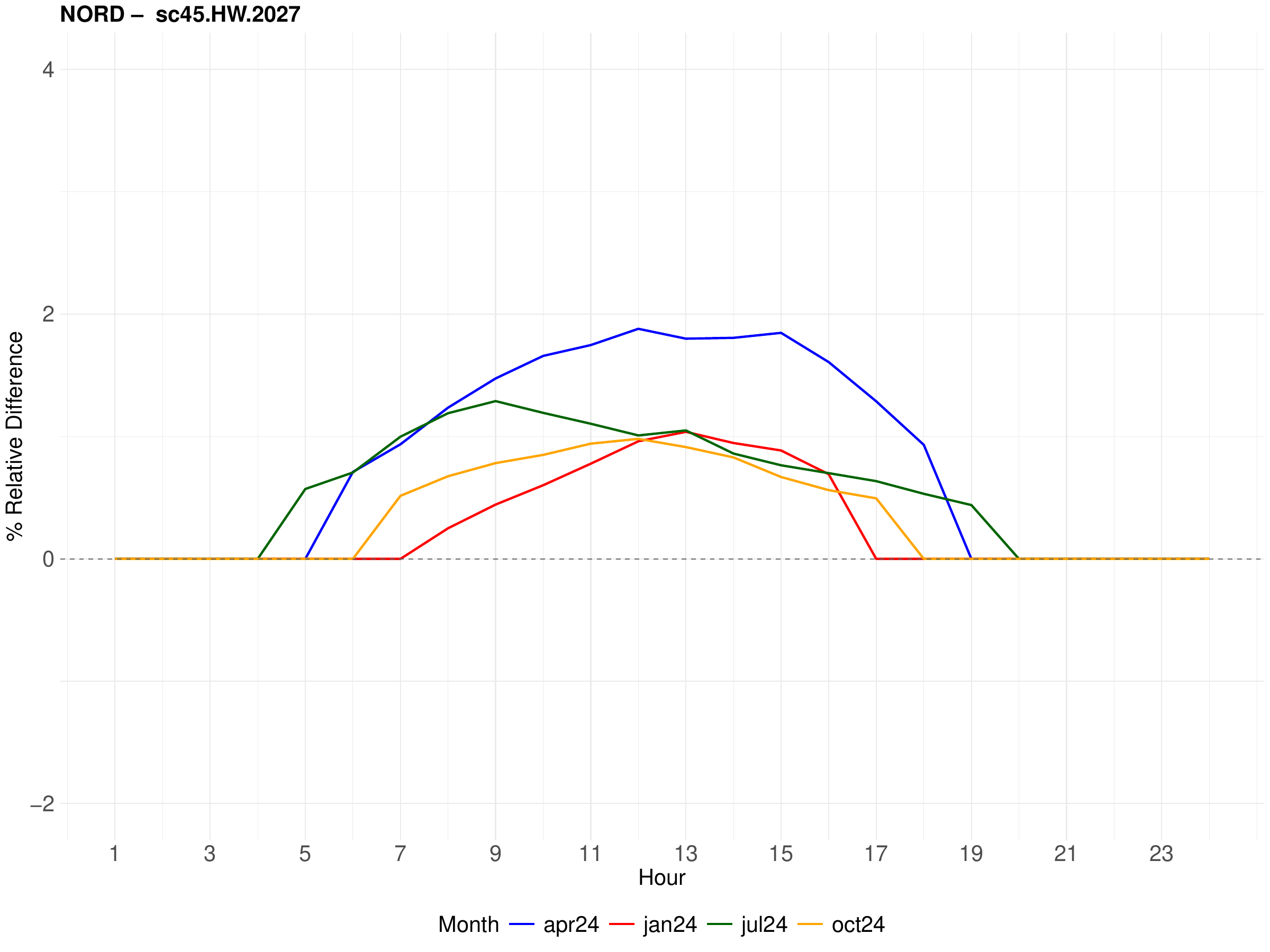}
\caption{\label{fig:NORD_HW_rel_diff_by_month} Percentage relative difference between actual and counterfactual average hourly quantities in the NORD for each month, by hour. Scenario: \textit{sc45.HW.2027}.}
\end{figure}

In detail, Figure \ref{fig:NORD_BAU_rel_diff_by_month} displays results of the BU scenario for the selected months. The chart shows that April reports the largest effect of RECs on actual market quantities. As regards summer time (July), the results show a relatively stable impact of RECs on market quantities across settlement periods, averaging nearly 1$\%$ during peak hours. Similar patterns are evidenced both for October and January, with a slight prevailing impact in October, mostly due to slightly longer daylights (with consequent higher production from PV) compared to January. Figure \ref{fig:NORD_HW_rel_diff_by_month} shows comparable results for the HW scenario: in July, the impact of RECs on market quantities emerges earlier (4–6 a.m.), mirrors the April effect until 8 a.m., then weakens during the day, and resurfaces around 7 p.m. due to extended sunlight - similarly to the BU scenario. \\

\begin{figure}[htbp]
\centering
\includegraphics[width=0.7\linewidth]{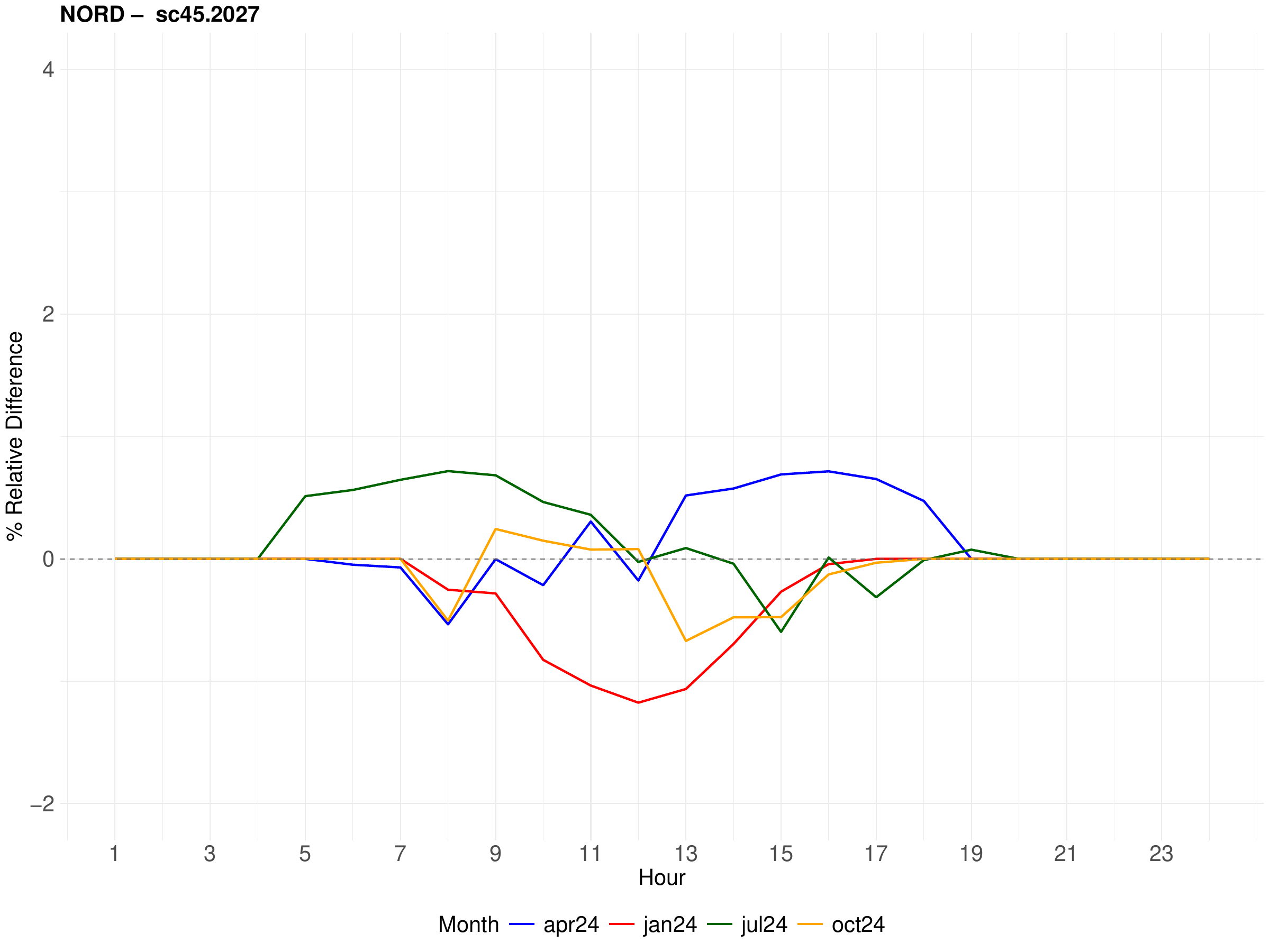}
\caption{\label{fig:NORD_POLICY_rel_diff_by_month} Percentage relative difference between actual and counterfactual average hourly quantities in the NORD for each month, by hour. Scenario: \textit{sc45.2027}.}
\end{figure}

As for the Policy scenario (Figure \ref{fig:NORD_POLICY_rel_diff_by_month}), the impact of RECs in April would oscillate around the zero-line resembling "zig-zags" until midday, thus witnessing a balanced effect between lower market demand (due to larger self-consumption by RECs) and higher electricity supply by RECs. Instead, the impact of RECs would be more pronounced during the rest of the day. Interestingly, the effect of RECs in July is the opposite, indicating a prevailing effect of energy injected by RECs until 1 p.m., to then switch back to the dominance of the effect of reduced market demand resembling "zig-zags", which could be explained by the strong need of power consumption for space cooling during summer. Indeed, we notice that the green line constantly falls before dropping below zero after 2 pm. We observe that the effect of reduced market demand (explicitly evident within the Policy scenario) mimics the hourly self-consumption rates of different prosumers' categories. For example, in July, self-consumption rates in the afternoon (1pm - 7pm) for all prosumers' categories are higher than the morning self-consumption rates (6am - 12pm).\footnote{Public prosumers raise self-consumption rates on average at 2.3\%, SME prosumers at 2.7\%, residential prosumers at 18.5\%, and NPO prosumers at 1.1\%.} In contrast, in April, self-consumption rates are, on average, higher in the morning than in the afternoon. We think that the specific composition of prosumers in RECs (Figure \ref{fig:avg_capacity_shares}) can also affect the impact's dynamics, \textcolor{black}{which we investigate in the sensitivity sub-section.} \\

\begin{figure}[htbp]
\centering
\includegraphics[width=0.7\linewidth]{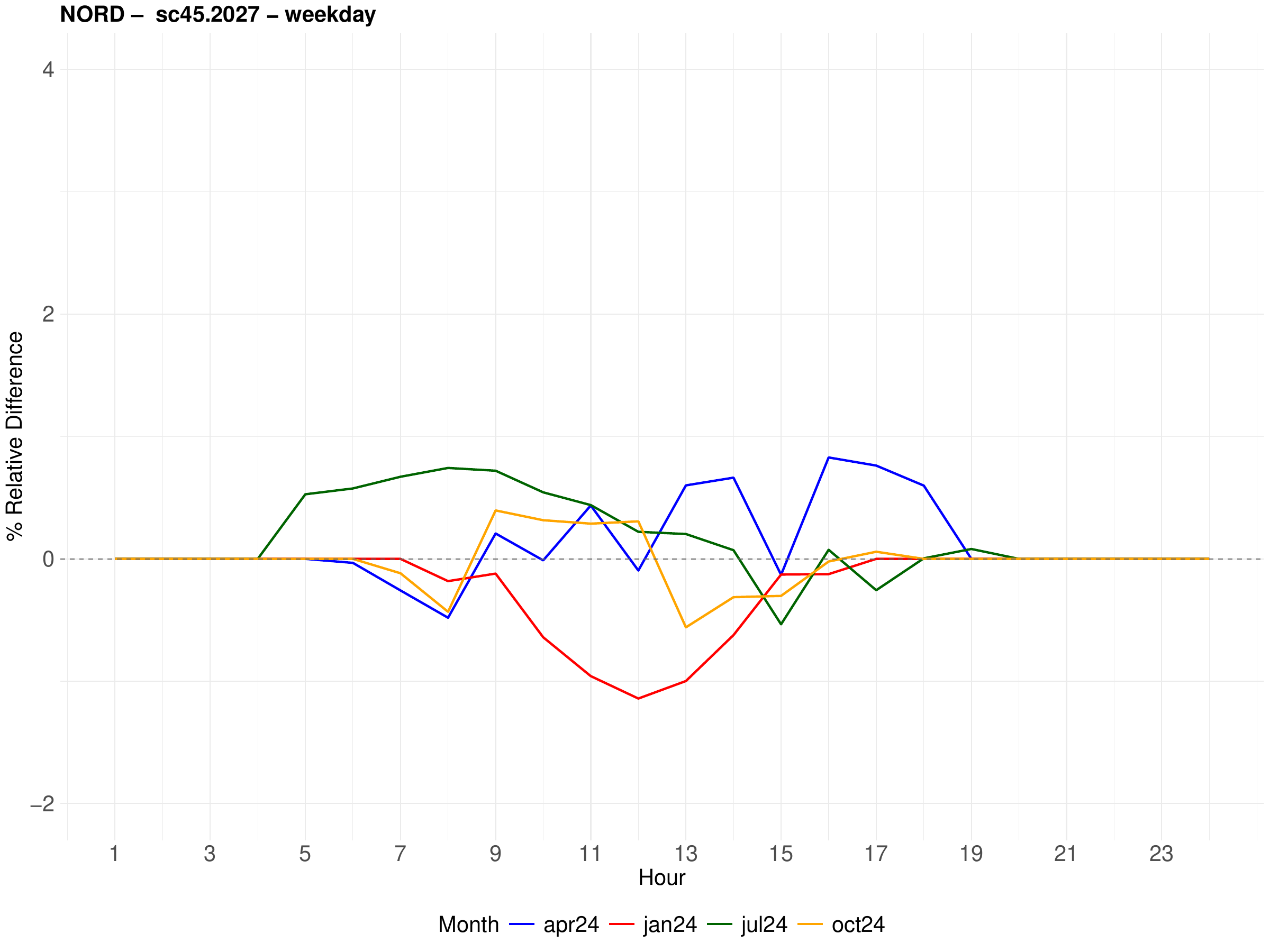}
\caption{\label{fig:NORD_POLICY_rel_diff_by_month_weekday} Percentage relative difference between actual and counterfactual average hourly quantities in NORD for each month during weekdays. Scenario: \textit{sc45.2027}.}
\end{figure}

\begin{figure}[htbp]
\centering
\includegraphics[width=0.7\linewidth]{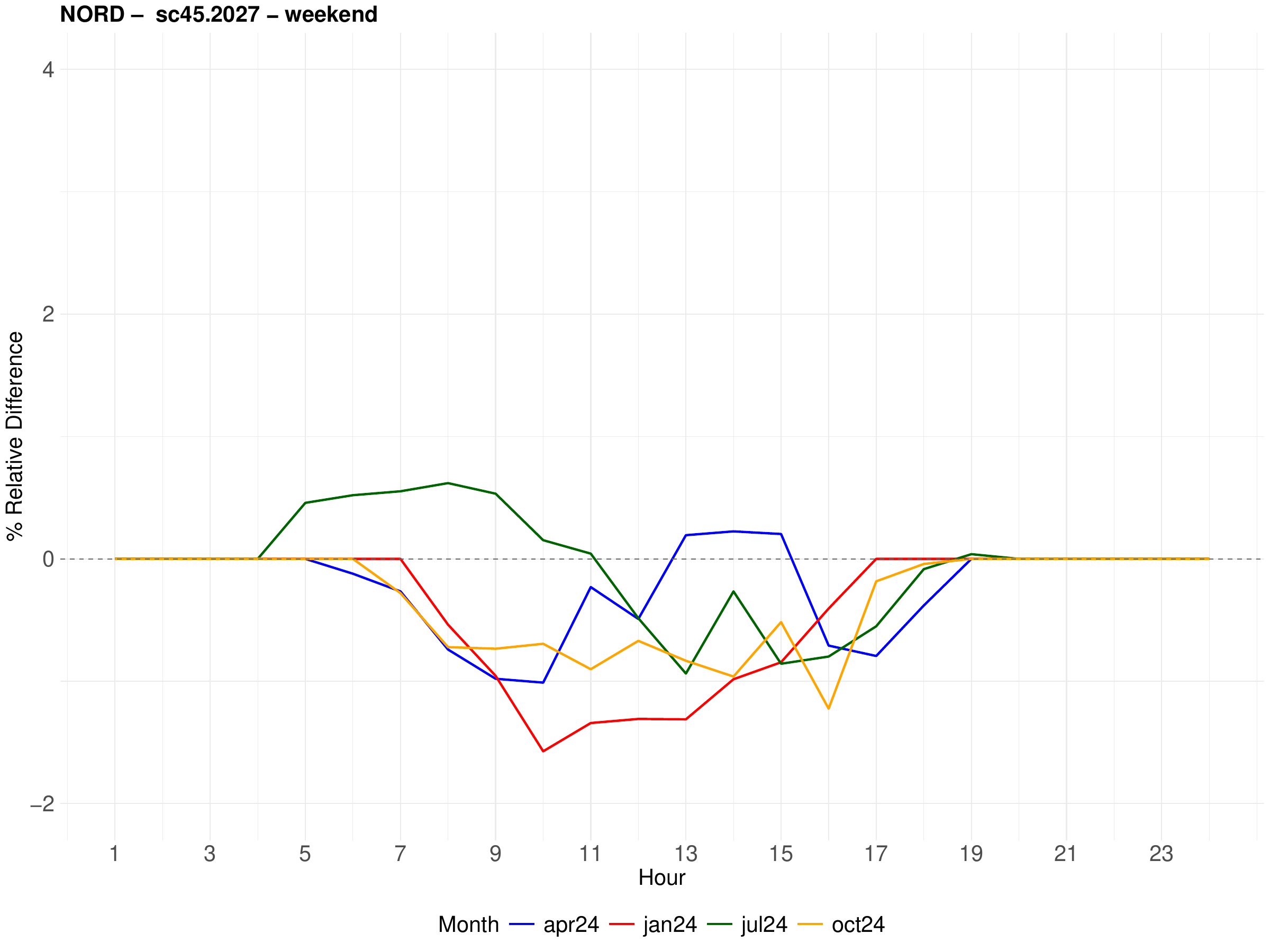}
\caption{\label{fig:NORD_POLICY_rel_diff_by_month_weekend} Percentage relative difference between actual and counterfactual average hourly quantities in NORD for each month during weekends. Scenario: \textit{sc45.2027}.}
\end{figure}

Lastly, we break down the results by showcasing the effects on weekdays and weekends. Concerning the BU scenario (\nameref{Appendix E}), the monthly patterns for weekdays (Fig. \ref{fig:NORD_BAU_rel_diff_by_month_weekday}) are consistent with aggregate findings for the full sample of Figure \ref{fig:NORD_BAU_rel_diff_by_month}. For weekdays, the positive impact of RECs on market quantities rather surpasses the 3$\%$ threshold. However, during weekends (Fig. \ref{fig:NORD_BAU_rel_diff_by_month_weekend}), the relative percentage difference is attenuated. Similarly, the results for weekdays under the HW scenario (Fig. \ref{fig:NORD_HW_rel_diff_by_month_weekday}) are consistent with outcomes reported in Figure \ref{fig:NORD_HW_rel_diff_by_month}. Regarding weekends (Fig. \ref{fig:NORD_HW_rel_diff_by_month_weekend}), the relative increasing effect of RECs on market quantities takes place in April, despite only after 10 a.m. Similarly, weekdays' results for the Policy scenario (Fig. \ref{fig:NORD_POLICY_rel_diff_by_month_weekday}) are consistent with aggregate outcomes of Figure \ref{fig:NORD_POLICY_rel_diff_by_month}. As for weekends (Fig. \ref{fig:NORD_POLICY_rel_diff_by_month_weekend}), the diminishing effect on market quantities due to RECs would turn out to be even more influential across seasons.\\

\subsubsection{Sensitivity analysis}
\begin{figure}[htbp]
\centering
\includegraphics[width=0.95\linewidth]{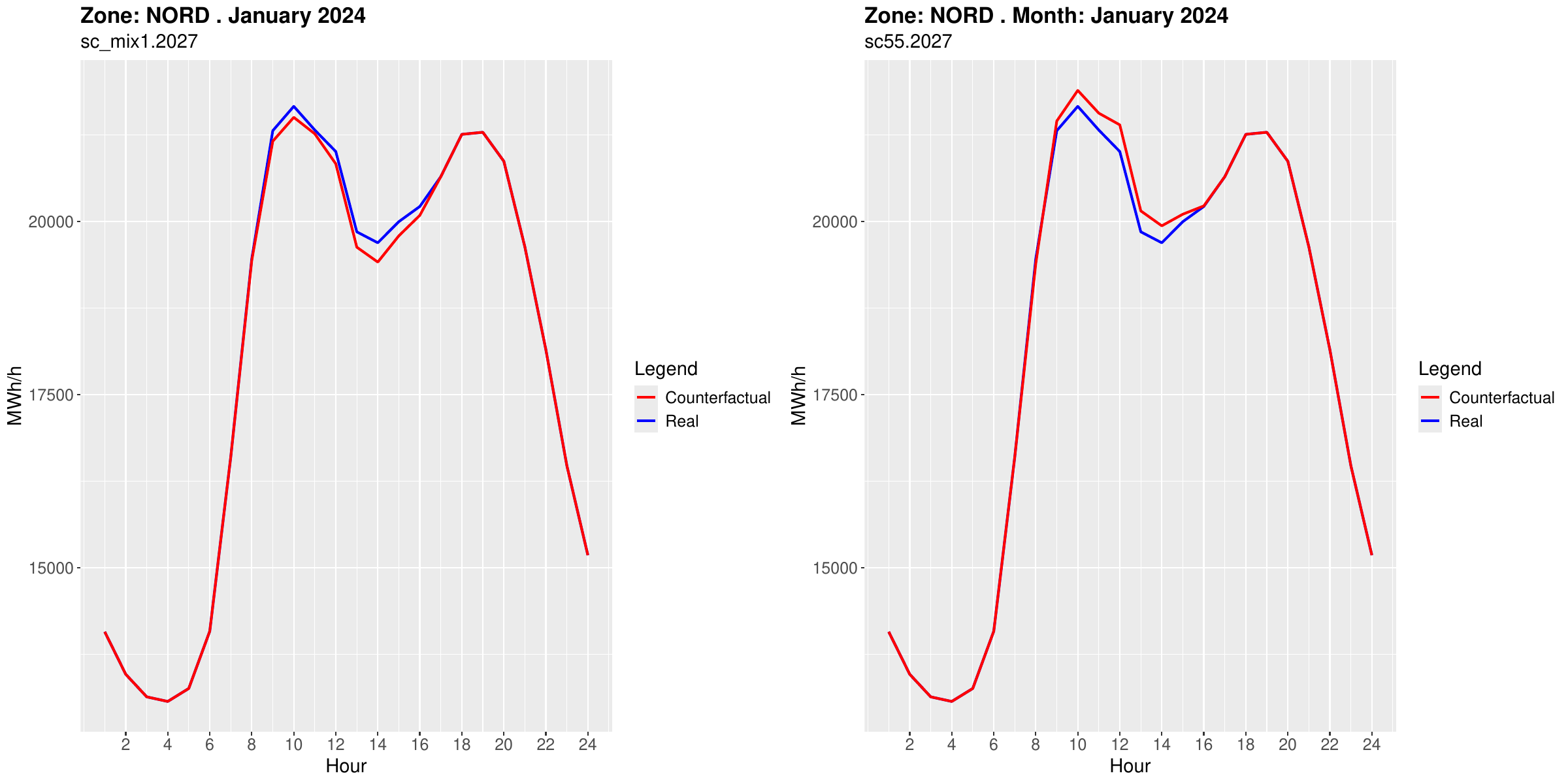}
\caption{\label{fig:NORD_ADP_jan24_mix1_vs_mix2} Profile of average hourly quantities in NORD for both actual and counterfactual scenarios. Period: January 2024. Comparison between mixed scenario and policy scenario: \textit{sc$\_$mix1.2027} vs. \textit{sc55.2027}.}
\end{figure}

\begin{figure}[htbp]
\centering
\includegraphics[width=0.99\linewidth]{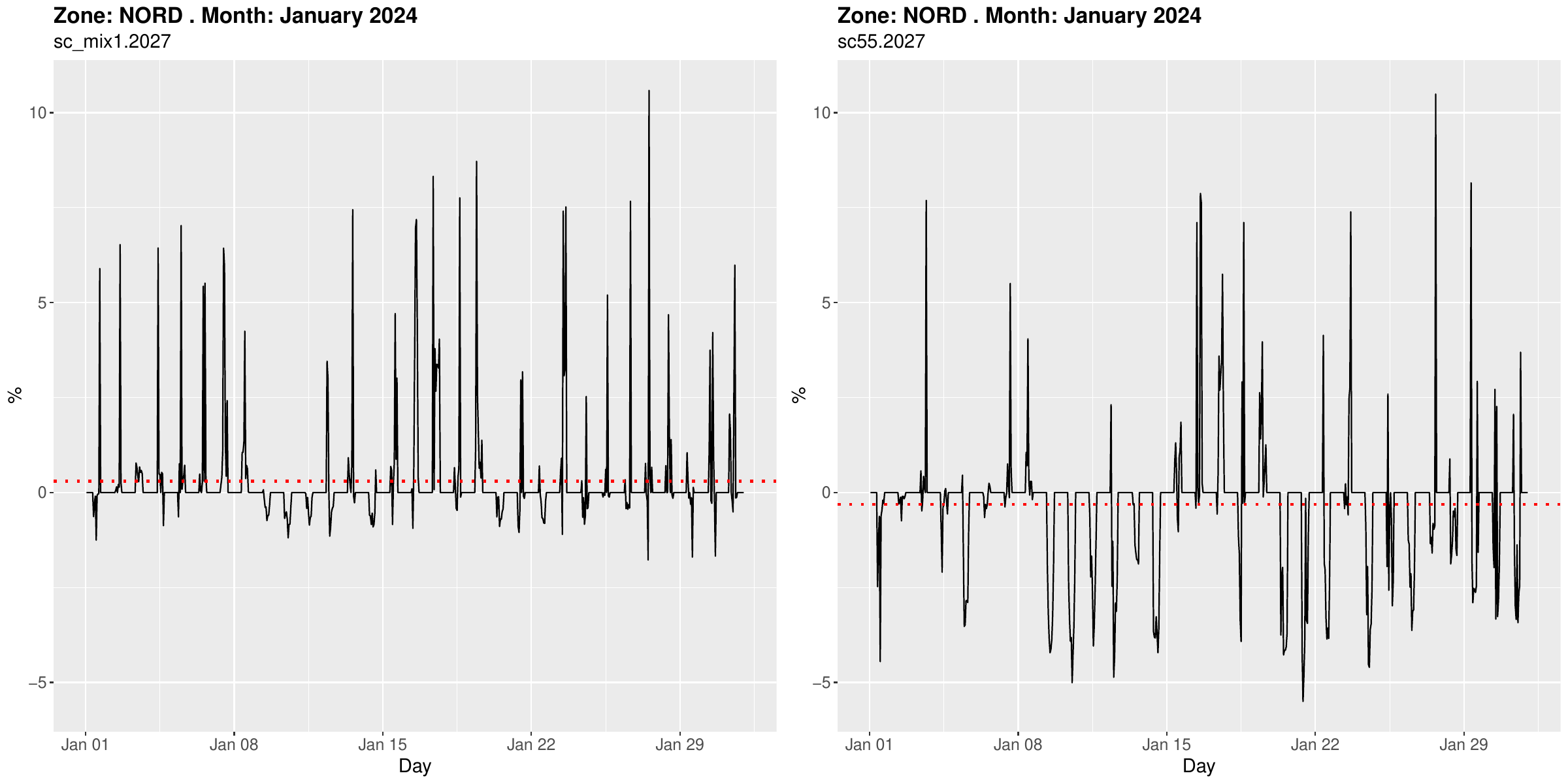}
\caption{\label{fig:sc_mix1_mix2_NORD_Jan24}Hourly percentage impact on equilibrium quantities by RECs: comparison between 1st mixed scenario and policy scenario:  \textit{sc$\_$mix1.2027} vs \textit{sc55.2027} for NORD. Period: January 2024.}
\end{figure}

\begin{figure}[h]
\centering
\includegraphics[width=0.95\linewidth]{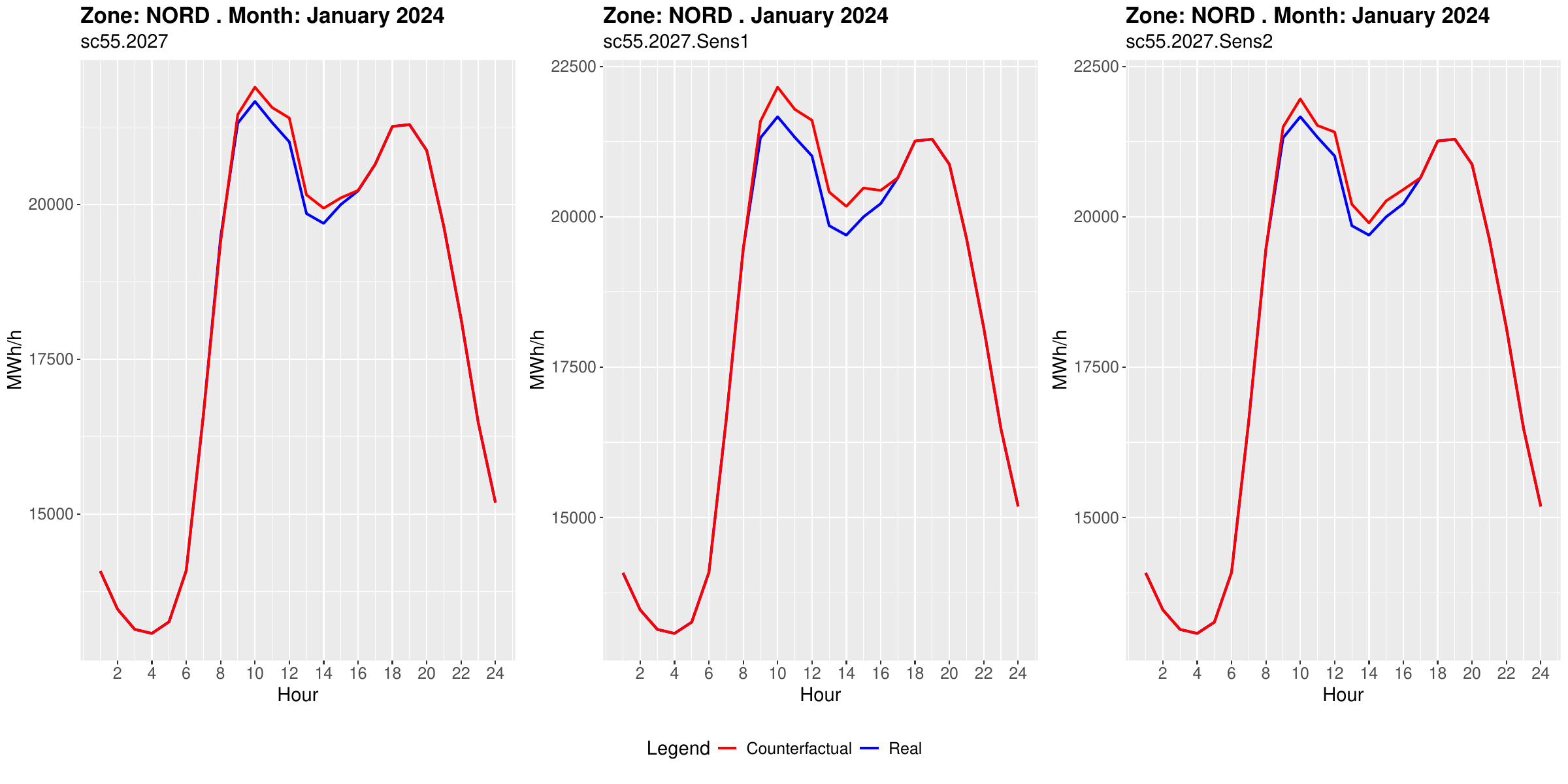}
\caption{\label{fig:NORD_ADP_jan24_mix2vsSens1vsSens2} Profile of average daily profile in NORD for both actual and counterfactual scenarios. Period: January 2024. Comparison between the policy scenario, the Sens1 and Sens2 sensitivity scenarios: \textit{sc55.2027} vs. \textit{sc55.2027.Sens1} vs. \textit{sc55.2027.Sens2}.}
\end{figure}

\begin{figure}[h]
\centering
\includegraphics[width=0.95\linewidth]{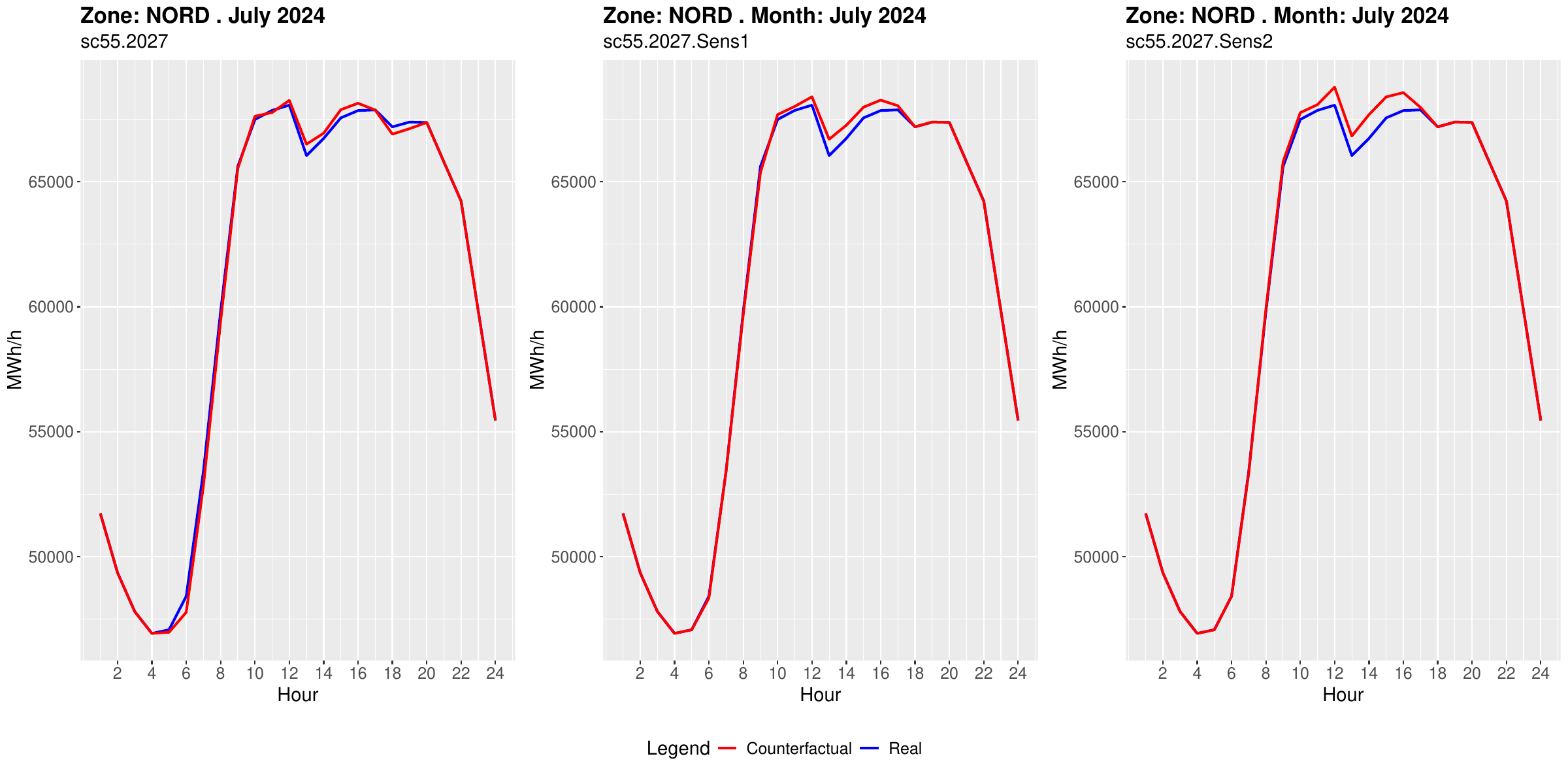}
\caption{\label{fig:NORD_ADP_jul24_mix2vsSens1vsSens2} Profile of average daily profile in NORD for both actual and counterfactual scenarios. Period: July 2024. Comparison between the policy scenario, the Sens1 and Sens2 sensitivity scenarios: \textit{sc55.2027} vs. \textit{sc55.2027.Sens1} vs. \textit{sc55.2027.Sens2}.}
\end{figure}

\textcolor{black}{For the sensitivity analysis, we test our findings using  two strategies applied to the North zone. First, we relax the assumption of the homogeneous self-consumption rate for all categories of prosumers and introduce heterogeneous self-consumption rates. Second, we modify the scaling parameters, derived from the actual deployment in Italy, by changing shares of diverse prosumer categories within a prototypical REC.}
\begin{itemize}[noitemsep]
    \item Heterogeneous self-consumption rates are indicated as the 10th "mixed" scenario in Table \ref{tab:scenarios}. We compare it with the 3d policy scenario in Table \ref{tab:scenarios}. As shown in the left panel of Figure \ref{fig:NORD_ADP_jan24_mix1_vs_mix2} representing January, the "mixed" scenario (\textit{sc$\_$mix1.2027}) with heterogeneous self-consumption rates indicates that the increasing effect on market quantities due to the energy injected by RECs would outweigh the diminishing effect on market demand given by their self-consumption, thus leading to relatively higher market quantities during peak-hours. Instead, the effect works in the opposite direction under the 3d policy scenario (\textit{sc55.2027}), as evidenced in the right panel of the chart, which models the scenario of a massive deployment of RECs in Northern Italy by assuming a homogeneous 55\% rate of self-consumption. The self-consumption rates in the 3d policy scenario (right panel) are higher than the rates of mixed scenario (left panel) for all  categories except for SMEs. The same intuition applies to Figure \ref{fig:sc_mix1_mix2_NORD_Jan24}, showing that the range of the overall effect of RECs in NORD for January would situate, in average terms, between 0.29$\%$ (left panel) and -0.3$\%$ (right panel) under our sensitivity analysis framework.\\

  \item \textcolor{black}{The actual prosumer composition of a typical REC for NORD is illustrated in Figure \ref{fig:avg_capacity_shares}.\footnote{\textcolor{black}{Public: 41.5\%, Residential: 11.2\%, SME: 16.6\%, NPO: 9.1\%, Standalone: 21.5\%.}} We calibrated these shares by creating a synthetic public-dominant REC (\textit{Sens1}) and a synthetic residential-dominant REC (\textit{Sens2}). In \textit{Sens1}, the capacity deployed by the public prosumers prevails (79.7\%), while SME prosumers and standalone facilities are eliminated. By contrast, in \textit{Sens2}, the share of residential prosumers prevail (74.2\%), while public prosumers and standalone facilities are eliminated. The shares of the other prosumer categories are left unchanged. Figure \ref{fig:NORD_ADP_jan24_mix2vsSens1vsSens2} compares the impact of REC deployment on equilibrium volumes for the actual prosumer composition (left panel) with the public-dominant composition (center panel) and the residential-dominant composition (right panel) in January. The residential-dominant composition does not exhibit significant differences from the actual REC composition. In contrast, the deployment of public-dominant RECs yields a substantial reduction in equilibrium volumes compared to the actual REC composition throughout all solar hours. This can be explained by the greater electricity loads of public prosumers during winter compared to residential prosumers (see Figure \ref{fig:avg_daily_energy_jan}). Figure \ref{fig:NORD_ADP_jul24_mix2vsSens1vsSens2} shows a similar sensitivity configuration, albeit for July. Both public-dominant RECs (center panel) and residential-dominant RECs (right panel) deepen the reduction in equilibrium volumes compared to the actual composition (left panel). In the actual composition plot, we also observe that at 6 a.m., market volumes increase by 644 MWh, while at 7 a.m., they increase by 590 MWh. In the evening, market volumes increase by 291 MWh at 6 p.m. and by 265 MWh at 7 p.m. Notably, these raises in market volumes observed during early morning hours and evening peak hours disappear in public-dominant and residential-dominant compositions, possibly due to the elimination of standalone facilities in both \textit{Sens1} and \textit{Sens2}. This suggests that standalone facilities within RECs may be important for satisfying steep load increases during early summer mornings and summer evenings, while their absence eliminates this possibility. In contrast to January, Figure \ref{fig:NORD_ADP_jul24_mix2vsSens1vsSens2} shows that residential-dominant RECs deepen volume reductions more than public-dominant RECs. This difference may be attributed to the beginning of the summer break for the modeled prosumer category (i.e., schools), which decreases the self-consumed energy on a daily average.}
\end{itemize}

\newpage

\section{Discussion}\label{sec:disc}

First, our results show that the impact of RECs on power market is focused primarily in the NORD and CSUD market zones, with current negligible or absent effects in the other five physical market zones. This finding is closely related to the current distribution of REC projects, which are predominantly located in such two areas (see Fig.~\ref{fig:avg_capacity_one}). Relatively slow and complex authorization and acceptance processes, institutional burdens, and limited technical support in other areas prevent broader deployment, confirming the need for targeted support policies to unlock REC potential across all regions. The similar finding has been reported by Zhu et al. \citeyearpar{zhuItalianRenewableEnergy2025} and Musolino et al. \citeyearpar{musolinoThreeCaseStudies2023}.\\

Second, the temporal dimension of REC impacts reveals strong seasonal and hourly patterns, driven by the interaction between solar production profiles and the assumed self-consumption rates. Our simulations for BU and HW scenarios indicate that certain early morning and evening hours - especially in spring and summer months - exhibit a relatively significant increase in market quantities, thereby highlighting the potential for RECs to better fulfill peak demand pressures in a decarbonizing system.  Similarly, although including BESS, Riaz et al. \citeyearpar{riazGenericDemandModel2019a} modeled a summer week in the Australian wholesale market. They investigated the aggregate effect of a large number of prosumers on the load profile and found that increased prosumer participation with BESS flattens demand profiles, enhancing voltage stability and reducing the need for gas peaking plants. However, the exception in their study is the scenario with low demand and excess RES generation. It leads to increased aggregate demand due to battery charging, thus decreasing the stability margins of the system. Boccard and Goetz \citeyearpar{boccardPowerTradingEnergy2025} argue that excessive RES generation from REC prosumers entails increased power exchanges with the grid, although they do not specify in which direction. \\

\textcolor{black}{Third, the choice of self-consumption rates in our scenario design reflects plausible and realistic efficiency levels for different types of REC participants. The purpose is to address the robustness through the comparison of three representative self-consumption levels (45$\%$, 50$\%$, and 55$\%$), which span conservative to more favourable REC operating conditions.} These values are supported by empirical evidence on current adoption of battery energy storage systems (BESS) in Italy, which remains lower than the EU average.

Nevertheless, as shown in recent studies \citep{veronese2024, secchi2021}, optimal sizing of BESS can significantly increase self-consumption levels, particularly for small-scale prosumers. Several studies \citep{soiniImpactProsumerBattery2020a, schickRoleImpactProsumers2022, chenRecoveringFixedCosts2023a} generally agree that individual prosumers equipped with BESS could reduce large thermal generation as well as replace less efficient pumped-hydro storage only if optimized based on electricity system and market needs. Despite obvious benefits for REC members from a community BESS, its role for volumes of renewable energy supplied to the wholesale market is much more nuanced. Sarfarazi et al. \citeyearpar{sarfaraziAggregationHouseholdsCommunity2020}  show that while a profit-maximizing community BESS leaves prosumer RES dispatch unchanged, a self-sufficiency-maximizing BESS reduces RES injections and increases marginal thermal generation, highlighting the need for flexible, price-responsive demand to support self-consumption. Nevertheless, without BESS integration into communities, the system may face volatility risks and missed opportunities for optimization - particularly during high RES generation hours, when curtailment or negative prices may occur. \\

Fourth, our results for the policy scenario indicate that RECs, by enabling higher self-consumption, could also reduce the need for costly investments in grid infrastructure. This finding aligns with the conclusions by \textcolor{black}{\citet{fuentesgonzalezCommunityEnergyProjects2022}}, who modeled the effect of energy communities on transmission grid costs in Chile and found substantial system-wide savings. Backe et al. \citeyearpar{backeImpactEnergyCommunities2022} came to similar conclusions regarding the expansion of projected transmission capacity of the grid in six EU countries by 2060. Studies by Boccard and Goetz \citeyearpar{boccardPowerTradingEnergy2025} and Sarfarazi et. al. \citeyearpar{sarfaraziAggregationHouseholdsCommunity2020} reveal that RECs could reduce pressure on the distribution grid too. Nonetheless, this positive effect is contingent upon the size of a PV system and composition of REC members in the former study and pricing strategies and community BESS operation in the latter study. While our study does not explicitly focus on optimization of grid costs, the observed market impacts also suggest that RECs may alleviate pressure on the grid. \\

Fifth, nuanced PV sizing will be essential to fully realize the potential of REC deployment not only to reduce stress on the grid but also wholesale market prices. Boccard and Goertz \citeyearpar{boccardPowerTradingEnergy2025} found that the different sizing of PV systems in RECs could induce opposite impacts on the distribution system. For example, RECs with homogeneous profiles of residential prosumers, which install PV systems that cover 25\%--50\% of their load, reduce power exchanges with the grid, whereas those who install PV systems that cover 50\%--100\% of their load increase power exchanges with the grid due to the mismatch between energy injecting and self-consuming. The latter situation could create reverse flows and thus force additional costs for the DSOs that eventually will be shifted onto the non-members of RECs. In contrast, RECs with greater number of residential heterogeneous prosumers and RECs with public prosumers, which install PV systems that cover 25\% of their load, increase power exchanges with the grid, while those who install PV systems that cover 50\%--100\% of their load reduce power exchanges with the grid. This drastically opposite impacts on power exchanges hint at importance of modeling not only diverse PV system sizes but also diverse categories of prosumers, as we do. In our simulations, we used real-world averaged PV sizes of REC prosumers and producers in Italy (Figure \ref{fig:avg_capacity_one}) differentiated by categories and electricity zones.  \\

\section{Conclusions and Policy Implications}\label{sec:concl}

\textcolor{black}{In this study, we aimed to investigate the current impact (BU scenario) and projected impact (HW and Policy scenarios) of REC deployment on wholesale market equilibrium. The study began with extensive data collection to simulate representative technical models of energy communities across different Italian market zones. The energy injected and self-consumed by these representative RECs were then projected under the assumed deployment scenarios using real-world distribution shares derived from the compiled database. We subsequently conducted a counterfactual analysis by incorporating the aggregated energy injected and self-consumed by RECs into a short-run economic model, which uses real 2024 market bids and closely resembles the day-ahead market algorithm employed by the Italian Market Operator.} Our results show that REC deployment generates non-negligible effects in specific zones of the market (notably North and Central-South), while its overall system-wide impact remains quite moderate. Seasonal and hourly patterns confirm that self-consumption is a key driver of economic benefits from REC deployment at the macro scale. Specifically, the estimated impacts on market volumes lie within a narrow range, from a minimum of -0.3\% (January, Policy Scenario) to a maximum of +1.16\% (April, BU Scenario). \textcolor{black}{The robustness of the results with regard to the assumptions concerning self-consumption is supported by the 45\%, 50\% and 55\% values, which represent a plausible range of operating conditions for prosumers of RECs.} While both BU and HW scenarios yield consistently positive but limited net effects (up to +1.16\% and +0.89\%, respectively), the Policy scenario reveals mixed evidence: the impacts remain close to zero in most of the months considered, but turn slightly negative in January, suggesting that higher REC self-consumption can, under certain conditions, reduce the volumes available in the wholesale market. This observation becomes especially relevant when examining \textcolor{black}{daytime} peak/off-peak dynamics. The reduction in equilibrium volumes during \textcolor{black}{daytime} off-peak hours due to institutional prosumers within RECs (public, SMEs, NPOs) \textcolor{black}{may} reduce the need for RES curtailment while still having additional RES capacity deployed. \textcolor{black}{Conversely}, increased volumes during peak hours in spring and summer reduce the need for expensive thermal peaking plants. In contrast, when increases in volumes occur during \textcolor{black}{daytime} off-peak hours, as in scenarios with lower deployed capacity (BU and HW), RECs may adversely affect market outcomes. \\

Indeed, by extending the analysis to seasonal and intra-day variations, our results confirm that the magnitude of REC impacts varies substantially across months and hours. Under the BU scenario, April emerges as the most responsive period, as market volumes with RECs surpass volumes without RECs by up to 3\% during weekdays, while July exhibits a steadier impact close to 1\% during peak hours. Conversely, cold months display more muted effects. The main factors behind this trend are the lower solar irradiation and the greater heating needs supplied by electricity during cold months. Lastly, when increasing the self-consumption rates for all categories of prosumers (from 45\% to 55\%) in line with Italian policy targets for RECs (\textit{sc55.2027}), the results yield an overall average reduction of about -0.3\% on the equilibrium volumes of the Italian DA market, which is on 0.11\% greater than the average reduction of -0.19\% in the 45\% self-consumption scenario. This finding indicates that the increase in self-consumption rates for RECs deepens the reduction in equilibrium quantities.\\

Standalone systems are included in our REC modeling corresponding to the actual deployment status in Italy. The reducing effect of RECs' self-consumption on zonal market quantities also suggests that both the volumes of energy injected into the grid from renewable standalone plants and volumes from prosumers' plants within a REC could be offset by the self-consumption of REC prosumers. This finding contributes to the debate on the effects of the renewable-dominated system with high-electrified energy demand (incl. heating, cooling and transport) on wholesale market outcomes \citep{bottgerWholesaleElectricityPrices2022a,backeImpactEnergyCommunities2022,riazGenericDemandModel2019a}. Moreover, our sensitivity analysis shows that the standalone plants in REC configurations are responsible for the increase in energy injected during the summer early mornings and summer evening peaks, which has a positive effect on the overall system. Therefore, since the potentially negative impact of injections from standalone plants during periods of lower system demand \textcolor{black}{(i.e. daytime off-peak hours)} can be offset by prosumer self-consumption, we conclude that standalone facilities may have a \textcolor{black}{net} beneficial effect on the system \textcolor{black}{when incorporated into RECs.} \\

\underline{Based on these insights, we outline several policy recommendations:}
\begin{enumerate}
\item  The greater the capacity deployed by RECs, the more pronounced the impact of institutional members (public entities, small and medium enterprises, and non-profit organizations), due to their traditionally higher volumes of self-consumed energy during daytime off-peak hours compared to residential members (Figures \ref{fig:avg_daily_energy_jan}, \ref{fig:avg_daily_energy_apr}). The current Italian policy incentivizes smaller-capacity installations over larger ones by applying rigid capacity thresholds \textcolor{black}{(200kW--600kW--1MW,} see Figure \ref{fig:scheme_benefit_CER} and \nameref{sec: Appendix B}), irrespectively of a prosumer category. However, rather than injecting energy into the grid and \textcolor{black}{negatively affecting the system} during daytime off-peak hours, rooftop installations by institutional members can greatly benefit the system by covering the \textcolor{black}{aggregated loads of REC members, including prosumers but also consumers.} Therefore, the level of incentivization could be tailored to the extent of aggregated PV coverage of \textcolor{black}{the loads of all REC members}, rather than relying on rigid capacity thresholds. Specifically, higher premium tariff could be granted to RECs that install \textcolor{black}{aggregated} PV capacity that matches \textcolor{black}{aggregated loads of its members}. This measure could yield welfare gains \textcolor{black}{at both wholesale market and transmission grid levels}.
\item Institutional REC members inject more energy into the grid during peak hours in spring and summer than residential members, who tend to self-consume more energy during this period. Consequently, institutional REC members help meet peak demand in the system. In addition to public prosumers and non-profit organizations, small and medium-sized enterprises also exhibit load profiles that differ from those of residential prosumers. However, the Italian incentive scheme for shared energy does not make participation in RECs economically attractive for SMEs (see Koltunov et al. \citeyearpar{koltunov2025financing}). The 50\% reduction in the premium tariff for SMEs that benefit from capital subsidies constitutes one of the main obstacles. If such restrictive rules are relaxed, system-wide benefits from new RECs could increase.
\item According to the logic of Italian policy \citep{DecretoCACER2023} and regulation \citep{GSE_CACER_2025, schiavo6VirtualModel2022}, the economic systemic rationale behind RECs lies in the avoided use of the grid resulting from the simultaneous local injection and consumption of energy. We also demonstrate that reductions in market equilibrium volumes—particularly pronounced during the cold months—could help the system avoid RES curtailment when greater REC-driven capacity is deployed (i.e., 5 GW or more). This additional systemic rationale should be recognized by policy and thus valued/remunerated accordingly.
\item The \textcolor{black}{excessive} injection of energy from both standalone facilities and rooftop solar installations during daytime off-peak hours can be offset by local self-consumption \textcolor{black}{in winter and summer. Moreover, standalone facilities can be beneficial for the overall system during the steep increase of load in early morning hours and evening peak hours in summer.} Since feed-in tariffs and feed-in premiums for large-scale renewables have been progressively phased out in the EU, the incentive for energy communities can be used to support the roll-out of standalone plants with capacities of up to 1 MW, but in a more location-efficient and socially inclusive manner. Moreover, policymakers may consider raising the maximum capacity threshold (above 1 MW) for individual REC configurations in order to enhance economic viability for REC members as well as for third-party entities \textcolor{black}{that may operate standalone facilities.}\footnote{Participation in RECs is not open to large companies or companies whose primary activity belongs to the energy sector. Therefore, utilities and ESCOs tend to participate in "energy sharing" scheme as third-party entities. It means they cannot directly benefit from the capital subsidy and incentive on shared energy, but they can request lease payments if they own a renewable installation and/or charge configuration management fee to RECs. \textcolor{black}{Additionally, profits from selling energy into the grid are usually retained by these third-party entities.}}
\item  By adopting battery storage within energy communities, even greater reductions in wholesale market equilibrium quantities can be achieved. The discharge of community-owned batteries to meet the needs of REC members generates substantial systemic benefits during peak hours. In this way, costly and polluting thermal generation can be displaced by clean and social energy. Still, as was discussed by many authors \citep{soiniImpactProsumerBattery2020a, schickRoleImpactProsumers2022, chenRecoveringFixedCosts2023a, sarfaraziAggregationHouseholdsCommunity2020}, the specific modality of a storage system ultimately determines the scale and direction of this systemic impact. In Italy, capital subsidies are limited to 40\% of the costs incurred by RECs, a design feature that discourages the installation of costly storage. To avoid additional public spending while still promoting systemic efficiency, policy could specifically incentivize integrated photovoltaic–battery systems in regions with excess renewable generation (e.g., Southern Italy).
\item  RECs have historically been regarded as collective action initiatives created primarily by citizens and for citizens (e.g., energy cooperatives). Recently, Italian policy has also allowed institutional actors to participate in RECs. The load profiles of institutional prosumers enable greater systemic benefits from REC deployment - particularly when appropriately aggregated with residential loads - than those associated with multiple residential prosumers. Consequently, other countries may also consider permitting institutional actors to participate in RECs. \textcolor{black}{Policymakers could also stimulate the specific design of RECs in which the composition of institutional and residential prosumers, as well as standalone facilities, would neutralize negative wholesale market effects and deliver systemic benefits.}
\item \textcolor{black}{Finally, RECs create significant value by aggregating heterogeneous prosumer loads and generation, particularly when orchestrated by energy management systems. Local flexibility markets may unlock this value to a greater extent than wholesale markets. Therefore, policymakers and regulators should design flexibility markets that position RECs as central actors.}
\end{enumerate}

This paper is not exempt from limitations. First, the calibration of scenarios relies on publicly available data on REC uptake and profiles, standard self-consumption rates, and the announced policy targets by 2027. Second, behavioral aspects and demand-side management responses are not fully captured, although they may play a decisive role in the future and in dynamic adjustments in demand profiles, mirroring peculiar weather conditions and real-time modifications. Third, the projected impact of REC deployment is evaluated under the assumption that the capacities of other RES technologies remain fixed. This allows us to isolate the marginal contribution of the REC deployment to market equilibrium, although the actual system impacts could differ if broader RES additions were accounted for. Fourth, we disregard consumers who participate in RECs, although theoretically they might amend their consumption profiles to maximize the incentive for "shared energy" when joining RECs. Fifth, we do not explicitly model the evolving interaction between RECs and wholesale markets through aggregation or active market participation mechanisms. However, as stated in Introduction \ref{sec:intro}, this degree of coordination has not yet been implemented in the Italian energy communities, moreover, at a substantial scale. \textcolor{black}{Fifth, we do not explicitly model the evolving interaction between RECs and wholesale markets through aggregation or active market participation mechanisms. This simplifying assumption is consistent with the short-term, \textit{ceteris paribus} nature of our counterfactual framework, which isolates the one-directional impact of RECs on market equilibrium, but may not fully capture future market configurations with more active REC participation.}
\textcolor{black}{Sixth, the use of hourly granularity and standard load profiles inevitably introduces simplifications, especially in the representation of self-consumption patterns and imbalances. However, we argue that, at the level of aggregation considered, such simplifications do not materially affect the interpretation of the main results. Instead, they allow for a coherent comparison across scenarios and highlight meaningful differences across zones and time periods.}\\

Despite limitations, our methodology has a broad replication potential. The synthetic counterfactual approach can be adapted to other EU (and non-EU) electricity markets to assess the role of RECs (or other heterogeneous distributed energy resources) on day-ahead market equilibrium. This would allow for a broader comparative insight into how local RECs can reshape the aggregate electricity demand and supply, price signals, market resilience, and grid requirements. \textcolor{black}{Moreover, future modeling
exercises should address the impact of RECs on the potential flexibility market that are currently in design phase across EU countries. Future research should also explicitly consider the interaction between increased temporal granularity and higher penetration of self-consumption. In particular, once suitable high-frequency data become available, empirical strategies such as difference-in-differences or regression discontinuity designs could be employed to identify the causal effects of the transition to 15-minute settlement intervals. \textcolor{black}{Besides, the analysis of price effects is left for future research, where the impact of RECs can be assessed conditional on a broader set of exogenous factors influencing price formation.} The modeling framework proposed in this paper can be extended to accommodate such developments, providing a basis for more granular and forward-looking analyses.} \textcolor{black}{An extension of this work concerns the representation of demand profiles. Although standard load profiles are adequate for scenario-based analyses at an aggregate level, future research should incorporate 15-minute interval measurement data or sets of stochastic load profiles to better capture intra-hourly variability.} A robustness check under an “accelerated battery adoption scenario” could further refine our estimates, offering a valuable sensitivity benchmark for future policy analysis. Similarly to Boccard and Goertz \citeyearpar{boccardPowerTradingEnergy2025}, our future research could also adopt sensitivity tests based on differentiating PV sizes, which could suggest insights related to PV sizing for energy communities and its consequences for changes in a wholesale equilibrium. Future research can also address the differential impact of diverse prosumer categories on market outcomes in more depth. \textcolor{black}{Indeed, our methodology is applicable not only to the REC deployment but also to prosumers in general.} As was evident from our sensitivity analysis, this underexplored direction could deliver interesting insights. \\

Finally, our findings provide the timely inputs for energy and environmental policy-making. The diffusion of RECs could foster systemic efficiency, energy justice and independence, particularly in current times of geopolitical uncertainty and high energy costs, as highlighted during the 2022 energy crisis. \textcolor{black}{The policy innovation encouraging institutional members to participate in energy communities -- previously considered solely as groups of private citizens -- can improve day-ahead market efficiency as well as generate additional benefits for private citizens and other market stakeholders.} Nevertheless, REC's legacy will depend on complementary measures: efficient policy and regulation, harmonized governance frameworks, stronger support for storage and demand-side flexibility, and access to diverse financing resources.

\section*{Supplementary material}
\label{supplementary}
Our database of Italian RECs is available at the following GitHub link: \newline
https://github.com/maksym-koltunov/energy-communities-Italy.git

\section*{Acknowledgements}
The authors thank Prof. Alessandro Massi Pavan, from the Department of Engineering and Architecture and the Interdepartmental Center for Energy, Environment and Transport “Giacomo Ciamician”, University of Trieste, and Dr. Adriano Bisello from EURAC Research for their excellent organizational and networking support during this study. Furthermore, the authors gratefully acknowledge the stakeholders consulted for providing valuable insights on the participation of aggregators and ESCOs in the Italian power market, namely an ESCO managing several RECs (ForGreen Spa Società Benefit), as well as the Italian market operator (GME).

\newpage

\section*{Appendices}
\subsection*{Appendix A}
\begin{center}
    \textbf{\large Glossary}
\end{center}
\label{sec: Appendix A}
\begin{table}[htbp]
    \centering
    \begin{tabular}{c c}
    \hline \hline
    \textbf{Acronym} & \textbf{Description} \\
    \hline
    BESS & Battery Energy Storage System \\
    DA & Day-ahead \\
    CO\textsubscript{2} & Carbon dioxide \\
    DER & Distributed Energy Resources \\
    DSO & Distribution System Operator \\
    EU & European Union \\
    GW & Gigawatt \\
    GSE & Gestore Servizi Energetici \\
    IEM & Internal Electricity Market Directive \\
    MW & Megawatt \\
    MWh & Megawatt per hour \\
    LCOE & Levelized Cost Of Energy \\
    MOE & Merit order effect \\
    NGO & Non-governmental Organisation \\
    NPO & Non-profit Organisation \\
    PV & Photovoltaic \\
    REC & Renewable Energy Community \\
    RED-II & Second Renewable Energy Directive\\
    RES & Renewable Energy Sources \\
    SME & Small and medium-sized enterprises \\
    TSO & Transmission System Operator \\
    USA & United States of America \\
    JARSC & Jointly acting renewable self-consumers \\
    \hline
    \end{tabular}
\end{table}

\subsection*{Appendix B}
\label{sec: Appendix B}
\begin{center}
    \textbf{\large Shared energy and cashback calculation}
    \end{center}

In Italian regulation, shared energy refers to the minimum amount between the total energy injected into the public grid by all EC members within a given hour and the total energy withdrawn (consumed) by all EC members within the same hour.

The generalized formula for calculating shared energy is:
\begin{equation}
E_{\mathrm{sh}}(h) = \min \left\{ \sum_{i=1}^{n} E^{\mathrm{in}}_{i,h} \; ; \; \sum_{i=1}^{n} E^{\mathrm{out}}_{i,h} \right\}
\end{equation}
where:
\begin{description}[noitemsep]
  \item[$E_{\mathrm{sh}}$:] the shared energy,
  \item[$E^{\mathrm{out}}_{i,h}$:] the energy withdrawn from the grid in a given hour $h$ by the $i$-th member,
  \item[$E^{\mathrm{in}}_{i,h}$:] the energy fed into the grid in a given hour $h$ by the $i$-th member.
\end{description}

The cashback logics is given by ARERA \citeyearpar{TIAD2022} and Technical Rules from GSE \citeyearpar{GSE_CACER_2025}. It can be described as:
\begin{equation}
\mathrm{Cashback} = E_{\mathrm{inj}} \, P_{\mathrm{sell}} \;+\; E_{\mathrm{sh}} \, TIP \;+\; V_{\mathrm{agus}} \;-\; \mathrm{GSE\;costs}
\end{equation}
where:
\begin{description}[noitemsep]
  \item[$P_{\mathrm{sell}}$:] price for which GSE is purchasing renewable energy from eligible subjects. Regulated by ARERA 280/07 and equivalent to the \emph{Prezzo Minimo Garantito} (PMG).
  \item[TIP :] premium tariff for shared energy.
  \item[$V_{\mathrm{agus}}$:] valorization of avoided grid usage due to self-consumption.
  \item[GSE costs:] administrative costs incurred by GSE related to cashback calculation and remuneration.
\end{description}

It is critical to consider how distinct components of the cashback are calculated. The premium tariff itself is not a single value; the basic premium tariff for an EC consists of fixed and variable components:
\begin{equation}
\mathit{TIP} =
\begin{cases}
  \min\bigl(80 + \max(0;\, 180 - P_{z});\, 120\bigr) & \text{if } P \leqslant 200,\\[2pt]
  \min\bigl(70 + \max(0;\, 180 - P_{z});\, 110\bigr) & \text{if } 200 < P \leqslant 600,\\[2pt]
  \min\bigl(60 + \max(0;\, 180 - P_{z});\, 100\bigr) & \text{if } P > 600.
\end{cases}
\end{equation}

\noindent{where}
\begin{itemize}[leftmargin=1.5em]
    \item 80-70-60 are \textit{fixed components} of a premium tariff that according to CER Decree stands at 60--80~\euro/MWh depending on the asset capacity (if $P < 200$ equals 80; if $200 < P < 600$ equals 70; if $P > 600$ equals 60). PV systems in central and northern Italy get an extra premium of 4 and 10~\euro/MWh respectively.

    \item $\max(0;\,180 - P_{z})$ is a \textit{variable component} with $P_{z}$ being the hourly zonal price of electricity. In the event of zonal prices being higher than 180~\euro/MWh the \textit{variable component} is counted as 0~\euro/MWh. This is made to prevent excessive premium during the periods of wholesale price spikes.

    \item 120-110-100 are components that vary by installed capacity. If $P > 600$ equals 100~\euro/MWh, if $200 < P < 600$ equals 110~\euro/MWh, if $P > 600$ equals 120~\euro/MWh. In the event of extremely low wholesale price $P_{z}$ this parameter prohibits exceeding the premium tariff beyond 100--120~\euro/MWh.
\end{itemize}
The valorization of avoided grid usage due to self-consumption \textit{Vagus} includes two components: the reduction in grid costs due to self-consumption and the reduction in costs related to decreased grid losses from self-consumption. JARSCs are compensated for both components, whereas RECs receive compensation only for the reduction in grid costs due to self-consumption. This distinction arises because RECs still rely on the distribution grid for energy sharing, thereby incurring grid losses. In contrast, JARSCs operate within a single condominium, bypassing the distribution grid. The \textit{Vagus} for JARSCs and RECs can be calculated using the following formulas:

\begin{equation}
\text{JARSC: } V_{agus} = T + L
\end{equation}
\begin{equation}
\text{REC: } V_{agus} = T
\end{equation}

where
\begin{itemize}[noitemsep]
    \item $T$ is reduction of grid costs due to self-consumption
    \item $L$ is reduction of costs related to decreased grid losses due to self-consumption
\end{itemize}
Reduction of grid costs due to self-consumption (T) is calculated \citep{schiavo6VirtualModel2022} by formula:
\begin{equation}
T = E_{sh} \times \bigl( TRASe + \max(BTAUm) \bigr)
\end{equation}
where
\begin{itemize}[noitemsep]
    \item \textit{TRASe} is a sum of transmission tariff for LV household consumers,
    \item \textit{BTAUm} is a distribution tariff for LV commercial and industrial consumers
\end{itemize}

Reduction of costs related to decreased grid losses due to self-consumption ($L$) is calculated \citep{schiavo6VirtualModel2022} by formula:
\begin{equation}
L = \sum_{i,h} \bigl( E_{sh} \times c_{agl} \times P_{z} \bigr)
\end{equation}
where
\begin{itemize}[noitemsep]
    \item $c_{agl}$ is a coefficient for avoided grid losses of the voltage level to which generator/s are connected (2.6\% for generators in LV grid, 1.2\% for generators in MV grid),
    \item $P_{z}$ is the hourly zonal energy price of the day-ahead market,
    \item $E_{sh}$ is the energy shared by the $i$-th member
\end{itemize}

By identifying all components of the cashback, we can calculate the actual economic benefit for EC members:
\begin{equation}
\textit{Benefit} = E_{sc}(P_{buy}) + [E_{inj}(P_{sell}) + E_{sh}(P_{r}) + V_{agus} - GSE_{costs}]
\end{equation}

where
\begin{itemize}[noitemsep]
    \item $E_{sc}$ is an amount of self-consumed energy by common loads for JARSCs and by prosumers for EC,
    \item $P_{buy}$ is a retail price that a member of EC pays to its retailer.
\end{itemize}

\subsection*{Appendix C}
\label{sec:Appendix C}
\begin{table}[H]
\begin{threeparttable}
\centering
\caption{Economic impact of RECs deployment on other market stakeholders}
\footnotesize
\renewcommand{\arraystretch}{1.3}
\begin{tabular}{|p{10cm}|p{2cm}|p{4cm}|}
\hline
\textbf{Specific impacts of RECs on market stakeholders} &
\textbf{Direction of impact} &
\textbf{Reference} \\ \hline
\multicolumn{3}{|c|}{\textbf{On Generators}} \\ \hline
RECs deployment reduces the need for additional large-scale generation due to a merit-order effect (MOE) of renewables therefore reducing conventional producers’ surplus. & Detriment & Robinson \& Guayo \citeyearpar{robinson5AlignmentEnergy2022}, Backe et al. \citeyearpar{backeImpactEnergyCommunities2022} \\ \hline
RECs operating BESS could displace fossil fuel generators at a faster scale than large renewable generators without BESS due to MOE at peak demand periods, therefore reducing conventional producers’ surplus. & Detriment (generators) / Benefit (system) & Robinson \& Guayo \citeyearpar{robinson5AlignmentEnergy2022} \\ \hline
RECs operating BESS reduce RES curtailment due to self-consumption at peak production congested times which allows to decrease supply of a distributed generation when grid needs it most (in turn, large renewable generators do not need to be curtailed thus preserving revenues). & Benefit & Backe et al. \citeyearpar{backeImpactEnergyCommunities2022} \\ \hline

\multicolumn{3}{|c|}{\textbf{On Retailers}} \\ \hline
Aggregator RECs operating BESS achieve cost reduction by directly participating in a wholesale market bypassing retailers thus shrinking their potential revenues. & Detriment & Faia et al. \citeyearpar{faiaProsumerCommunityPortfolio2021} \\ \hline
When a REC acts as an intermediary between consumers and retailer, the latter lose ability to diversify risks by differentiating over tariffs to different consumer groups. & Detriment & Biggar \& Hesamzadeh \citeyearpar{biggar8EnergyCommunities2022} \\ \hline
Diminished sales from REC members elicit tariffs inflating for remaining customers, which in turn trigger latter to become more self-sufficient, too. Situation leads to a financial ``death spiral'' of retailers. & Detriment & Parag \& Sovacool \citeyearpar{paragElectricityMarketDesign2016a}, Sarfarazi et al. \citeyearpar{sarfaraziAggregationHouseholdsCommunity2020} \\ \hline
If a community energy storage is owned by a retailer, then the real-time pricing optimized based on behavior of REC members could yield profits for the retailer while not increasing costs for any type of REC members and delivering profits for flexible REC members. However, real time tariffs need implementation of EMS unavailable at scale in many countries. & Benefit & Sarfarazi et al. \citeyearpar{sarfaraziAggregationHouseholdsCommunity2020} \\ \hline

\multicolumn{3}{|c|}{\textbf{On DSOs}} \\ \hline
When a REC acts as a united entity supplied only by a single external retailer, effect of a non-simultaneity\tnote{a} of contracted capacity is eliminated, which in turn reduces overall payment to DSOs\tnote{b}. & Benefit & Biggar \& Hesamzadeh \citeyearpar{biggar8EnergyCommunities2022}, Robinson \& Guayo \citeyearpar{robinson5AlignmentEnergy2022} \\ \hline
Electrotechnical criterion for defining perimeter of connection based on connection to same HV/MV substation (appr. 10000 PODs) and geographical criterion based on zip-code (appr. 1000 PODs) simplify DSO interaction with RECs. & Benefit & Del Pizzo et al. \citeyearpar{delpizzo18ItalianEnergy2022} \\ \hline
Electrotechnical criterion of defining perimeter of connection based on connection to same MV/LV substation (appr. 70 PODs) and geographical criterion based on municipality belonging (appr. 10 to 850000 PODs) impedes DSO interaction with numerous or heterogeneous RECs. & Detriment & Del Pizzo et al. \citeyearpar{delpizzo18ItalianEnergy2022} \\ \hline
Presence of RECs in some markets can cause operability issues and grid disruption due to a more complicated control and management schemes. & Detriment & Parag \& Sovacool \citeyearpar{paragElectricityMarketDesign2016a} \\ \hline
Possibility to automatically detect and respond to actual and emerging grid problems through aggregated RECs (similar to VPPs), that may increase system’s resilience and decrease renewable energy oversupply concerns. & Benefit & Parag \& Sovacool \citeyearpar{paragElectricityMarketDesign2016a} \\ \hline

\multicolumn{3}{|c|}{\textbf{On non-member consumers}} \\ \hline
DSOs can incur revenue losses due to decreased volumetric (decreased electricity purchasing by RECs) and/or fixed (non-simultaneity effect) network payments as well as possible subsidized exemption of RECs from the network payments. These revenue losses would typically be shifted onto non-member consumers. & Detriment & Biggar \& Hesamzadeh \citeyearpar{biggar8EnergyCommunities2022}, Del Pizzo et al. \citeyearpar{delpizzo18ItalianEnergy2022}, Robinson \& Guayo \citeyearpar{robinson5AlignmentEnergy2022}, Sarfarazi et al. \citeyearpar{sarfaraziAggregationHouseholdsCommunity2020} \\ \hline
Take-up of renewable energy supplied by RECs and all the subsequent MOE decrease energy prices, thereby non-members can greatly benefit. & Benefit & Biggar \& Hesamzadeh \citeyearpar{biggar8EnergyCommunities2022} \\ \hline
If RECs obtain implicit subsidy from a government, the costs are usually cross subsidized to non-members through retail bills. & Detriment & Robinson \& Guayo \citeyearpar{robinson5AlignmentEnergy2022} \\ \hline
\end{tabular}
\begin{tablenotes}[flushleft]
\footnotesize
\item[a] Example: if 10 consumers each contract a 10-kW capacity-based network charge, the total contracted capacity would be 100 kW. However, due to non-simultaneity, regulation typically requires that the system only meets a combined demand of 75 kW, effectively overcharging consumers by 25 kW, which covers fixed network costs. When consumers form a REC, they contract only for 75 kW, benefiting from the non-simultaneity effect themselves and avoiding the 25-kW surcharge \citep{robinson5AlignmentEnergy2022}.
\item[b] This impact can backfire to non-member consumers because DSOs, as regulated monopolies, would typically shift the reduced revenues to non-member consumer bills.
\end{tablenotes}
\end{threeparttable}
\end{table}

\subsection*{Appendix D}
\label{Appendix D}
{\textbf{\large\centerline{Impact of prosumers on electricity systems}} \\
We found several studies that analyze the systemic impact of prosumers. While two studies are theoretical review works that provide an extensive narrative of benefits and challenges \citep{robinsonChapter11Why2023, simshauserChapter3Sunshine2023}, the remaining studies rely on optimization models.\\
\\
Among theoretical works, Simshauser et al. \citeyearpar{simshauserChapter3Sunshine2023} illustrate the case of Queensland state, Australia, which has the highest PV rooftop adoption rate in the world, while Robinson and Arcos-Vargas \citeyearpar{robinsonChapter11Why2023} present both positive and negative effects of prosumer proliferation, focusing on Spain. Both studies outline many implications of prosumer proliferation, which often coincide with REC impacts. Some of the most pronounced implications include an adverse effect on conventional generators and challenges faced by DSOs due to a voltage rises (e.g., damage to customers’ electrical appliances). Another drawback is that retailers rapidly lose market share as prosumers penetration increases. Studies also point to an ambiguous impact on non-prosumers, who on one hand experience rising bills due to cross-subsidies and increased distribution charges, and on the other hand benefit from the decrease in the energy component of their retail bill as the result of the MOE and reductions in fuel costs. For instance, new PV installations in Spain resulted in a reduction of the wholesale electricity price of 0.01 euro for every 25 MWp installed, which, when aggregated across the overall market in 2021, led to user savings of more than 100,000 euros per year \citep{robinsonChapter11Why2023}. In Australia, another positive impact on non-prosumers was significantly lower installation costs due to the substantial growth of PV installation companies. In addition, DERs can lead to an increased need for ancillary services, which in turn reduces the price of these services due to greater market liquidity. Robinson and Arcos-Vargas \citeyearpar{robinsonChapter11Why2023} argue that when prosumers and producers are aggregated (similar to RECs), costs of the distribution grid can drop because aggregated agents can provide flexibility services to the network — something that is almost impossible with disorganized individual installations. The proliferation of individual prosumers can decrease ohmic losses in the distribution grid, but only up to a certain level; beyond that, significant reverse flows occur, increasing losses again. Nevertheless, losses in such situations remain lower than before any DERs were deployed (2023, p. 135). An ambiguous impact, as underscored by Robinson and Arcos-Vargas \citeyearpar{robinsonChapter11Why2023}, occurs in terms of the security of supply when DERs do not utilize storage and are not aggregated. In this scenario, network costs can increase if a system is planned from an N-1 deterministic perspective; however, if it is planned from a probabilistic perspective, prosumers could improve security of supply even without storage or aggregation.\\
\\
The first group of empirical studies on the systemic impact of prosumers attempts to quantify the associated challenges. For example, Schick and Hufendieck \citep{schickAssessmentRegulatoryFramework2023} investigate the distributional spatial effect of the German feed-in-tariff during the period 2000-2021. Aggregated across Germany, the feed-in-tariff led to a cost shift of more than 500 million euros onto traditional consumer households. In 2021, maximization of self-consumption accounted for approximately half of this total effect. Tsybina et al. \citeyearpar{tsybinaEffectProsumerDuality2023a} explore strategic behaviour of prosumers (exercising market power) and their response to the allocation of network losses—either to demand-side or the supply-side—as well as the impact of net metering policies. The authors determine that prosumers sell more electricity when losses are allocated to the demand base, whereas when losses are allocated to the supply base, prosumers sell less electricity. Another key observation is that lower wholesale equilibrium prices occur when network losses are allocated to the demand side due to two main factors. First, incorporating losses into the retail price (demand side) keeps selling prices higher and incentivizes prosumers to inject electricity into the grid. Second, higher retail prices encourage both consumers and prosumers to adjust their consumption patterns, leading to a more efficient use of energy, thereby reducing peak demand and the need for expensive peaking plants. Chen et al. \citeyearpar{chenRecoveringFixedCosts2023a} compare net-metering, net-billing, and benchmark policies\footnote{Benchmark policy in this study assumes prosumers selling at a wholesale equilibrium price and buying at a retail price, therefore same tariffs are implied for prosumers as for other agents.}, examining their differential effects on various aspects of system welfare. The authors find that social surplus under the benchmark policy is significantly higher than under the other policies, making the benchmark policy the most welfare-enhancing. Under net metering, the transmission tariff would be 33\% higher compared to the benchmark case, whereas the transmission tariff under net billing is closer to the benchmark case, making net billing the second-best solution. Another key finding is that wholesale social surplus (excluding prosumer surplus) deteriorates when prosumers saturate the node due to a greater number of consumers converting to prosumers—an indication of the “death spiral” effect. Chen et al. \citeyearpar{chenDeathSpiralTransmission2023a} refine the optimization model used in their previous study to investigate the impact of prosumers on transmission charges and social surpluses under the benchmark case alone. The authors assume different deployment levels of aggregated prosumers and analyse scenarios of perfect and imperfect competition, with the latter allowing prosumers to exhibit market power. The study reveals that wholesale prices decline under both scenarios due to increased renewable dispatch and reduced demand by prosumers. However, under imperfect competition — where prosumers strategically maximize their individual economic optimum — a significant increase in transmission charge occurs at all levels of deployment, particularly in scenarios with a high saturation of prosumers at nodes. This, in turn, reduces overall welfare. \textcolor{black}{Finally, Galves-Garcia et al. \citeyearpar{galvez2023prosumers} investigate prosumer exchange with the market through aggregators. Authors test their optimization model on a 5-nodal simplified example of the US PJM market. They found that the peak demand is reduced by 4.19\% and 3.14\% with flexible prosumer demand that is managed by an aggregator.}\\
\\
The second group of empirical studies examines solutions to mitigate network costs shifting onto non-prosumers. Schick et al. \citeyearpar{schickImpactNetworkCharge2021} demonstrate that network allocation schemes based on peak-coincident network capacity utilization can more effectively incentivize distribution network-oriented behaviour while ensuring a fairer distribution of financial burden between prosuming and non-prosuming households compared to volumetric network charges. A subsequent study by the same authors \citeyearpar{schickRoleImpactProsumers2022} finds that higher self-consumption, when operated at least partially in a grid-beneficial manner (e.g., coupled with storage capable of providing flexibility service), can enhance RES integration and reduce CO\textsubscript{2} emissions while avoiding cost shifting onto consumers.  This finding suggests that dispersed prosumers could contribute more effectively to the grid if coordinated through a REC. On the other hand, when prosumers focus solely on maximizing their individual economic optimum— without considering system economic optimum (e.g., when storage operation is entirely inflexible)—RES integration could decrease, leading to a substantial rise in system costs and CO\textsubscript{2} emissions \citep{schickRoleImpactProsumers2022}. These findings align with those of Chen et al. \citeyearpar{chenRecoveringFixedCosts2023a}.\\
\\
The third group of studies explores future scenarios characterized by a high penetration of renewables, including DERs. Böttger and Härtel \citeyearpar{bottgerWholesaleElectricityPrices2022a} investigate hypothetical German power day-ahead market in 2050, assuming the deployment of carbon-neutral electricity/heat/transport systems. Importantly, their study considers the role of various novel demand-electrification technologies, which contribute to both supply and demand. The authors find that variable RES market values can be stabilised by power demand from diverse electrification applications, including flexible storage, power-to-gas, and power-to-heat (heat pumps). Consequently, a fully renewable future does not necessarily imply the “cannibalization effect” and highly volatile wholesale prices. Soini et al. \citeyearpar{soiniImpactProsumerBattery2020a} investigate the impact of prosumers’ BESS on power supply costs  in the Swiss electricity market for 2030, comparing it to the status quo in 2015. Their findings indicate that when BESS operation is optimized from a power system perspective—through time-of-use tariffs, grid charging, power exchange minimization, and households’ aggregation via RECs—substantial cost savings can be achieved. These savings primarily result from the reduced generation requirements and the substitution of pumped-hydro storage with more efficient BESS. Conversely, when fully independent households optimize their self-consumption, costs increase. This outcome is aligned with the findings of both Schick et al. \citeyearpar{schickRoleImpactProsumers2022} and Chen et al. \citeyearpar{chenRecoveringFixedCosts2023a}.
Finally, Riaz et al. \citeyearpar{riazGenericDemandModel2019a} analyze the effect of large-scale prosumer aggregation (including BESS) on wholesale demand positions and load profiles, with a specific focus on loadability\footnote{Loadability - maximum amount of a load that a system can support before it collapses.} and voltage stability. Their study reveals that the increased prosumer-BESS participation  smooths demand profiles, enhances loadability and voltage stability, and reduces gas power plants utilization—-thus lowering wholesale electricity prices. However, in scenarios of low demand and excess RES generation, these benefits do not occur. In contrast, a higher RES penetration without BESS leads to reverse flows and a reduction in reactive power support capability, ultimately lowering system stability margins.

\subsection*{Appendix E}
\label{Appendix E} {\textbf{\small{Percentage relative difference
between actual and counterfactual average hourly quantities in NORD
for each month during weekdays and weekends for BU and HW
scenarios.}}
\begin{figure}[htbp]
\captionsetup{font=footnotesize}
\centering
\includegraphics[width=0.6\linewidth]{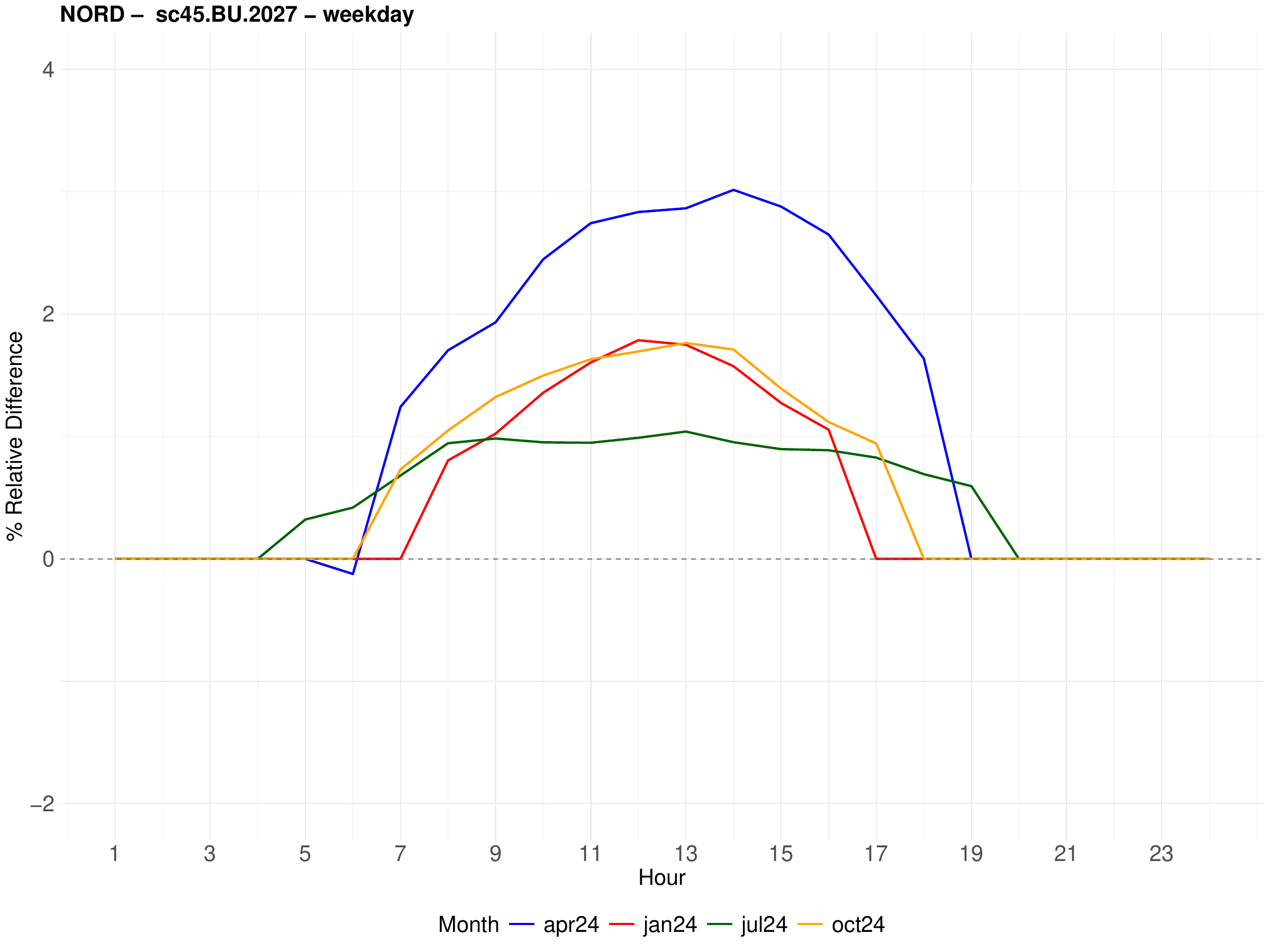}
\caption{\label{fig:NORD_BAU_rel_diff_by_month_weekday} Percentage relative difference between actual and counterfactual average hourly quantities in NORD for each month during weekdays. Scenario: \textit{sc45.BU.2027}.}
\end{figure}

\begin{figure}[htbp]
\captionsetup{font=footnotesize}
\centering
\includegraphics[width=0.6\linewidth]{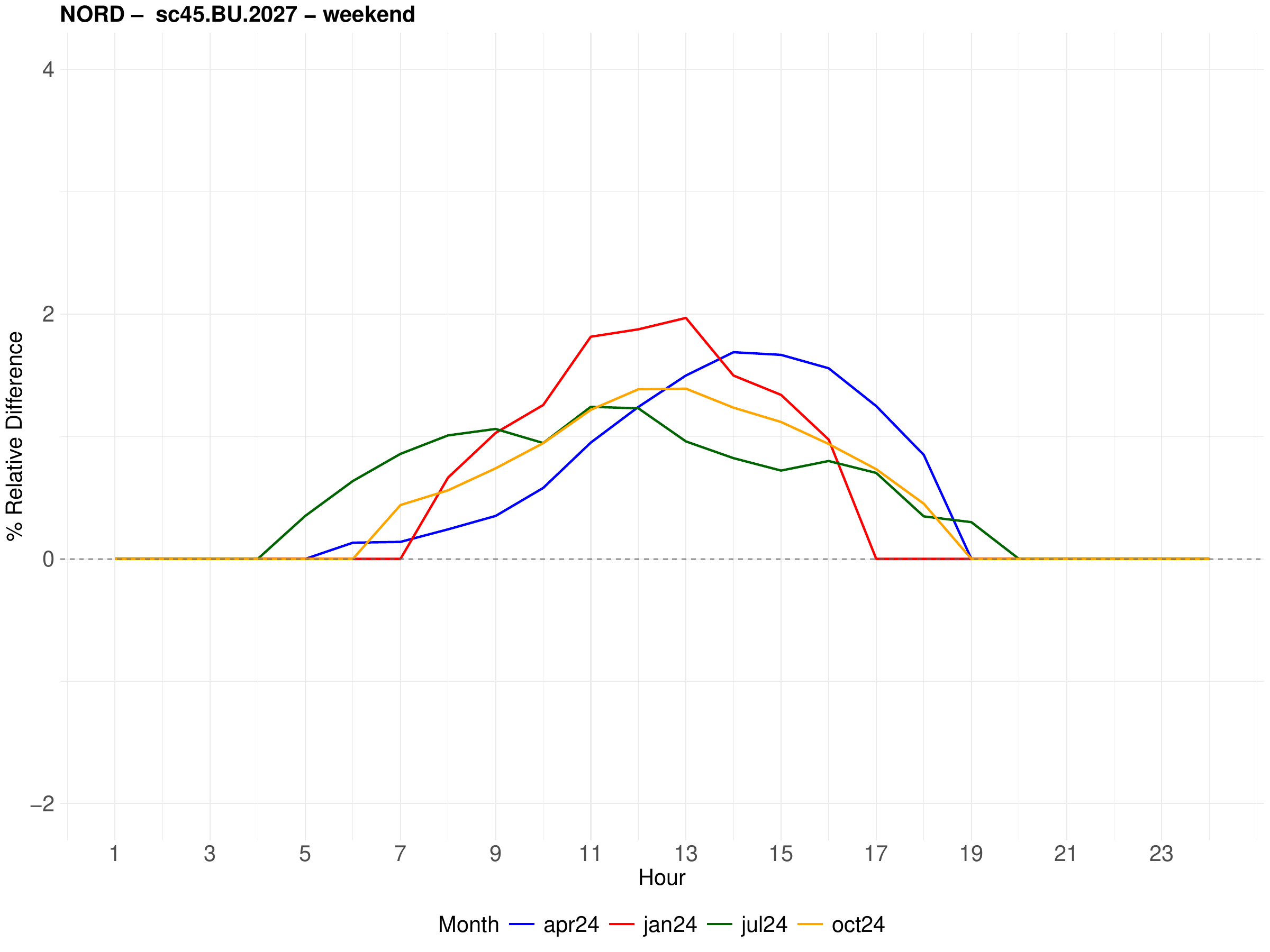}
\caption{\label{fig:NORD_BAU_rel_diff_by_month_weekend} Percentage relative difference between actual and counterfactual average hourly quantities in NORD for each month during weekends. Scenario: \textit{sc45.BU.2027}.}
\end{figure}

\begin{figure}[htbp]
\captionsetup{font=footnotesize}
\centering
\includegraphics[width=0.6\linewidth]{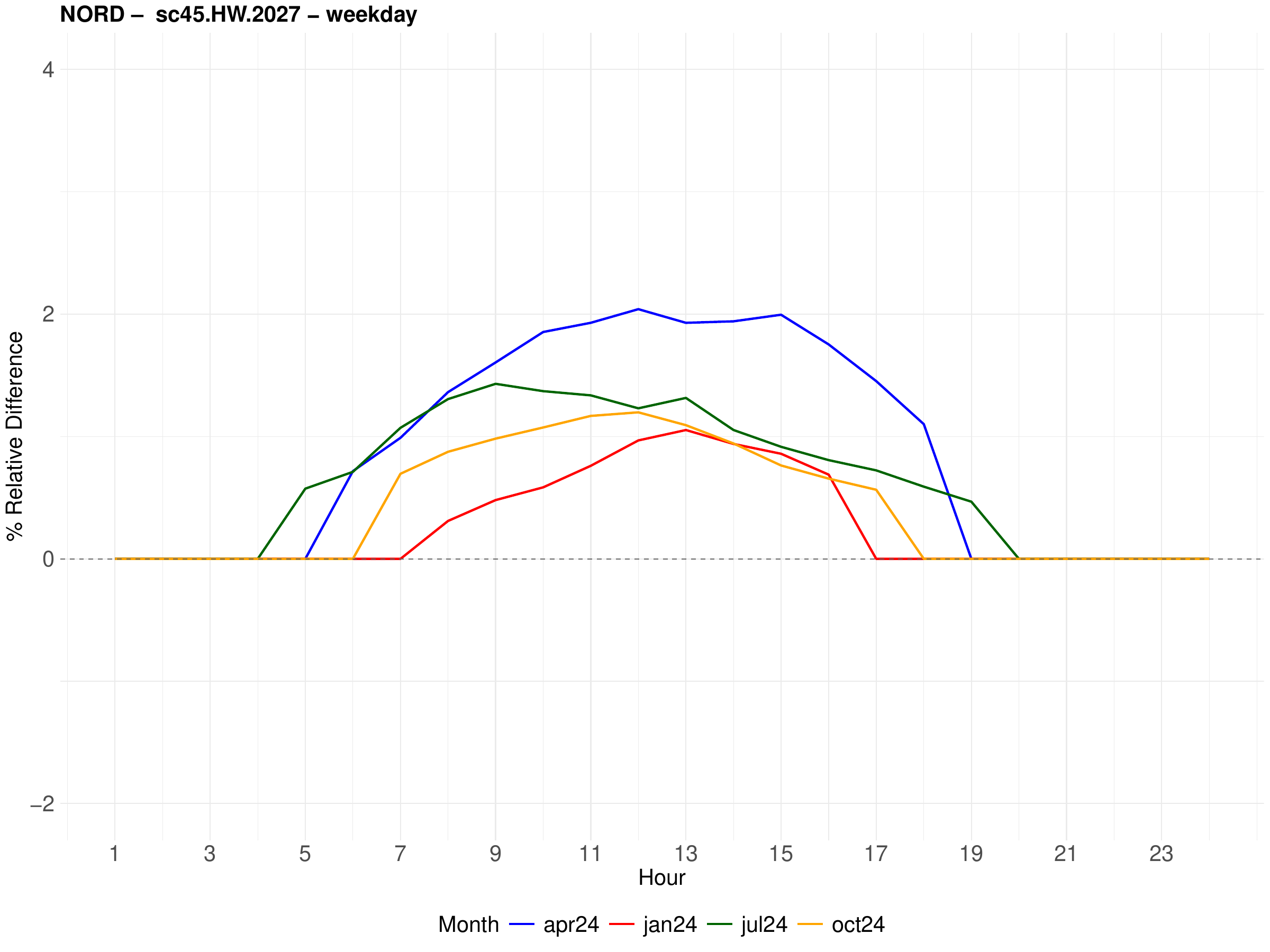}
\caption{\label{fig:NORD_HW_rel_diff_by_month_weekday} Percentage relative difference between actual and counterfactual average hourly quantities in NORD for each month during weekdays. Scenario: \textit{sc45.HW.2027}.}
\end{figure}

\begin{figure}[htbp]
\captionsetup{font=footnotesize}
\centering
\includegraphics[width=0.6\linewidth]{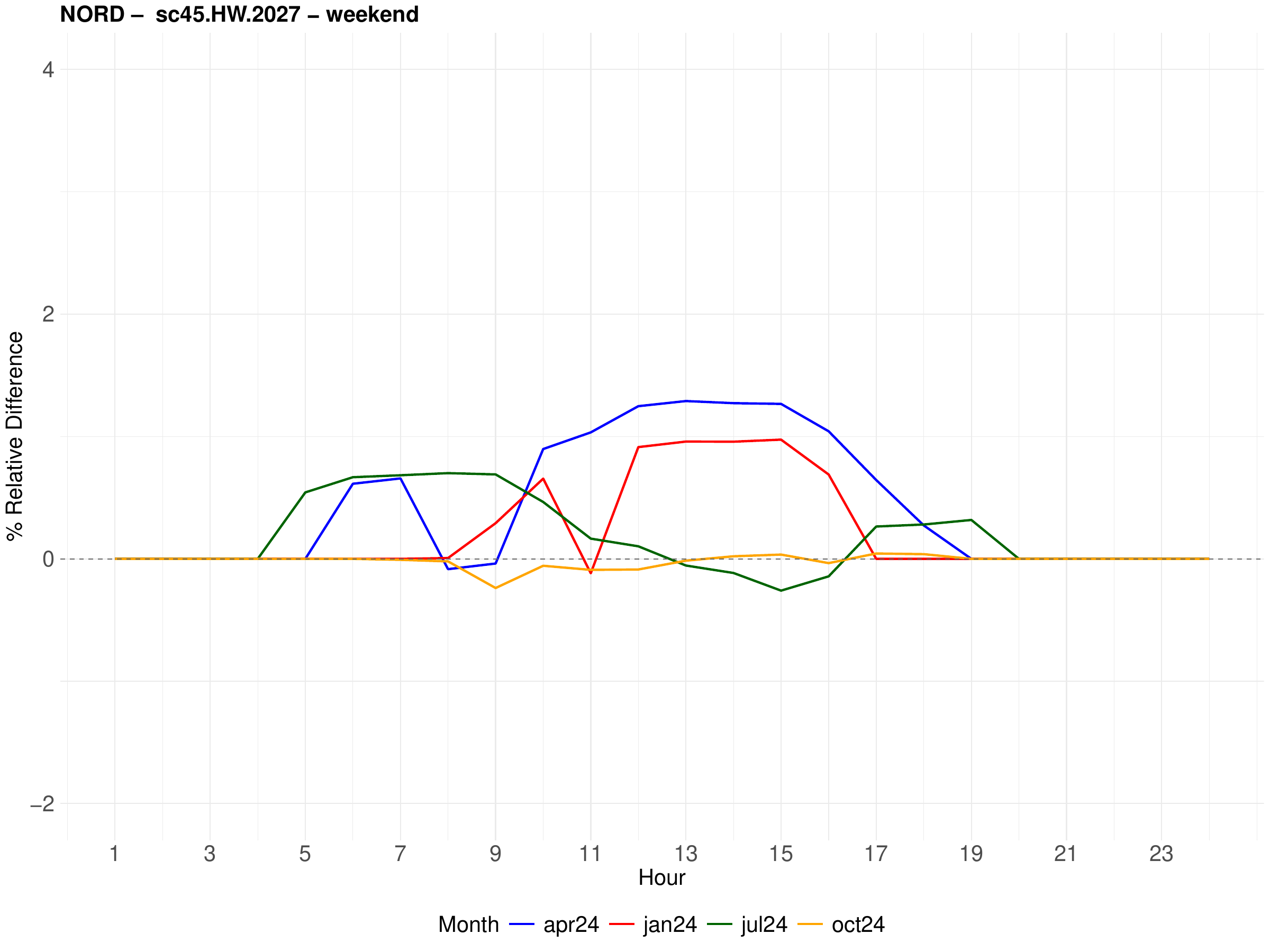}
\caption{\label{fig:NORD_HW_rel_diff_by_month_weekend} Percentage relative difference between actual and counterfactual average hourly quantities in NORD for each month during weekends. Scenario: \textit{sc45.HW.2027}.}
\end{figure}

\FloatBarrier
\bibliographystyle{elsarticle-harv}
\bibliography{references}
\end{document}